\documentclass[prd,aps,nofootinbib,preprintnumbers,floatfix]{revtex4}
\usepackage{amsmath,amssymb}
\usepackage{epsfig}
\usepackage{graphicx}
\usepackage{subfigure}

\def\lsim{\mathrel{\hbox{\rlap{\hbox{\lower4pt\hbox{$\sim$}}}\hbox{$<$}}}}
\def\gsim{\mathrel{\hbox{\rlap{\hbox{\lower4pt\hbox{$\sim$}}}\hbox{$>$}}}}

\def\be{\begin{equation}}
\def\ee{\end{equation}}
\def\bea{\begin{eqnarray}}
\def\eea{\end{eqnarray}}

\newcommand{\Eqref}[1]{Eq.~(\ref{#1})}

\newcommand{\bk}{{\bf k}}

\newcommand{\HH}{H}
\newcommand{\Om}{\Omega}
\newcommand{\OmG}{\Omega_{gw}}

\newcommand{\al}{\alpha}
\newcommand{\rh}[1]{\rho_{#1}}
\newcommand{\vev}[1]{\langle #1 \rangle}

\begin{document}
\title{Cosmological Backgrounds of Gravitational Waves and eLISA/NGO:\\
Phase Transitions, Cosmic Strings and Other Sources}
\author{Pierre Bin\'etruy$^{1}$, Alejandro Boh\'e$^{2}$, Chiara Caprini$^{3}$, Jean-Fran\c{c}ois Dufaux$^{1}$}
\affiliation{$^{1}$APC, Univ. Paris Diderot, CNRS/IN2P3, CEA/Irfu, Obs. de Paris, Sorbonne Paris Cit\'e, France\\
$^{2}$UPMC-CNRS, UMR7095, Institut d'Astrophysique de Paris, F-75014, Paris,  France\\
$^{3}$Institut de Physique Th\'eorique, CEA, IPhT, CNRS, URA 2306, F-91191Gif/Yvette Cedex, France}

\begin{abstract}
We review several cosmological backgrounds of gravitational waves accessible to direct-detection experiments, with a special emphasis on those backgrounds due to first-order phase transitions and networks of cosmic (super-)strings. For these two particular sources, we revisit in detail the computation of the gravitational wave background and improve the results of previous works in the literature. We apply our results to identify the scientific potential of the NGO/eLISA mission of ESA regarding the detectability of cosmological backgrounds.
\end{abstract}

\maketitle

\tableofcontents


\section{Introduction}

Gravitational waves represent one of the most promising messengers to probe the 
Universe in some of its most fundamental manifestations, in particular the 
violent phenomena associated with the last stages of the evolution of stars, or 
with the collisions of galaxies and the early history of the Universe. Various 
frequency windows of gravitational radiation are now being targeted by different
experimental means, in particular $[10^{-18}$Hz, $10^{-16}$Hz$]$ through the polarization of 
the Cosmic Microwave Background, $[10^{-9}$Hz, $10^{-7}$Hz$]$ through millisecond 
pulsar timing, $[10^{-4}$Hz, $10^{-1}$Hz$]$ with space interferometers and 
$[10$Hz, $10^3$Hz$]$ with ground interferometers. The LISA program has been 
pursuing the goal of detecting gravitational waves in the sub-Hz region for 
more than ten years as a joint venture between ESA and NASA. A new development 
in this program was the decision by Europe early 2011 to move forward on its own. 
The new Europe-only mission, named New Gravitational wave Observatory (NGO) by ESA 
and eLISA (evolved LISA) by the community\footnote{In what follows, we use the latter 
name.}, departs significantly from the previous ESA-NASA mission. It is thus important 
to assess its scientific program. This important endeavour has been undertaken by 
various groups in the last months, covering the full spectrum of the scientific program 
of the new mission~\cite{yellowbook}. We report here on  the assessment of the scientific 
potential of the new mission regarding the measurement of cosmological backgrounds. Although 
not the main motivation for the mission, the detection of such backgrounds would be a major 
discovery. Indeed, gravitational waves (GW) penetrate all of cosmic history, which allows GW 
detectors to explore scales, epochs, and new physical effects not accessible in any other 
way~\cite{allen, hoganRev,maggiore,buonanno}. Two important mechanisms for generating stochastic 
backgrounds are phase transitions in the early Universe and cosmic strings.

Gravitational waves produced after the Big Bang form a fossil radiation:
expansion prevents them from coming in thermal equilibrium with the other 
components because of the weakness of the gravitational interaction. Important 
information on the first instants of the Universe is thus imprinted in these 
relics and can be decoded. The mechanical effect of expansion is simply to 
redshift the corresponding frequency (and amplitude). Assuming that the wavelength 
is set by the apparent horizon size $c/H_* = c \, (a/\dot a)_*$ at the time of production 
(when the temperature of the Universe is $T_*$), the redshifted frequency today is
\begin{equation}
\label{eq:1}
f \, \approx \, 10^{-4} \, \hbox{Hz} \, \sqrt{ H_*(t) \times \frac{1 \hbox{mm}}{c}} 
\, \approx \, 10^{-4} \, \hbox{Hz} \, \left(\frac{k_B T_*}{1 \hbox{TeV}}\right)\,.
\end{equation}
Thus, eLISA frequency band of about $0.1$ mHz to $100$ mHz
today corresponds to the horizon at and beyond the
Terascale frontier of fundamental physics.  This
allows eLISA to probe bulk motions at times about
$[3\cdot 10^{-18}, 3\cdot 10^{-10}]$ seconds after the Big Bang, a period not
directly accessible with any other technique.  Taking a typical broad
spectrum into account, eLISA has the sensitivity to detect
cosmological backgrounds caused by new physics active in the range of
energy from $0.1$ TeV to $1000$ TeV, if more than a modest fraction
of about $10^{-5}$ of the total energy density is converted to
gravitational radiation at the time of production. A
standard example of new physics is a first-order phase transition
resulting in bubble nucleation and growth, and subsequent bubble
collisions and turbulence.

Phase transitions also often lead to the formation of one-dimensional 
topological defects known as cosmic strings. Among possible topological
defects, cosmic strings are unique from a cosmological point of view because,
whereas their relative energy density should grow with the expansion, they 
interact and form loops which decay into gravitational waves. Thus cosmic
strings tend to form networks with a typical scaling behaviour, losing
energy mainly through gravitational radiation with a very broad
and uniquely identifiable spectrum. Besides topological defects in field theories,
cosmic strings may also occur as fundamental objects of string theory, 
the theory that is aiming to provide a unified framework for all particles and 
forces of nature. Indeed, although fundamental strings were devised as
sub-microscopic objects, it has been progressively realized \cite{CMP} that some of 
these strings could be stretched to astronomical size by the cosmic expansion. eLISA
will be a very sensitive probe for these objects and so offers the possibility of detecting 
direct evidence of fundamental strings.

We review in what follows the different cosmological backgrounds that eLISA could potentially probe, with a special attention to first-order phase transitions and cosmic strings. Because the study of these backgrounds has made a lot of progress over the years, we review the literature in a rather detailed way in order to motivate our own choices of parameters and dynamical processes at work, choices which may not coincide with previous analyses. The aim of this work is to provide reliable and precise detection forecasts for the GW signal coming from these two sources. In order to do that, we have improved and completed the models of the GW spectra available in the literature in the following aspects.

In the case of first order phase transitions, the GW signal has been evaluated in previous analyses under the assumption that the broken phase bubbles propagate through a Jouguet detonation. This assumption simplifies considerably the evaluation of the signal but is very restrictive and not justified in the vast majority of phase transition scenarios. Here this approximation is relaxed: as a consequence, we find that the amplitude of the GW signal depends not only on the temperature, strength, and duration of the phase transition, but also on a new parameter representing the friction exerted on the bubble walls by the interaction with the surrounding plasma. Moreover, we combine for the first time the GW spectra from bubble collisions and magnetohydrodynamic turbulence in a fully consistent way, and provide detection forecasts for the total GW signal. In previous works on the estimation of the GW signal from first order phase transitions, the strength and the duration of the phase transition have often been considered as independent parameters. However, a relationship among them is to be generically expected for thermal first-order phase transitions: we account for this fact in the present evaluation of the GW signal, and point out that it has relevant consequences as far as detection prospects are concerned. 

In the case of cosmic (super-)strings, two main methods have been used in the literature to evaluate the GW background. The first one is based on models for the GW spectrum emitted by each individual loop, while in the second method the background is obtained as a superposition of GW bursts produced by specific loop configurations called cusps and kinks. We show that the two methods lead to similar results up to two differences: an overal normalisation of the GW background and the effect of removing the rare bursts that do not contribute to the gaussian and stationary background. We study carefully how these two differences affect the results. We also show that the GW background today depends weakly on the particular spectrum emitted by each individual loop. For instance, contrary to what is often stated in the literature, cusps do not lead to any significant increase of the background amplitude. On the other hand, the signal depends strongly on properties of the string network evolution that are still uncertain, in particular the characteristic size of the loops when they are produced. We therefore study in detail how these uncertainties affect the results. Compared to previous works, we also improve the modelling of the cosmological evolution and we study how the results depends on the thermal history of the very early universe. Indeed, since cosmic strings emit GW continually during the cosmological history, the GW background today depends also on the details of the cosmological evolution.

Readers interested only in the scientific potential of eLISA may go directly to Section~\ref{estimate} for first-order phase transitions and to Sections~\ref{compaS} and \ref{compaL} for cosmic strings. For a brief but exhaustive review of other cosmological sources of GW, see Section~\ref{Other}.

The rest of the paper is organized as follows. We briefly review general aspects of cosmological GW backgrounds in sub-section~\ref{notations}, introducing our notations, and discuss their detection with eLISA in sub-section~\ref{ElisaStoch}. Section~\ref{PT} is dedicated to first-order phase transitions. After an general overview of the GW sources and spectrum
(sub-section~\ref{general}), we review the literature about the calculation of the GW background (sub-section~\ref{literature}) and about particle-physics models of first-order phase transitions at the electroweak scale (sub-section~\ref{models}). In sub-section~\ref{EWPTsignal}, we then use the latest results available in the literature to evaluate the GW spectra in two models of first-order phase transition, and to assess more generally the detection prospects for eLISA. Section~\ref{Strings} is dedicated to cosmic (super-)strings. We start in sub-section~\ref{IntroStrings} with an introduction and overview of the literature, where we review the main aspects that must be modeled in order to compute the GW background. In sub-section~\ref{MethodStrings}, we detail our own method to calculate the GW spectrum and compare it to previous methods used in the literature. The cases of small and large initial loop sizes are then discussed separately in sub-sections~\ref{Sloops} and \ref{Lloops}, where we check the dependence of the results on the model assumptions and determine the regions of the parameters space that are accessible to eLISA, as well as ground-based experiments and pulsar timing observations. In Section~\ref{Other}, we review several other cosmological sources of GW, that operate during inflation (sub-section~\ref{during}), just after inflation (sub-section~\ref{justafter}) and during the subsequent thermal evolution (sub-section~\ref{thermevol}). Finally, Section \ref{conclu} contains our conclusions.

\subsection{General Aspects and Notations}
\label{notations}

In a cosmological context, GW may be represented by a tensor perturbation $h_{ij}$ ($i, j = 1, 2, 3$) of the 
Friedmann Robertson-Walker metric
\be
ds^2 = -dt^2 + a^2(t) \, (\delta_{ij} + h_{ij}) \, dx^i dx^j
\ee  
which is transverse and traceless
\be 
\label{TT}
\partial_i h_{ij} = h_{ii} = 0 \, ,
\ee
where we assume flat spatial sections, $a(t)$ is the scale factor, $t$ denotes the physical time and repeated latin indices are summed. The transverse-traceless condition (\ref{TT}) leaves only two independent degrees of freedom, which are the only ones that propagate and carry energy out of a source. In Fourier space, their linearized equation of motion given by Einstein equations is\footnote{From now on, we set $k_B=c=1$.}
\be
\label{gweq}
\ddot{h}_{ij}(\bar{\mathbf{k}}, t) + 3 H \, \dot{h}_{ij}(\bar{\mathbf{k}}, t) + 
\frac{\bar{k}^2}{a^2} \, h_{ij}(\bar{\mathbf{k}}, t) = 16\pi G \, \Pi_{ij}^{(TT)}(\bar{\mathbf{k}}, t)\,,
\ee
where $G$ is Newton constant, $H$ is the Hubble rate, a dot denotes derivative with respect to $t$, and 
$\Pi_{ij}^{(TT)}$ is the transverse-traceless part of the anisotropic stress $\Pi_{ij}$. The latter is given 
by $a^2 \, \Pi_{ij} = T_{ij} - p \, a^2 \, (\delta_{ij} + h_{ij})$, where $T_{ij}$ denotes the spatial components 
of the energy-momentum tensor and $p$ is the background pressure. In Eq.~(\ref{gweq}) and in the following, we use 
$\bar{k} = |\bar{\mathbf{k}}|$ to denote the \emph{comoving} wave-number of the GW, while we will use $k = \bar{k} / a$ 
to denote the \emph{physical} wave-number at a given moment of time. For a source operating at 
sub-Hubble scales ($k \gg H$), Eq.~(\ref{gweq}) may be approximated as a standard wave-equation~\footnote{The terms in 
$3 H \, \dot{h}_{ij}$ and $16\pi G \, p \, h_{ij}$ in Eq.~(\ref{gweq}) are negligible compared to $k^2\,h_{ij}$ for 
$k \gg H$.} $\ddot{h}_{ij} + k^2 \, h_{ij} = 16\pi G \, T_{ij}^{(TT)} / a^2$. Another mechanism of GW production is provided by the parametric amplification of quantum fluctuations during inflation. In that case, the source term is negligible in Eq.~(\ref{gweq}), but super-Hubble tensor perturbations are generated by the very fast expansion of the universe. These tensor perturbations become standard GW once they re-enter the Hubble radius during the post-inflationary evolution. 

Once produced in the early universe, GW propagate essentially freely until today, being simply redshifted by the expansion of the universe. Their energy density today can be written as~\cite{maggiore}
\be
\rho_{gw} = \frac{\langle \dot{h}_{ij} \, \dot{h}_{ij} \rangle}{32 \pi G} = 
\int \frac{d f}{f} \, \frac{d \rho_{gw}}{d \log f}\,,
\ee
where $f = \bar{k} / (2 \pi \, a_0)$ is the present-day GW frequency ($a_0$ denoting the scale factor today) and 
$\langle \rangle$ denotes ensemble average. The superposition of GW produced by a large number of unresolved sources 
in the early universe form a stochastic background that is assumed to be statistically isotropic, stationary and nearly 
Gaussian~\cite{allen}. Its main properties are then described by its power spectrum. The quantity that is usually considered to characterize cosmological backgrounds is the spectrum of energy density per logarithmic frequency interval divided by the critical density $\rho_c$ today 
\be
\label{spectrum3}
h^2 \, \Omega_{gw}(f) = \left(\frac{h^2}{\rho_c} \, \frac{d \rho_{gw}}{d \log f}\right)_0
\ee
where the subscript $0$ refers to the present epoch and $h$ parametrizes the small uncertainty in the value of the Hubble constant today, $H_0 = 100\,h\,$km/s/Mpc. 

A source that operates at sub-Hubble scales at some time $t_*$ after inflation emits GW with a characteristic physical 
wave-number $k_* = \bar{k} / a_*$ that is larger than the Hubble rate $H_*$ at that time: $k_* = H_* / \epsilon_*$ with 
$\epsilon_* \leq 1$. The characteristic GW frequency today is then given by $f_c = (k_* / 2\pi) \, (a_* / a_0)$. For GW produced in the radiation era when the plasma temperature is $T_*$, and assuming a standard adiabatic thermal history 
for the evolution of the universe after GW production, the characteristic frequency today can be written as
\be
\label{kstarg}
f_c \simeq \frac{1.6 \times 10^{-4} \, \mathrm{Hz}}{\epsilon_*} \, \left(\frac{T_*}{1 \, \mathrm{TeV}}\right) \, 
\left(\frac{g_*}{100}\right)^{1/6}
\ee
where $g_*$ is the number of relativistic degrees of freedom at temperature $T_*$. The parameter $\epsilon_* \leq 1$ depends on the dynamics of the particular GW source under consideration. For a first-order phase transition, one may have for instance $\epsilon_* \sim 10^{-3} - 1$. In that case, Eq.~(\ref{kstarg}) shows that GW produced around the electro-weak scale are potentially interesting for detection with eLISA. On the other hand, cosmic string loops produce GW continually (over a wide range of values of $T_*$), so that the present-day spectrum covers a very wide range of frequencies. For a given frequency of observation, the signal is dominated by the emission at some time $t_*$ from loops of size smaller than the Hubble radius by many orders of magnitude, so the typical values of $\epsilon_*$ are much smaller in that case. As another example, Eq.~(\ref{kstarg}) can also be applied to GW from inflation, but with $T_*$ the temperature when a given tensor mode re-enters the Hubble radius after inflation, hence with $\epsilon_* = 1$. The resulting GW spectrum covers also a very wide frequency range, since tensor modes continually re-enter the Hubble radius.

Before moving on to eLISA, let us briefly review current constraints on cosmological GW backgrounds in different frequency ranges, see e.g.~\cite{buonanno} and references therein. If GW contributed too much to the total energy density of the universe at the epoch of Big Bang Nucleosynthesis (BBN), when $T \sim$ MeV, they would spoil the succesfull predictions for the light element abundances. This leads to a constraint on any GW background that is present at that time, which can be written in terms of the present-day GW energy density as 
$h^2 \, (\rho_{gw} / \rho_c)_0 < 5.6 \times 10^{-6} \, (N_\nu - 3)$, where $N_\nu$ is the effective number of neutrino species. Although this bound applies to the total energy density in GW, neglecting the unlikely possibility of a GW spectrum with a very narrow peak one can take as a rule of thumb $h^2 \Omega_{gw} < 10^{-5}$ for $f \gsim 10^{-10}$ Hz 
($f \sim 10^{-10}$ Hz corresponds to the comoving Hubble scale at the time of BBN). Cosmological GW at very low 
frequencies are also constrained from their contribution to the temperature anisotropies of the Cosmic Microwave Background (CMB). This leads to the so-called COBE bound for GW that entered the Hubble radius after the time of last scattering, corresponding to frequencies in the range $f \sim 10^{-18} - 10^{-16}$ Hz. The strongest bound on $h^2 \Omega_{gw}$ is obtained at $f \sim 10^{-16}$ Hz: $h^2 \Omega_{gw} < \mathrm{few} \times 10^{-14}$. A GW background with a too large amplitude can also perturb the very accurate timing of msec pulsar. The resulting constraint obtained in \cite{jenet} is $h^2 \Omega_{gw} < 2 \times 10^{-8}$ at $f = 1 / (8 {\rm yr})$ (up to a weak dependence on the slope of the spectrum). Finally, the observational upper bound $h^2 \Omega_{gw} < 3.6 \times 10^{-6}$ at $f \approx 100$ Hz has been obtained from the S5 LIGO run~\cite{ligo}.

\subsection{Looking for Stochastic Backgrounds with eLISA}
\label{ElisaStoch}

\begin{figure}
\includegraphics{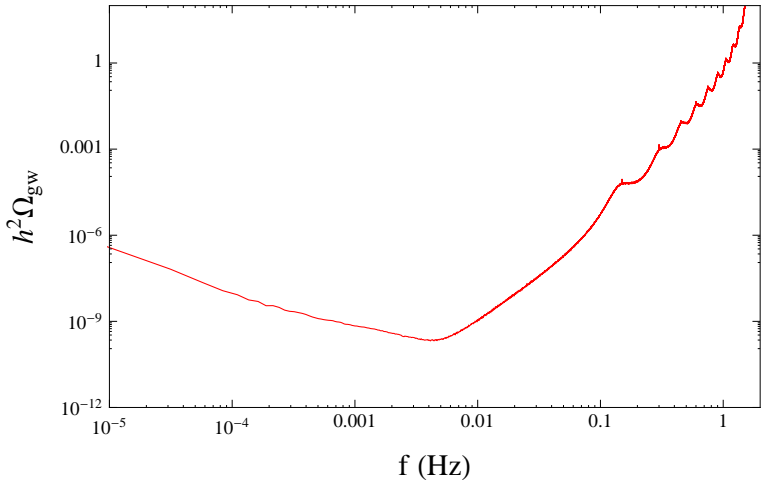}
\caption{Sensitivity curve of eLISA computed using the expected instrumental noise and the confusion noise generated by unresolved galactic binaries \cite{yellowbook}.}
\label{fig:sensitivitycurve}
\end{figure}

NGO/eLISA consists, like the original LISA mission, of a triangular constellation 
of 3 satellites. But whereas LISA had 6 laser links joining the three 
satellites (two in opposite directions for each side of the triangle), eLISA 
only has four lasers, which link one satellite (called Mother) to the other two  
(called the Daughters). Hence, eLISA consists of a single Michelson-type 
experiment. The inter-spacecraft distance is reduced to one million kilometers
(down from 5 millions in the case of LISA) and the nominal mission lifetime to 
2 years (extendable to 5 years). The eLISA sensitvity curve in terms of 
$h^2 \Omega_{gw}$ is given in Figure \ref{fig:sensitivitycurve}.   

In order to distinguish backgrounds of gravitational waves from
waves emitted by point sources, it is essential to make use of the
successive positions of eLISA around the Sun, and thus to wait a
sufficient amount of time (of the order of a few months). Moreover, it will be more
difficult for eLISA, with respect to LISA, to disentangle an isotropic cosmological (or astrophysical)
background from an instrumental one, all the more because the eLISA 
Mother-Daughter configuration, providing only two measurement arms, 
does not allow to use Sagnac calibration \cite{Hogan:2001jn}. Luckily 
as we will see, in the case of phase transitions as well as
cosmic strings, the spectral dependence of the signal is well
predicted and may allow to distinguish cosmological backgrounds as
long as they lie above the eLISA sensitivity curve.

\section{First-Order Phase Transitions}
\label{PT}

\subsection{General Overview: Characteristic Frequency and Amplitude of the GW Signal}
\label{general}

In the course of its adiabatic expansion, the universe might have undergone several phase transitions (PTs) driven by the temperature decrease. The nature of the primordial PTs depends on the particle theory model, but if they are first-order they proceed through the nucleation of broken phase bubbles, which is a very violent and inhomogeneous process capable of sourcing GW. In this section we concentrate specifically on GW generation by processes which are related to bubble nucleation during a first-order PT. We study the characteristics of the GW power spectrum and its dependence on the PT parameters, and provide a complete analysis of the GW signal improving the results present in the literature, in particular concerning bubble propagation and the energy balance of the PT. 

The electro-weak phase transition (EWPT) in the standard model is a crossover, and it is not expected to lead to any appreciable cosmological signal; however, deviations from the standard model in the Higgs sector or the introduction of supersymmetry can lead to a first-order EWPT. Similarly, the QCDPT is also predicted to be a crossover by lattice simulations but it can become first-order if the neutrino chemical potential is sufficiently large \cite{arXiv:0906.3434}. 
GW detection would help to probe the nature of these PTs, and provide interesting information on the underlying particle theory. In particular, a sufficiently strong first-order EWPT is potentially very interesting, because it could have produced a relic GW signal right in the frequency band of eLISA (c.f. section \ref{notations}). 

Towards the end of the PT, the true vacuum bubbles collide and convert the entire universe to the broken phase. The collisions break the spherical symmetry of the bubble walls, generating a non-zero anisotropic stress which acts as a source of GW (see Eq.~\eqref{gweq}). Moreover, bubble collision causes an injection of energy in the primordial plasma, which has a very high Reynolds number (of the order of $10^{13}$ at 100 GeV and at the typical scale of the 
bubbles \cite{arXiv:0909.0622}): the energy injection leads to the formation of magnetohydrodynamic (MHD) turbulence, which sources GW through the anisotropic stresses of the chaotic fluid motions. MHD turbulence also leads to the amplification of small magnetic fields generated by charge separation at the bubble wall \cite{arXiv:1001.3694,hep-ph/9507429}, which have non-zero anisotropic stress and also source GW. 

There are in summary two processes which can lead to the production of GW towards the end of a first-order PT: bubble collision and MHD turbulence (where the anisotropic stress is due both to the velocity field and to the magnetic field). They are related to the collision of bubbles, which involves two quantities: the duration of the PT, commonly denoted by the parameter $\beta^{-1}$, and the typical size of the bubbles at the moment of collision, $R_* \simeq v_b \,\beta^{-1}$, where $v_b$ is the bubble wall velocity. The characteristic frequency of the GW generated by the two processes can correspond either to the duration of the PT or to the bubble size: $k_*\simeq \beta$ or $R_*^{-1}$, depending on the details of the time evolution of the source (c.f. section \ref{literature}). These two parameters can therefore both determine the value of $\epsilon_*$ in \Eqref{kstarg}. If the growth of the bubble proceeds at a highly relativistic speed, the two time/length-scales are equal. Assuming for the moment $k_*\simeq \beta$, one obtains the following order of 
magnitude estimate of the characteristic frequency today (c.f. \Eqref{kstarg}):
\be
f_c \simeq 10^{-2}\,\frac{\beta}{H_*}\,\frac{T_*}{100\,{\rm GeV}}\, \left( \frac{g_*}{100} \right)^{\frac{1}{6}} {\rm mHz}\,.
\label{kstar}
\ee
The parameter $\beta/\HH_*$ is the ratio of the Hubble time to the duration of the PT. Since the entire universe must be converted to the broken phase, the PT must complete faster than a Hubble time, so in general we expect $\beta/\HH_*>1$. From Eq.~(\ref{kstar}) it appears that the characteristic frequency of GW emitted at the EWPT at 100 GeV falls in the frequency range of eLISA for values $1\lesssim \beta/\HH_* \lesssim 10^4$. As another example, we see from the above formula that GW production at the QCDPT at $T_*\simeq 100$ MeV can fall into the frequency range of detection with pulsar timing array, $f\geq 10^{-8}$Hz (see e.g. \cite{arXiv:1007.1218} and references therein). The precise value of 
$\beta/\HH_*$ has to be determined in the context of a given model for the first-order PT (c.f. section \ref{EWPTsignal}).

A simple estimate of the GW amplitude, which shows how the result scales with the duration and the energy density of the GW source, can be given through the following heuristic argument. We rewrite the GW equation of motion (\ref{gweq}) for 
$\beta / H_* > 1$ simply as $\beta^2 h \sim 16\pi G \,T$, where $h$ denotes the amplitude of the tensor perturbation, $T$ the energy-momentum tensor of the source, and we inserted $1/\beta$ as the characteristic time on which the perturbation is evolving (we have dropped indices for simplicity). This suggests that $\dot h \sim 16\pi G \,T/\beta$, and the GW energy density at the time of production can then be estimated as $\rho_{gw} \sim \dot h^2/(32\pi G) \sim 8 \pi G\, T^2/ \beta^2$. Dividing by the total energy density $\rho_{tot} = 3 H_*^2 / (8 \pi G)$ at the time of GW production, we can write 
$\rho_{gw} / \rho_{tot} \sim (H_* / \beta)^2 \, (\rho_s / \rho_{tot})_*^2$ where $\rho_s \sim T$ denotes the part of the energy density available in the source for the GW generation. Accounting for the fact that the PT takes place in the 
radiation-dominated universe and that the GW energy density is diluted like radiation, one then obtains, for the peak amplitude of the GW spectrum (\ref{spectrum3}) today
\be
\Om_{gw} \sim \Om_{R} \left(\frac{\HH_*}{\beta}\right)^2 \left(\frac{\rho_s}{\rho_{tot}}\right)_*^2\,,
\label{OmgwPT}
\ee
where $\Om_{R} = \rho_R / \rho_c$ denotes the radiation abundance today and $(\rho_s / \rho_{tot})_*$ the fraction of energy density that contributes to GW generation at the time of production. The above equation shows that the GW energy density scales like the square of the ratio of the GW source duration and the Hubble time, and the square of the energy density in the source. As a rule of thumb, given $h^2\Om_{R} \simeq 4.15 \times 10^{-5}$, a GW signal above the lowest sensitivity of eLISA ($h^2\Om_{gw}\gtrsim 10^{-10}$) can be generated if 
$(\HH_*/\beta) (\rho_s / \rho_{tot})_* \gtrsim 3 \times 10^{-3}$. Therefore, detectable signals arise from very energetic processes, which involve a sizable fraction of the total energy density in the universe, and at the same time slow processes, which minimise the value of $\beta/\HH_*$ (in compatibility with the value of the characteristic frequency, \Eqref{kstar}, which should not exit the detection frequency band - c.f. the discussion in section \ref{EWPTsignal} 
and the spectra in Fig.~\ref{HKdim6}).   

The slope of the GW spectrum at wave-numbers $k$ smaller than the Hubble radius $H_*$ at the time of production can also 
be determined on general grounds, valid for any transient stochastic source after inflation. This is a consequence of the fact that the causal process generating the GW signal cannot operate on time/length-scales larger than $1/\HH_*$. Therefore, the anisotropic stresses $\Pi_{ij}^{(TT)}(\bk, t)$ sourcing the metric perturbations in \Eqref{gweq} 
are not correlated for $k<\HH_*$, and the anisotropic stress power spectrum is expected to be flat (white noise) up to the wavenumber $k_*$ (beyond $k_*$, the power spectrum can instead decay with a slope that depends on the details of the source, see e.g. \cite{arXiv:1005.5291} and references therein). If the anisotropic stress power spectrum is flat, so is expected to be the GW power spectrum $|\dot{h}_{ij}(\bk, t)|^2$, because the causal horizon sets also the maximal time-scale of the correlation, and \Eqref{gweq} cannot lead to any extra correlation for $k<\HH_*$. This in turn implies that the spectrum of GW energy density per logarithmic frequency interval must grow as $k^3$, because 
$d\rho_{gw} / d\log k \propto k^3 |\dot{h}_{ij}(\bk,t)|^2$. Thus the infra-red tail of the present-day GW spectrum behaves as $h^2 \Omega_{gw} \propto f^3$ for scales that were super-Hubble at the time of production.

The rest of this section \ref{PT} on GW generation by first-order PTs is organised as follows: we start with an overview of the literature, in order to present the main achievements and resume the state of the art of the subject 
(sections \ref{literature} and \ref{models}). We then explain how the results of the literature can be improved, in particular concerning two aspects: the shape of the GW spectrum arising from MHD turbulence, and the estimation of the PT parameters, especially about the physics of bubble propagation (section \ref{EWPTsignal}). We point out that restrictive formulas have been used for the bubble wall velocity and the kinetic energy of the bubble wall, and use new results of the literature to correct this problem (section \ref{PTpar}). In the last section, \ref{estimate}, we evaluate the GW spectra in two examples of first-order PTs, and then discuss the prospects for detection with eLISA in a way as model-independent as possible.

\subsection{Overview of the Literature: GW Production by Bubble Collision and MHD Turbulence}
\label{literature}

\subsubsection{Bubble Collision}

The fact that a first-order PT occurring explosively, through the nucleation of fast broken phase bubbles, can be a source of GW, has been first pointed out in Witten 1984 \cite{Witten:1984rs}. This work estimates the characteristic frequency and intensity of GW emitted during a first-order QCDPT, finding that the frequency falls in the sensitivity range of pulsar timing array observations: if $R_*$ denotes the radius of the typical broken phase bubbles, \cite{Witten:1984rs} demonstrates for the first time that the wavelength of the emitted GW corresponds to $R_*$, and that the energy density is proportional to $\OmG \propto (R_*\HH_*)^2$, with $\HH_*$ the Hubble time at the PT (c.f. \Eqref{kstar} and \Eqref{OmgwPT}).

Afterwards, Hogan 1986~\cite{hogan86} re-examines the problem in more detail, generalising the analysis to the EWPT as well, and providing for the first time an estimate of the shape of the GW spectrum. It is argued that the spectrum of GW energy density per logarithmic frequency interval must grow as $k^3$ on frequencies lower than the characteristic frequency $k_*$, because the source is uncorrelated on time-scales larger than $1/\beta$ and on length-scales larger 
than $R_*$ (see section \ref{general}). From arguments based on the dynamics of the pressure waves generating the GW signal, \cite{hogan86} also argues that the GW spectrum at frequencies higher than $k_*$ should fall off at least as 
fast as $1/k$. 

A few years later, Turner and Wilczek 1990~\cite{Turner:1990rc} study GW production by bubble collision in the context of extended inflation, where the transition between the inflationary and the radiation-dominated phases occurs through bubble nucleation and percolation. Again, it is found that the wavelength of the GW signal corresponds to the typical size of the bubbles when they collide, $R_*\sim 1/\HH_*$ (which in this case is of the order of the Hubble time at the end of inflation), and that the GW energy density is $\OmG\sim 10^{-5} (R_*\HH_*)^2$ (c.f. \Eqref{OmgwPT}, where 
$\rho_s = \rho_{tot}$ in that case).

In the above mentioned works, the GW signal was estimated using dimensional arguments and the quadrupole approximation, which gives the correct scaling but in principle breaks down for highly relativistic processes, for which higher multipoles are also important. Kosowsky et al 1992~\cite{FERMILAB-PUB-91-323-A} for the first time go beyond an analytical estimate of the GW signal, computing numerically the GW radiation produced in the collision of two broken phase 
bubbles in vacuum and Minkowski space. The result still depends only on the grossest features of the PT: the duration 
$1/\beta$, which here corresponds to the initial bubble separation and which is inserted as a time cutoff to break the 
$O(2,1)$ symmetry of the problem; and the fraction of false vacuum energy liberated by the collision, i.e. the energy available in the source for GW generation (which in \cite{Witten:1984rs,hogan86,Turner:1990rc} was taken to be of the order of the total energy density in the universe). The GW energy estimated from the collision simulation (before red-shifting to today) is $E_{gw}/E_{vac}\simeq 10^{-3} (\HH_*/\beta)^2$, confirming the usual dependence on the time/length-scale of the PT (c.f. \Eqref{OmgwPT}, and note that this is also comparable to the factor $(R_*\HH_*)^2$ since in this case 
$v_b=1$). The GW power spectrum shows a low frequency tail which is fitted in the paper to $k^{2.8}$ (we know by causality that it should vary as $k^3$ at very small frequencies, see section \ref{general}), and an high frequency tail which is fitted as $k^{-1.8}$. The spectrum peaks at $k_{peak}\simeq 3.8\, \beta$. Moreover, \cite{FERMILAB-PUB-91-323-A} realises for the first time that, since it is the bulk motion of the bubble walls that produce most of the GW signal, it is possible to infer the GW emission by simulating only the motion of the bubble walls as a simple propagation of spherical, infinitely thin shapes instead of using the Klein-Gordon equation to evolve the scalar field. In this approximation, the gravitational radiation produced by bubble collision is due only to the TT part of the energy-momentum tensor of the uncollided envelope of the spherical bubbles, ignoring the interaction region. This `envelope approximation' greatly simplifies the numerical simulation since it dispenses with the detailed dynamics of the scalar field and reduces the required computational power. It therefore allows to simulate the nucleation and collision of many bubbles, a more realistic model of the PT. 

This is first exploited in Kosowsky and Turner 1993~\cite{astro-ph/9211004}, where the PT is simulated with the collision of 20-200 bubbles in vacuum. The nucleation rate is taken to be exponential in time with characteristic inverse time 
$\beta$, $\Gamma\propto \exp(\beta t)$ (c.f. section \ref{EWPTsignal}), and the resulting GW spectrum agrees well with the two bubble case, except for a few features. First, the efficiency of GW generation is higher, leading to 
$E_{gw}/E_{vac}\simeq 0.06 (\HH_*/\beta)^2$. Second, the peak of the spectrum is at somewhat lower frequency, 
$k_{peak}\simeq 1.6 \beta$, which corresponds to the peak of the energy weighted distribution of bubble sizes. Third, the spectrum has a slower high frequency decay, due to the late nucleation of smaller bubbles which is not present in the two-bubble case. However, the low numerical accuracy does not allow a conclusive statement on the slope of the GW spectrum at high frequency.

Kamionkowski et al 1994~\cite{astro-ph/9310044} generalises the analysis of \cite{astro-ph/9211004} to a PT happening in a thermal environment. The bubbles are described by means of the theory of relativistic spherical combustion, following the work of Steinhardt 1982~\cite{Steinhardt:1981ct}. The velocity of the bubble wall is therefore not any longer equal to the speed of light, and the energy momentum tensor sourcing the GW is no longer the one of the scalar field, but the one of the relativistic fluid: $T_{ij}=(\rho+p) v_i v_j $, where $\rho$ and $p$ are the energy density and the pressure of the relativistic fluid, and $v_i$ is the velocity field describing the growth of the bubble. In Ref.~\cite{Steinhardt:1981ct} it is demonstrated that, in the case of bubble propagation by Jouguet detonations, is it possible to establish a very simple relation between the bubble wall velocity $v_b$ and the ratio $\alpha$ of the vacuum energy density to the radiation energy density:
\be
v_b(\alpha)=\frac{\sqrt{1/3}+\sqrt{\al^2+2\al/3}}{1+\al}\,, \qquad \qquad \al=\frac{\rh{vac}}{\rh{rad}^*}\,, 
\label{vbdeto}
\ee
(where $\rh{rad}^*$ denotes the radiation energy density at the moment of the PT). This formula is re-derived in 
\cite{astro-ph/9310044} and it will be also widely used in the following literature on GW from PTs, which therefore implicitly assumes that bubble propagation is well described by a Jouguet detonation; however, this need not be the case, see section \ref{EWPTsignal}. 

A second relevant quantity for GW production is first identified and evaluated in 
\cite{astro-ph/9310044}: the fraction $\kappa$ of vacuum energy that goes into bulk kinetic energy of the fluid as opposed to thermal energy. When the PT happens in vacuum, the liberated vacuum energy is transferred entirely into the propagation of the bubble wall (the gradient of the scalar field); for a thermal environment instead, the vacuum energy initially available is distributed during the PT among thermal energy of the plasma, bulk fluid motions at the bubble walls, and gradient energy of the scalar field. The latter need not be the dominant one, as in the case of vacuum bubbles; for moderately strong PT, it can actually be negligible with respect to the other components. The parameter $\kappa$ represents, in~\cite{astro-ph/9310044}, an efficiency factor quantifying the amount of bulk kinetic energy which is 
the relevant source of the GW for a PT happening in a thermal bath. Using the velocity profile of a Jouguet detonation front, Ref.~\cite{astro-ph/9310044} evaluates the energy momentum tensor $T_{ij}$ representing the propagation of one bubble, and provides a simple analytical fit of the efficiency factor as a function of $\al$:
\be
\kappa(\alpha)=\frac{\rho_{kin}}{\rho_{vac}}=\frac{0.715\alpha+4/27\sqrt{3\alpha/2}}{1+0.715\alpha}\,.
\label{kappakamion}
\ee
Moreover, \cite{astro-ph/9310044} analyses the spatial distribution of the energy momentum tensor and finds that for values of $\al\gtrsim 0.01$ the stress energy is concentrated in a thin shell nearby the bubble wall. It is then concluded that, even in the case of a PT in a thermal bath proceeding through detonation, one can use the envelope approximation to simulate the PT and evaluate the GW signal. This is how \cite{astro-ph/9310044} proceeds to get the following formulas for the amplitude of the radiated GW energy density at the peak of the spectrum, and for the peak frequency ($k_{peak}$ denotes the physical wavenumber at the moment of the PT, while 
$f=(k/2\pi) (a_*/a_0)$ is the redshifted frequency today):
\bea
& &\left. h^2 \Omega_{gw} \right|_{peak} \simeq 10^{-6}\,\kappa^2\left(\frac{\HH_*}{\beta}\right)^2 \left(\frac{\alpha}{1+\alpha}\right)^2 \left(\frac{v_b^3}{0.24+v_b^3}\right)\left(\frac{100}{g_*}\right)^{\frac{1}{3}}\,, 
\label{Omkos} \\
& & k_{peak}\simeq 1.9 \beta\,, \qquad \qquad f_{peak}\simeq 5.2\cdot 10^{-3} {\rm mHz}\left(\frac{\beta}{\HH_*} \right) \left( \frac{T_*}{100 \,{\rm GeV}}\right) \left( \frac{g_*}{100}\right)^{\frac{1}{6}}\,.
\label{peakkos}
\eea
In this work as well, no conclusive statement is made on the slope of the GW spectrum at high frequency. 

The above formulas, together with \Eqref{vbdeto} and \Eqref{kappakamion} have been widely used in the following literature on the subject, because they provide a very simple estimate of the GW signal, which is completely determined once the parameters $\al$, $\beta$ and $T_*$ are known. However, the situation can be more complicated, since this simple result strictly applies only in the case of Jouguet detonations, and in the case in which the envelope approximation is well justified. For example, it does not account for friction acting on the bubble wall, which could slow down the wall propagation into a deflagration \cite{Espinosa:2010hh}. Assuming Jouguet detonations amounts to setting the value of the fluid velocity at the inner boundary of the velocity profile in the wall frame to the speed of sound. There is no justification for this choice, although it has been used in the literature for simplicity. Setting arbitrarily some boundary condition for the velocity leads to inconsistencies and corresponds to ignoring the constraints from the equation of motion of the Higgs field, in other words, corresponds to ignoring the friction term that restrains the bubble expansion. In Espinosa et al 2010~\cite{Espinosa:2010hh} it is demonstrated that $\kappa$ and $v_b$ depend not only on $\alpha$, but also on the friction $\eta$, and numerical solutions for these parameters are provided, as well as analytical fits (in its appendix A). Moreover, the analysis of~\cite{Espinosa:2010hh} allows to extend the definition of the parameter $\kappa$ to represent the full energy balance of the PT, including not only the bulk kinetic energy of the fluid, but also the gradient energy of the Higgs field (see section \ref{PTpar}). We will see that this can be exploited to evaluate the GW signal continuously for any strength of the PT. In the following, we will adopt the results of~\cite{Espinosa:2010hh} for our analysis of the GW spectra in section \ref{estimate}. 

An attempt to go beyond the envelope approximation and find a result applicable also to the case of deflagrations, where the bulk fluid motions are expected to be extended over a wider region around the bubble wall, has been carried on in Caprini et al. 2008~\cite{Caprini:2007xq}. Here, an analytical method is developed to determine the GW spectrum, which exploits the intrinsic randomness of the nucleation process: the GW spectrum is found by solving analytically \Eqref{gweq} where the power spectrum of the GW source, given by the anisotropic stresses of the bulk fluid motions, is evaluated by Fourier transforming the two-point spatial correlation function of the velocity field describing the bubble. GW production from bubble collision is modeled in this approach by using Wick's theorem on the four-point velocity correlator which determines the anisotropic stress power spectrum~\cite{arXiv:0901.1661}. The resulting GW power spectrum has a well defined shape. It is here re-established that at low frequency the spectrum must increase as $k^3$, because the fluid motions are uncorrelated at large scales. The GW spectrum is found to peak at a wavenumber corresponding to the characteristic length-scale of the process, the average size of the bubbles $R_*$, and to decay at frequencies higher than the peak as 
$k^{-1.8}$. However, in this work the anisotropic stress grows with time (as it should, because of bubble collisions), but it is abruptly switched off at the end of the PT. Caprini et al. 2009~\cite{arXiv:0901.1661} have shown that the peak and the slope at high frequency of the GW spectrum strongly depend on the time structure of the source, and inserting a discontinuity in time can lead to spurious effects which influence the shape of the GW spectrum. In \cite{arXiv:0901.1661} it is demonstrated that the correct time modeling of the GW source when accounting for bubble collision is that the source must be always correlated in time (completely coherent), and it must be switched on and off in a continuous but not differentiable way. This leads to a peak in the power spectrum corresponding to the characteristic time of the source, i.e. the duration of the PT $1/\beta$ (as found in numerical simulations of bubble collision - for more details, c.f. \cite{arXiv:0901.1661}). 

Shortly after the analytical estimation of~\cite{Caprini:2007xq}, Huber and Konstandin 2008~\cite{arXiv:0806.1828} simulated again the collision of bubbles in the envelope approximation, with considerably improved  numerical accuracy, which allows to determine a larger portion of the spectrum and consequently a more careful analysis of the high frequency behavior with respect to \cite{astro-ph/9211004}. The bubble wall velocity is treated as a free parameter, so the analysis does not rely on the assumption of propagation by Jouguet detonation ($v_b$ must only be sufficiently close to the speed of light for the envelope approximation to hold). The GW spectrum is found to increase as $k^{2.8}$ at frequencies smaller than the peak ($k^3$ at very small frequencies), and to peak at a frequency which is somewhat smaller than the characteristic time-scale of the PT $1/\beta$, and slightly depends on the bubble wall velocity. The dependence of the peak amplitude with the parameters of the PT is broadly in agreement with \Eqref{Omkos}. The most interesting result of \cite{arXiv:0806.1828} is that the GW spectrum is found to decay as $1/k$ at high frequency, considerably slower than previously found. It is argued that this behavior is most probably related to the many small bubbles nucleated at a later stage of the PT, and it is potentially very interesting for detection prospects. In section \ref{EWPTsignal} we use the results of~\cite{arXiv:0806.1828} to evaluate the GW signal by bubble collision.

\subsubsection{MHD Turbulence}

GW production from turbulent motions has been first analysed in Kamionkowski et al 1994~\cite{astro-ph/9310044}, where it is pointed out that bubble collisions cause a stirring of the primordial fluid on a characteristic scale given by the bubble size at collision $R_*$, leading to the formation of turbulent fluid motions with a Kolmogorov spectrum. The GW signal has been estimated in this work using the quadrupole approximation; the peak frequency corresponds to the inverse largest eddied turnover time $\tau_R \sim R_* /v_R$, where $v_R$ is the typical velocity of the turbulent motions on scale $R_*$. The high frequency slope is found to be $k^{-9/2}$. The large-scale part of the spectrum is not calculated.

The problem has been re-analysed later on in Kosowsky et al 2002~\cite{astro-ph/0111483} and Dolgov et al 2002~\cite{astro-ph/0206461}, which both go beyond the quadrupole approximation directly integrating the GW equation (\ref{gweq}), and both remark that since the turbulence arises in a fully ionised plasma, it has to be treated as MHD turbulence: therefore one must add to the evaluation of the GW spectrum also the contribution from the turbulent magnetic field, besides the one from the turbulent velocity field. The procedure for the evaluation of the GW spectrum in \cite{astro-ph/0111483} is fully analytical (as a matter of fact, there have been no numerical simulations of GW production by MHD turbulence to date, contrary to the bubble collision case). Dolgov et al. 2002 \cite{astro-ph/0206461} uses the result of~\cite{astro-ph/0111483} for the GW spectrum, but parametrically introduces also power-law deviations from the Kolmogorov spectrum, and generalises the analysis to the case when the time during which turbulence is active is of the order of the Hubble time. Both analyses confirm the finding of \cite{astro-ph/9310044} in what concerns the peak frequency, $k_{peak}\simeq 1/\tau_R$, but the high frequency slope that they find is different: $k^{-7/2}$ in the Kolmogorov case. However, both these analyses have some fundamental problems. First of all, they only evaluate the GW signal arising from the Kolmogorov part of the turbulent velocity spectrum, which is that part of the spectrum with slope $k^{-11/3}$ in a range of wavenumbers comprised between the stirring scale $R_*$ and the dissipation scale. However, the power spectrum of the velocity field does not vanish on scales larger than the correlation scale (corresponding to the largest eddy size $R_*$), so the turbulent, as well as the GW spectrum, must cover the entire frequency range. Moreover, the extension of the Kolmogorov slope $k^{-11/3}$ up to the stirring scale $R_*$ causes an overestimation of the turbulent kinetic energy density, and therefore of the GW signal at the peak frequency. The second problem is that in these references the source is modeled as stationary in time, in a way that the two point correlator of the source at different times only depends on the time difference. However, by definition the MHD turbulence must decay in time, because the source of stirring (bubble collisions) is not eternal. Consequently, in this cosmological context, it is not fully justified to assume a stationary source. The third and most important problem, is an error in the dispersion relation of the emitted GW, which instead of satisfying the usual dispersion relation, by an artefact of the calculation ends up having the same dispertion relation of the turbulent source, i.e. $\omega\propto k^{-2/3}$. The formulas for the characteristic peak frequency and the GW amplitude at the peak based on these analyses have been collected in Nicolis 2004~\cite{gr-qc/0303084}, and have been widely used in the following literature:
\bea
& & \left. h^2 \Omega_{gw}  \right|_{peak} \simeq  10^{-4}\,v_R^5\,v_b^2 \left(\frac{H_*}{\beta}\right)^2 \left(\frac{100}{g_*}\right)^{\frac{1}{3}}\,,
  \label{Omkostur} \\
& & k_{peak}\simeq \frac{v_R}{R_*}\,, \qquad \qquad f_{peak} \simeq  3.4\cdot 10^{-3}{\rm mHz}\, \frac{v_R}{v_b}\left(\frac{\beta}{H_*}\right)\left( \frac{T_*}{100\,{\rm GeV}}\right) \left( \frac{g_*}{100}\right)^{\frac{1}{6}}\,.
\eea

An attempt to correct the above mentioned problems has been carried on in Caprini and Durrer 2006~\cite{astro-ph/0603476}. Here it is demonstrated that the large scale part of the turbulent velocity spectrum is indeed present, and it is characterised by a Batchelor spectrum (i.e. it grows as $k^2$): this allows to evaluate the full GW spectrum, not only the high frequency slope. The time dependence of the turbulent source is modeled not as stationary but based on an heuristic model of the oscillations of the random velocity field. This model is inspired by RichardsonÕs energy cascade through eddies of different sizes, motivated by the fact that the time dependence of the turbulent flow is not lost in the ensemble averaging. Moreover, the correct dispersion relation for the free propagation of the GW is assured, i.e. $\omega=k$ after the turbulent source ceases to be active (for more details, see Caprini et al. 2006~\cite{astro-ph/0607651}). The GW spectrum found in this work is quite different from the result of \cite{astro-ph/0111483,astro-ph/0206461}, with a peak frequency that corresponds to the size of the bubbles at collision $k_{peak}\simeq 1/R_*$, a low frequency slope as $k^3$ by causality, and a high frequency slope as $k^{-8/3}$. The GW spectrum from the magnetic field has the same characteristics but a different high frequency slope, $k^{-5/2}$, due to the different assumption for the time evolution of the magnetic and turbulent sources. The analysis ends with a comparison of the efficiency of the turbulent and magnetic sources, finding that since the magnetic field is a coherent source and it is active for a long time, there is a very effective conversion of magnetic energy in GW energy at horizon crossing. However, in this reference as well the Kolmogorov slope is extended up to the largest eddy size, so the amplitude of the GW signal at the peak is overestimated. This problem persists also in the analysis of Gogoberidze et al. 2007~\cite{arXiv:0705.1733}, which keeps the assumption of stationarity of the source, but corrects the problem of the dispersion relation. The results of \cite{astro-ph/0603476} have been used in Megevand 2008~\cite{arXiv:0804.0391} to evaluate the GW signal in the case of deflagrations, and it is shown that for some values of $\alpha<1$ and of the PT temperature, one can reach $\OmG \sim 10^{-11}$. 

The analysis of Kahniashvili et al. 2008~\cite{arXiv:0802.3524,arXiv:0809.1899} shows that helical turbulence, which could be produced during a first-order PT if there is macroscopic violation of parity \cite{hel}, generates circularly polarized GW. Kahniashvili et al. 2005~\cite{astro-ph/0505628} concludes that it is unlikely that this GW signal has sufficient amplitude to be detected by the planned GW detectors at that time, but this conclusion is changed by the subsequent analysis of \cite{arXiv:0802.3524,arXiv:0809.1899}, which also accounts for the presence of magnetic fields and for the time evolution of the MHD turbulence by inverse cascade. The MHD turbulence is assumed to be stationary and decorrelating in time, but an overall time dependence is inserted through the decay of the MHD spectrum with the turbulent cascades. It is found that, compared to the non-helical case, the spectrum of GW generated by MHD helical turbulence peaks at lower frequency with larger amplitude and could have been detected by LISA. However, here as well the Kolmogorov slope is extended up to the largest eddy size, leading to an overestimation of the signal. In Caprini et al. 2009~\cite{arXiv:0906.4976}, the same analysis has been carried on with similar models of the inverse cascade, but in the MHD turbulence spectrum the transition from the low wavenumber part of the spectrum to the inertial range is modeled in a more realistic way, as opposed to extending the Kolmogorov slope. This leads to an amplitude of the GW signal which is lower by nearly one order of magnitude. In Ref.~\cite{arXiv:0906.4976} it is argued that the different assumptions for the unequal time correlation of the source may also play a role in determining this difference (in \cite{arXiv:0802.3524,arXiv:0809.1899} the source is assumed to be stationary with exponential decorrelation, while in \cite{arXiv:0906.4976} the source is assumed to be completely coherent, which leads to an underestimation of the GW signal). 

The last analysis of GW emission from MHD turbulence generated during a first-order PT and freely decaying afterwards is the one of Caprini et al. 2009~\cite{arXiv:0909.0622}. This paper takes into account only non-helical MHD turbulence, but models the source in a time-continuous fashion, starting with zero kinetic energy and building up the Kolmogorov spectrum after one eddy turnover time (instead of instantaneously), and then accounting for the free decay of the source (since the stirring due to the PT lasts only for about one eddy turnover time). The source power spectrum is again modeled with an interpolation in wavenumber between the large and small scale behaviours, determined respectively by causality and by the Kolmogorov (or Iroshnikov-Kraichnan) theory. The time de-correlation of MHD turbulence mimics the model proposed by Kraichnan, as in \cite{arXiv:0802.3524,arXiv:0809.1899}. The claim of this analysis is that, in the case of MHD turbulence, neither the coherent \cite{arXiv:0906.4976} nor the stationary \cite{arXiv:0802.3524,arXiv:0809.1899} approximations previously used in the literature are the correct ones, but the time dependence of the source has to be modeled using a top hat ansatz, in order to best reproduce the Kraichnan decorrelation model. Moreover, it is demonstrated that it takes several Hubble times for the MHD turbulence to decay, due to the very low viscosity: therefore, MHD turbulence has to be modeled as a long lasting GW source, which causes an amplification of the GW spectrum at very large scales by about two orders of magnitude with respect to the short lasting case. The GW spectrum found in this work shows a peak frequency that corresponds to the size of the bubbles at collision $k_{peak}\simeq 1/R_*$, a low frequency slope as $k^3$ by causality, and a high frequency slope as $k^{-5/3}$ if the source has a Kolmogorov spectrum, and as $k^{-3/2}$ if the source has an Iroshnikov-Kraichnan spectrum. The analytical evaluation maintains a certain level of intrinsic uncertainty which probably can only be addressed by numerical simulations of relativistic MHD turbulence. We nevertheless use the result of \cite{arXiv:0909.0622} in our estimation of the GW signal in the eLISA frequency band, section \ref{EWPTsignal}.

\subsection{Overview of the Literature: GW Production in Models of the EWPT}
\label{models}

Parallel to the analyses mentioned above, which aim to find the form of the GW spectrum by bubble collision and MHD turbulence in terms of a few free parameters describing the source, there have been a number of works which consider specific models of the EWPT and evaluate the expected GW signal in the context of the given model. This allows to determine, for example, the parameters $\al$ and $\beta/\HH_*$ in connection with the symmetry breaking potential, and from this make definite predictions of the GW signal. Often the formulas of \Eqref{Omkos} and \Eqref{Omkostur} have been used, more or less implicitly assuming propagation by Jouguet detonation, the envelope approximation, the validity of the Kolmogorov spectrum up to the relevant eddy scale, and not correctly accounting for the GW dispersion relation in the case of turbulence.

Apreda et al. 2001~\cite{hep-ph/0102140} and Apreda et al. 2002~\cite{gr-qc/0107033} use \Eqref{Omkos} and \Eqref{Omkostur}, together with \Eqref{vbdeto} and \Eqref{kappakamion}, to evaluate the GW production during the EWPT in different supersymmetric models where, contrary to the case of the standard model (SM), the transition can be first-order. The GW energy density is found to be low in the minimal supersymmetric standard model (MSSM), but in some parameter range of the next to minimal supersymmetric standard model (NMSSM) can be as high as $\OmG\sim 4 \cdot 10^{-11}$. This happens since the cubic terms present in the NMSSM tree-level Higgs potential can lead to a much stronger PT. This analysis has been redone in Huber and Konstandin 2008~\cite{arXiv:0709.2091} for the nMSSM case (nearly minimal supersymmetric standard model), and considering also the SM augmented by dimension six operators (which can be viewed as the low energy effective description of some strongly coupled dynamics at the TeV scale). In this model, the barrier in the Higgs potential is also present at tree-level \cite{Grojean:2004xa}. The method used to determine the bubble configurations is improved with respect to \cite{gr-qc/0107033} (where it led to an overestimation of the signal). In~\cite{arXiv:0709.2091}, Eq.~(\ref{Omkos}) is used for the estimation of the GW signal by bubble collision, while for MHD turbulence both the results of \cite{astro-ph/0206461} and those of \cite{astro-ph/0603476} are used. Eqs.~(\ref{vbdeto}) and \Eqref{kappakamion} make the connection between the bubble wall velocity and the strength of the PT (implicitly assuming Jouguet detonations). The strong point of this work is that the parameters $\alpha$ and $\beta/\HH_*$ are evaluated directly from the three dimensional Euclidean action $S_3$ of the symmetry breaking theory, and are therefore directly related to the features of the potential. The connection is such that large values of $\alpha$ imply small values of $\beta/\HH_*$, because stronger PTs happen at lower temperature and create larger bubbles. This is bad for detection, since the peak of the spectrum is shifted to lower frequencies, out of the sensitivity range of LISA  (see also the discussion in section \ref{EWPTsignal}). As a consequence, \cite{arXiv:0709.2091} finds that the GW signal is in a large portion of the parameter space too small to be detected. This result is in qualitative agreement with Delaunay et al. 2008~\cite{arXiv:0711.2511}, which provides a more detailed and complete one-loop finite temperature calculation of the Higgs potential augmented by dimension-6 operators, as well as the corresponding GW signal.

Another model leading to a strong first-order PT is provided by extensions of the SM with hidden sector scalars coupled to the Higgs (Espinosa and Quiros 2007~\cite{hep-ph/0701145}, Espinosa et al. 2008 \cite{arXiv:0809.3215}). The GW signal from the PT in the context of this model is calculated, by means of \Eqref{Omkos}, \Eqref{vbdeto} and \Eqref{kappakamion}, but it is found that no observable traces are expected for LISA. Kehayias and Profumo 2010~\cite{arXiv:0911.0687}, on the other hand, calculates the GW signal from a generic SM-like potential: using a semi-analytic approximation for the Euclidean three-action $S_3$, approximated expressions are derived for the tunneling temperature, $\al$ and $\beta/\HH_*$, as functions of the parameters in the potential, which allow to fully specify the GW spectrum. Singlet, triplet, higher dimensional operators and top-flavor extensions to the Higgs potential are considered, and it is found that the addition of a temperature independent cubic term in the potential, arising from a gauge singlet, is the most promising model to enhance the GW signal. The main difference with the previous analyses is the application of the semi-analytic approximation for the three-action. 

More promising for detection are GW arising from PTs in the context of nearly conformal dynamics at the TeV scale, and, by holography, of extra-dimensional warped geometries. This has been first proposed in Randall and Servant 2007~\cite{hep-ph/0607158}, where it is demonstrated that the Randall-Sundrum 1 (RS1) model can provide a strong signature in GW, generated during the first-order PT from the AdS-Schwarzshild phase to the RS1 geometry, which proceeds through the nucleation of brane bubbles. In this work as well, the GW signal is estimated using Eqs.~(\ref{Omkos}, \ref{Omkostur}) and (\ref{vbdeto}, \ref{kappakamion}), reducing the problem to the unique determination of the parameters $\al$ and $\beta/\HH_*$. It is found that the PT can be very strongly first-order and gives an interesting GW signal in the LISA frequency band. Konstandin et al. 2010~\cite{arXiv:1007.1468} confirms this finding, improving the analysis on the model side, by partially accounting for the back-reaction effect from the scalar field undergoing the transition on the metric, and on the GW estimation side, using the formulas given in the most recent simulations of \cite{arXiv:0806.1828}. In Konstandin and Servant 2011~\cite{arXiv:1104.4791} it is emphasised that a detectable GW signal would be a smoking gun of nearly conformal dynamics at the TeV scale. The reason is twofold: on one side, due to the large supercooling, these models give rise to GW signals which are typically larger in amplitude than for an EWPT arising from a generic  polynomial potential (for which a fine-tuning of parameters is required to have a large supercooling); on the other side, the PT results from dynamics at the TeV scale, and therefore the peak frequency is about one order of magnitude larger than for a generic EWPT, which improves the prospects for observation.

We conclude our review of the literature with the work of Nicolis 2004~\cite{gr-qc/0303084} and Grojean and Servant 2007~\cite{hep-ph/0607107}: these papers aim to provide model independent analyses of the GW signal from a first-order PT, scanning the values in the parameter space ($\al$, $\beta/\HH_*$) which lead to an observable signal. However, these analyses need to be updated in at least two aspects. First of all, they use Eqs.~(\ref{Omkos}, \ref{Omkostur}) and (\ref{vbdeto}, \ref{kappakamion}) to estimate the GW signal: as we have mentioned, these equations are only valid in the case of Jouguet detonation and in the envelope approximation, and give a wrong estimate of the GW signal from turbulence. Secondly, they consider $\al$ and $\beta/\HH_*$ as independent parameters, while it is demonstrated, for example in Ref.~\cite{arXiv:0709.2091}, that this is not the case in the context of complete EWPT models. In the following, we aim to provide the most reliable possible analysis of the GW signal relevant for eLISA, given the present status of the research on this subject, correcting for these aspects which are biasing previous analyses.

\subsection{Possible GW Signal in the eLISA Frequency Band}
\label{EWPTsignal}

The GW signal from a first-order EWPT is given by the sum of the GW spectra due to bubble collision and MHD turbulence. Concerning bubble collision, Huber and Konstadin 2008 \cite{arXiv:0806.1828} provides the most recent simulation of the collision of bubbles in the envelope approximation. We therefore take the result of this work for our estimation of the signal in the eLISA frequency band. The GW spectrum resulting from the simulation is \cite{arXiv:0806.1828,arXiv:1007.1468}
\bea
\left. h^2 \OmG(f) \right|_{coll} &=& 1.67 \cdot 10^{-5}   \, \left( \frac{\HH_*}{\beta} \right)^2 \left( \frac{\kappa\alpha}{1+\alpha} \right)^2  \left(\frac{0.11\,v_b^3}{0.42+v_b^2}\right) \left( \frac{100}{g_*} \right)^{\frac{1}{3}} \frac{3.8 \, f_{peak}\, f^{2.8}}{ f_{peak}^{3.8}\,+\,2.8\, f^{3.8}}\,, \label{OmGbub} \\
f_{peak} &=& 16.5 \cdot 10^{-3} \, {\rm mHz} \left(\frac{0.62}{1.8-0.1v_b+v_b^2} \right) \left(\frac{\beta}{\HH_*} \right) \left(\frac{T_*}{100\,{\rm GeV}}\right) \left( \frac{g_*}{100} \right)^{\frac{1}{6}} \label{fbub}\,.
\eea
This result is comparable to the one of Eqs.~\eqref{Omkos} and \eqref{peakkos}, with the difference that the peak frequency depends somehow on the bubble wall velocity $v_b$, and also that the dependence of the GW amplitude on $v_b$ is slightly modified. The peak frequency approaches the one of Eq.~\eqref{peakkos} in the limit $v_b\rightarrow 0$, while it is slightly smaller for large velocities\footnote{In principle, in the limit of small bubble wall velocity the envelope 
approximation breaks down, because one enters the regime of propagation by deflagration.}. This corresponds to a peak wave-number of the order of $\beta$, the inverse characteristic time-scale of the PT. As mentioned before, the physical reason is that the bubble collision source can be approximated as totally coherent (see \cite{arXiv:0901.1661,arXiv:1005.5291}). The most notable aspect of the analysis of \cite{arXiv:0806.1828} is that, thanks to the improved numerical accuracy, 
it has been possible to determine the shape of the GW spectrum. In Eq.~\eqref{OmGbub} we have introduced the fit to the shape of the GW spectrum resulting from the numerical simulations, given in Eq.~(19) of \cite{arXiv:0806.1828}. Note that this fit is optimized for a frequency range close to the peak frequency, and does not correctly reproduce the asymptotic low frequency behavior, which should be $f^3$ by causality (see section \ref{general}). We use it nevertheless to determine the full spectrum, since anyway the interesting frequency region for detection is the one close to the peak. The above result is also in broad agreement with the rough estimates given in Eqs.~\eqref{kstar} and \eqref{OmgwPT}. In particular we see, from \eqref{OmgwPT}, that for the bubble collision case the relative energy density available in the source for the GW generation is given by $(\rho_s / \rho_{tot})_* = \rh{kin}/(\rh{rad}^*+\rh{vac}) = \kappa\,\alpha/(1+\alpha)$. 

Concerning MHD turbulence, for our estimation of the signal in the eLISA frequency band we use the result of the most recent analysis of Caprini et al. 2009~\cite{arXiv:0909.0622}. Here the GW spectrum is evaluated analytically; a simple fit to the result of \cite{arXiv:0909.0622} is given in Caprini et al. 2010~\cite{arXiv:1007.1218} and reads:
\bea
\left. h^2 \OmG(f) \right|_{turb} &=& 8 \, (h^2\Om_{R})\, \left(\frac{\kappa\,\alpha}{1+\alpha}\right)^{\frac{3}{2}}\,
\left( \frac{\HH_*}{\beta} \right) \,  \, v_b\, \frac{\left(\frac{f}{f_{peak}}\right)^3}{\left[ 1+ \frac{f}{f_{peak}} \right]^{\frac{11}{3}} \left(1+4 \frac{k_*}{H_*}\right)} \;,   \label{OmGturb} \\
f_{peak} &=& 2.7 \cdot 10^{-2} \, {\rm mHz} \, \frac{1}{v_b}\, \left(\frac{\beta}{\HH_*} \right) \left(\frac{T_*}{100\,{\rm GeV}}\right) \left( \frac{g_*}{100} \right)^{\frac{1}{6}} \label{fturb} \, ,
\eea
where $k_* = 2\pi\,f\,a_0 / a_*$ is the physical wave-number at the time of production.
Since these formulas are less commonly used in the literature than Eqs.~\eqref{OmGbub} and \eqref{fbub}, let us analyse them in some detail. In \cite{arXiv:0909.0622} it is argued that for MHD turbulence, both the velocity field and the magnetic field contribute to the GW source: the collision of the bubbles stirs the fluid leading to kinetic turbulence, and the kinetic energy of the turbulence is then converted partly into magnetic energy by the MHD cascade, until equipartition is reached, i.e. the same amount of energy is contained into turbulent motions and the magnetic field,  
$\rho_T\simeq \rho_B$. Because of the (well justified) assumption of equipartition\footnote{The tail of the MHD spectrum is not firmly established in the theory of MHD turbulence, so here we consider for simplicity that the magnetic field as well develops a Kolmogorov tail in its spectrum, just as the kinetic turbulence: $k^{-11/3}$. In \cite{arXiv:0909.0622} the Iroshnikov-Kraichnan shape was also considered, but the difference is irrelevant as far as detection prospects are concerned.}, we refer only to the kinetic energy density of the turbulent source to represent the total energy available in the MHD source, setting $\rh{T}+\rh{B} \simeq 2 \rh{T}$. In~\cite{arXiv:1007.1218}, the relative energy available in the source for the GW generation is given in terms of the parameter 
$\Om_s^*=(\rho_s / \rho_{tot})_*=2 \rh{T} /\rho_*$; in \Eqref{OmGturb}, we have expressed this in terms of 
$\kappa\,\alpha/(1+\alpha) = \rho_{kin}/(\rho_{rad}^*+\rh{vac})$. As for the bubble collision case, we adopt the parameter 
$\kappa$ quantifying the kinetic energy of the fluid to represent the total energy available in the MHD source. We are therefore assuming that the efficiency of converting the bulk kinetic energy of the fluid motions due to the bubble expansion into turbulent motions after collision is just one: all the energy of the bulk motions $\rh{kin}$ is converted to turbulent kinetic energy and is then transformed partially into magnetic energy by the cascade, $\rh{kin}\simeq 2\rh{T}$. A proper estimate of the efficiency with which the bulk motions due to bubble expansion transform into chaotic turbulent motions has never been performed; however, the kinetic Reynolds number of the universe at the time of the EWPT and at the scale of the bubble radius is of the order of ${\rm Re}(R_*)\sim 10^{13}$ \cite{arXiv:0909.0622}. Given this extremely high value of the Reynolds number, corresponding to an extremely low value of the kinetic viscosity, it seems very reasonable to consider that the bulk fluid motions are converted into kinetic turbulence without any dissipative loss, at least at the scale of the bubble radius. Furthermore, in Ref.~\cite{arXiv:0909.0622} it is demonstrated that, also due to the very small kinetic viscosity, the MHD turbulence dissipates slowly in the early universe: therefore, the GW source is in general active for many Hubble times after the completion of the PT. This fact needs to be taken into account when evaluating the GW spectrum. That is why, the relative energy density available in the source for the GW generation has been normalised in \cite{arXiv:0909.0622} to the total radiation energy density in the universe, $\Om_s^*=(\rho_s / \rh{tot})_*$ with 
$\rh{tot}=\rh{rad}$ after the completion of the PT. Note that this is not in contradiction with what we did above, since after the completion of the PT 
$\rh{tot}=\rh{rad}$, while during the PT $\rh{tot}=\rh{rad}^*+\rh{vac}$ (we remind that $\rh{rad}^*$ denotes the thermal energy density present in the universe at the moment of the PT). For the pre-factor in \Eqref{OmGturb} we are simply accounting for the fact that the vacuum energy plus thermal energy present at the moment of the PT is completely converted into radiation energy at the end of the PT (plus, obviously, the turbulent kinetic and magnetic energies, but these are subdominant with respect to the radiation energy density in order not to spoil the homogeneity and isotropy of the universe). 

The fact that the source is long lasting changes in \Eqref{OmGturb} the dependence of the GW energy density with 
$(\rho_s / \rh{tot})_*$ and $\HH_*/\beta$ with respect to the scaling given in \Eqref{OmgwPT}, and causes the slope of the MHD signal to turn from $f^3$ to $f^2$ at sub-horizon scales $a_* H_*/(8\pi a_0)<f< f_{peak}$ \cite{arXiv:0909.0622}. Moreover, in the case of MHD turbulence, the peak frequency corresponds to the inverse characteristic length-scale of the source, i.e. the bubble size towards the end of the PT, $R_*=v_b/\beta$. Again, the physical reason for this is related to the properties of temporal correlation of the GW source: in \cite{arXiv:0909.0622} is it argued that MHD turbulence should be modeled with a top hat temporal correlation, since this is the best way to reproduce the Kraichnan decorrelation model, which state that the turbulent velocity field on a given scale $\ell$ is not any longer correlated after a characteristic time-scale given by the eddy turnover time on the scale $\ell$, $\tau_\ell=v_\ell/\ell$. GW sources with a top-hat decorrelation structure typically lead to GW spectra peaked at the inverse characteristic length-scale of the source \cite{arXiv:1005.5291,arXiv:0901.1661}. 

From Eqs.~\eqref{OmGbub}, \eqref{fbub} and \eqref{OmGturb} and \eqref{fturb} it appears that five parameters must be known to determine the GW spectrum, namely $\alpha$, $\beta/\HH_*$, $\kappa$, $v_b$ and $T_*$. These are not independent parameters, and their evaluation, once a specific model for the first-order PT is given, requires involved numerical computations:  $\alpha$, $\beta/\HH_*$ and $T_*$ are in principle determined from the shape of the potential at finite temperature, while for $\kappa$ and $v_b$ one needs to know the details of the propagation of the bubble wall in the surrounding plasma. We now revise how the problem of evaluating these parameters has been tackled in some works in the literature, and explain the strategy we adopt to use the results of these analyses to calculate reliably the GW spectra. 

\subsubsection{Phase Transition Parameters}
\label{PTpar}

In the first analyses of the GW signal from PTs, a general approach was provided to calculate the parameters $\beta/\HH_*$ and $T_*$ \cite{Turner:1992tz,FERMILAB-PUB-91-323-A,astro-ph/9211004}. The approach is based on the Taylor expansion of the tunneling action around the time when the transition completes: 
\be
S(t)=S(t_*)-\beta(t-t_*)+\mathcal{O}((t-t_*)^2)\,, \quad \quad {\rm so~that} \quad \quad \frac{\beta}{\HH_*}= T_* \left. \frac{dS}{dT} \right|_{T_*}\,,
\ee
with $dT/dt=-T\HH$. The `temperature of the transition' $T_*$ is defined as the temperature at which the probability of nucleating one bubble per Hubble volume per Hubble time is equal to one: denoting $\Gamma(t)$ the bubble nucleation rate per unit volume and time, one has
\be
\Gamma(t) \simeq T^4 {\rm e}^{-S(t)}\,, \quad \quad \frac{\Gamma}{H^4}\sim 1 \quad \rightarrow \quad S(T_*)=-4\ln\frac{T_*}{m_{P}}\,,
\ee
where $m_P$ is the Planck mass. Since the Hubble time is the characteristic time-scale, Refs.~\cite{hogan86,astro-ph/9211004} set $\beta \sim \HH_* S_*$ as an order-of-magnitude estimate, so that $\beta/\HH_*\sim  -4 \ln (T_*/m_P)$. This provides a relation among the parameters $\beta$ and $T_*$, and for the EW scale gives the value 
$\beta/\HH_*\sim \mathcal{O}(10^2)$. Using this relation reduces the number of required parameters to determine the GW spectrum from five to four; furthermore, under the assumption of Jouguet detonations, this can be combined with Eqs.~(\ref{vbdeto}) and (\ref{kappakamion}), which give the bubble wall velocity and the efficiency factor as functions uniquely of $\al$. Therefore, in this approach the parameter space to fully specify the GW signal can in principle be reduced to only two parameters: either the couple $(\al\,,~T_*)$ or the couple $(\al\,,~\beta/\HH_*)$. As already mentioned, this has been exploited for example in Nicolis 2004~\cite{gr-qc/0303084} and Grojean and Servant 2007~\cite{hep-ph/0607107}. Since then, there have been other works which have analysed the problem using more accurate estimations of $\beta/\HH_*$. Note, however, that this necessarily implies a loss of generality, because one has to focus on a specific model of the PT. 

A more accurate way to proceed given a model, i.e. some extension of the SM of which one wants to study the EWPT, is explained for example in Huber and Konstandin 2008 \cite{arXiv:0709.2091} and Espinosa et al. 2008~\cite{arXiv:0809.3215}. In general, one needs to determine the bounce solution of the three-dimensional Euclidean action $S_3(T)$, which quantifies the probability of tunneling through $\Gamma \simeq T^4 [S_3(T)/ 2\pi T]^{3/2} \exp(-S_3(T)/T)$. One can then calculate the average number of bubble nucleations per Hubble volume $P(T)$, and the fraction of space that is covered by bubbles neglecting overlap, $f(T)$:
\be
P(T)=\int_T^{T_c} \frac{d \bar{T}}{\bar{T}} \frac{\Gamma(\bar{T})}{H^4}\,, \quad \quad f(T)= \frac{4\pi}{3} \int_T^{T_c} \frac{d \bar{T}}{\bar{T}} \frac{\Gamma(\bar{T})}{H} R^3(T,\bar{T}) \,,
\ee
where $T_c$ denotes the critical temperature of the PT and $R$ the bubble radius. This allows one to determine precisely the initial and final temperatures of the PT: the temperature at which the PT starts is defined by the nucleation of the first bubble, $P(T_n)= 1$, which leads to the well known condition $S_3(T_n)/ T_n \simeq 142$; the temperature at which the PT ends is defined as $f(T_f)=1$. Moreover, knowing the three action as a function of temperature, one can calculate $\beta/\HH_*=T\,d(S_3/T)/dT$, which is in general a function of temperature and has to be evaluated towards the end of the PT to represent, as a matter of fact, the `duration' of the PT \cite{arXiv:0709.2091}. Alternatively, it is possible to relate $\beta$ to the typical bubble size at the end of the PT through $v_b$, $\vev{R}\simeq 3v_b/\beta(T)$, where $\vev{R}$ can be estimated from the maximum of the bubble volume distribution and has to be evaluated at an intermediate temperature between $T_n$ and $T_f$ \cite{arXiv:0709.2091}. Without going into many details, here we only want to point out that from the point of view of estimating the GW signal, the importance of adopting the more accurate way to evaluate $\beta/\HH_*$ described above resides in the fact that this reveals a general relation among the parameters $\beta/\HH_*$ and $\alpha$: large values of $\alpha$, i.e. strong PTs, lead in general to small $\beta/\HH_*$ and therefore to a lower peak frequency. This is due to the characteristic shape of the three action as a function of temperature (see e.g. Fig.~5 of \cite{arXiv:0809.3215}), and has relevant consequences as far as detection prospects are concerned: while large $\alpha$ make the amplitude of the signal increase, the fact that this corresponds to small $\beta/\HH_*$ shifts the peak of the spectrum to lower frequencies, progressively out of the detectable frequency band (c.f. Fig.~\ref{HKdim6} and also the discussion in \cite{arXiv:0709.2091}).

We now turn to the analysis of the parameters $v_b$ and $\kappa$, which together with $\alpha$ and $\beta/\HH_*$ are the other two parameters that one would like to predict for a given particle physics model leading to a first-order PT at temperature $T_*$. As already mentioned, Eqs.~(\ref{vbdeto}) and (\ref{kappakamion}) do not hold in all generality, because depending on the characteristics of the PT the propagation of the bubble in the surrounding plasma need not be a Jouguet detonation. Other kinds of solutions for the bubble expansion are possible, such as weak detonations, deflagrations, hybrids and runway solutions. While in detonations and deflagrations the wall velocity is supersonic or respectively subsonic, leading to a rarefaction wave behind the bubble wall or a compression wave in front of it, in hybrid solutions both the rarefaction and the compression waves are present, and in runaway solutions the wall continuously accelerates, nearly without interaction with the surrounding plasma. Model independent studies of the hydrodynamics of bubble growth are performed for example in Refs.~\cite{Megevand:2009ut,Espinosa:2010hh,Leitao:2010yw}. We concentrate on the work of Espinosa et al. 2010~\cite{Espinosa:2010hh}, where it is shown that the knowledge of the parameter $\alpha$ alone is not enough to determine the bubble propagation mode. For a given $\alpha$, there are many possible values of the wall velocity $v_b$ and of the efficiency coefficient $\kappa$: to fix the solution, one has to specify the friction acting on the bubble wall because of its interaction with the surrounding plasma. The friction grows with the velocity of the expanding bubble and opposes the driving force due to the difference of the free energy across the wall, until the bubble growth reaches a steady state at a constant terminal velocity. The hydrodynamical analysis alone cannot completely determine the terminal wall velocity $v_b$. Rather than solving the complicated system of Boltzmann equations describing the distribution functions of all particle species and their interaction with the Higgs field, as done for example in \cite{Moore:1995ua,Moore:1995si}, Ref.~\cite{Espinosa:2010hh} uses a phenomenological description of the friction $\eta$ as a parameter independent of $v_b$ (see also \cite{Ignatius:1993qn,Megevand:2009ut,Megevand:2009gh}). Model-independent contour plots for $v_b$ and $\kappa$ are then given in the $(\alpha\,,~\eta)$ plane: from a given particle physics model, one can in principle evaluate $\alpha$ and $\eta$, and once these two quantities are known, the results of \cite{Espinosa:2010hh}  provide a way to determine $v_b$ and $\kappa$. 

Moreover, Espinosa et al. 2010~\cite{Espinosa:2010hh} evaluates the energy budget of the PT, i.e. the efficiency of the transfer of the vacuum energy into gradient energy of the Higgs field, bulk kinetic energy of the fluid motions at the bubble wall, and thermal energy. This is extremely important, since it allows to estimate the relative energy available in the source for the GW generation realistically, given a value for the parameter $\alpha$. In the following, we use the results of~\cite{Espinosa:2010hh} for the first time in the calculation of the GW signal. As far as production by bubble collision is concerned, GW can be sourced by both the gradient energy of the Higgs field and the bulk kinetic energy of the fluid, since both components lead to tensor anisotropic stresses. The efficiency parameter $\kappa$, appearing in \Eqref{OmGbub}, needs therefore to be divided into two components: $\kappa=\kappa_\phi+\kappa_v$, where $\kappa_\phi$ denotes the fraction of vacuum energy that goes into gradient energy of the Higgs field, while $\kappa_v$ denotes the fraction that goes into bulk kinetic energy. Inserting the total $\kappa$ in \Eqref{OmGbub}, one can use this equation to model in a continuous way the GW signal from both weak (for which $\kappa_v$ is important) and very strong (for which $\kappa_\phi$ is important) PTs. One has then $\kappa_\phi+\kappa_v+\rh{ther}/\rh{vac} =1$, where $\rh{ther}$ is the thermal energy liberated by the PT, and it is the only component that does not source GW\footnote{To summarise our notations, the total energy density in the universe through the PT can be expressed as $\rho_{tot}=\rho_*= \rh{vac}+\rho_{rad}^*$ just before nucleation, which becomes $ \rh{vac}+\rho_{rad}^*=\rh{kin}^\phi+\rh{kin}^v+\rh{ther}+\rh{rad}^*$ during bubble nucleation, and which becomes then $\rh{kin}^\phi+\rh{kin}^v+\rh{ther}+\rh{rad}^*=\rh{rad}$ at the end of the PT (after percolation, and neglecting the remaining energy density in MHD turbulence). Furthermore we have set $\alpha=\rh{vac}/\rh{rad}^*$, $\kappa_\phi=\rh{kin}^\phi/\rh{vac}$ and $\kappa_v=\rh{kin}^v/\rh{vac}$.}. On the other hand, the gradient energy of the Higgs field does not, by definition, source the MHD turbulence: the efficiency parameter $\kappa$ appearing in \Eqref{OmGturb} corresponds only to $\kappa_v$. 

The parameters $\kappa_\phi$, $\kappa_v$ and $\rh{ther}/\rh{vac}$ are calculated in \cite{Espinosa:2010hh} as a function of $\alpha$ and for two different values of the friction, $\eta=0.2$ and $\eta=1$ (c.f. Fig. 12 of \cite{Espinosa:2010hh}). As the PT gets stronger ($\alpha$ increases), the total kinetic energy $\rh{kin}^\phi+\rh{kin}^v$ naturally gains importance with respect to the thermal energy $\rh{ther}$. Ref.~\cite{Espinosa:2010hh} finds that, as long as the bubble expansion proceeds as a deflagration or a detonation, the gradient energy in the Higgs field is completely subdominant and $\kappa = \kappa_v$. But when $\alpha$ overcomes a certain value, the bubble propagation enters the runaway regime. This means that the PT is so strong, that the friction exerted on the bubble wall by the plasma is not sufficient to counterbalance the driving force due to the pressure difference in the broken and symmetric phases, and the wall continuously accelerates. When this happens, $\kappa_\phi$ is no longer negligible. The value of $\alpha$ after which one enters the runaway regime depends on the friction, and corresponds to $\alpha\gtrsim 0.25$ for $\eta=0.2$, and $\alpha \gtrsim 1$ for $\eta=1$: as expected, \cite{Espinosa:2010hh} finds that as the friction increases, the ratio between $\kappa_v$ and $\kappa_\phi$ increases for fixed $\alpha$, i.e. more energy is transferred to bulk motions of the fluid. On the other hand, for fixed friction and increasing $\alpha$, the gradient energy of the Higgs becomes larger and lager with respect to the bulk kinetic energy.  

The analysis of Ref.~\cite{Espinosa:2010hh} leads to the conclusion that only for a restrained set of values for $\alpha$ and $\eta$ the propagation solution is a detonation, while most likely solutions are deflagrations or runaway solutions. This needs to be taken into account when estimating the GW signal. Even though the value of the friction $\eta$ is very difficult to determine in a given EWPT model (it is in fact known only in the SM \cite{Moore:1995si} and in the MSSM \cite{John:2000zq}), still the analysis of Ref.~\cite{Espinosa:2010hh} provides a much more realistic estimate for the parameters $v_b$ and $\kappa$ than what is usually assumed using Eqs.~(\ref{vbdeto}) and (\ref{kappakamion}): we therefore use the result of this analysis to evaluate the GW signal. Moreover, introducing $\kappa_\phi+\kappa_v$ in \Eqref{OmGbub}, and $\kappa_v$ in \Eqref{OmGturb}, we can add the bubble collision signal and the MHD turbulence signal in a fully consistent way for any given value of $\alpha$.

\subsubsection{Estimate of the GW Signal}
\label{estimate}

From the discussion in the above section, it appears that one is left with four parameters that fully determine the GW signal, according to Eqs.~(\ref{OmGbub}-\ref{fturb}): $\alpha$, $\beta/\HH_*$, $\eta$ and $T_*$, since $\kappa$ and $v_b$ can be determined from $\alpha$ and $\eta$ \cite{Espinosa:2010hh}. For a given model of the PT these parameters are all known: $\alpha$, $\beta/\HH_*$, and $T_*$ from the finite temperature potential, $\eta$ from the particle content and the interactions of the theory. Rigorously, these parameters are not independent of each other, and one cannot in principle perform a completely model-independent study of the GW signal letting the parameters vary freely. Nevertheless, at the end of this section we present the detection prospects of the GW signal in the eLISA frequency band from PTs in the GeV-TeV energy range, trying to remain as model-independent as possible. We have also chosen two examples presented in the literature, which turn out to be the most promising for detection with eLISA and for which we calculate the GW spectra. 

\begin{figure}
\begin{center}
\includegraphics[width=8cm]{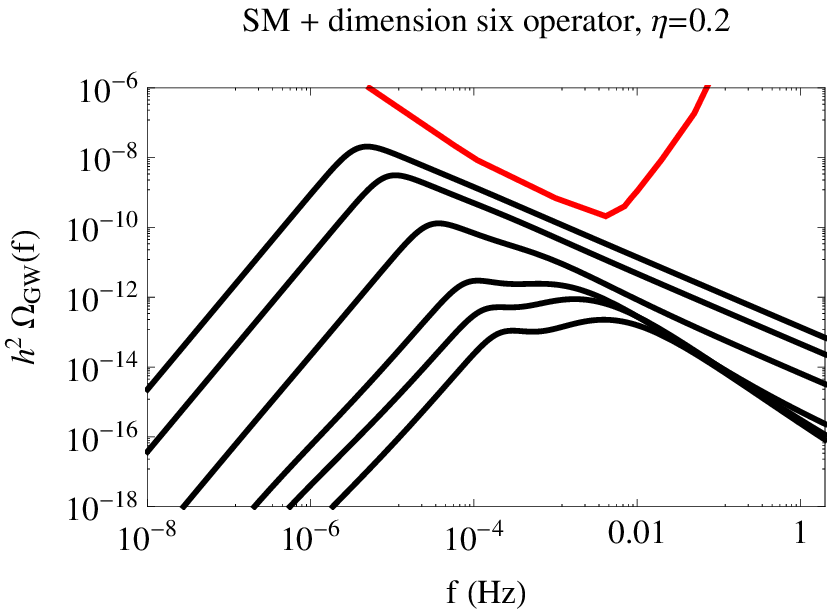}\hspace*{1cm}
\includegraphics[width=8cm]{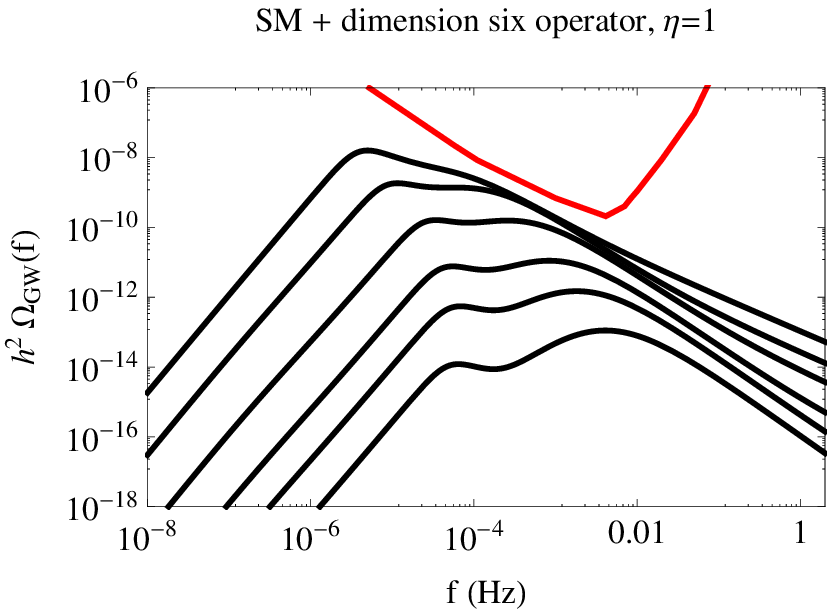}
\caption{GW spectra for the EWPT in the SM augmented by dimension six operators \cite{arXiv:0709.2091}. Left panel, $\eta=0.2$, Right panel, $\eta=1$. The red line shows the eLISA noise curve. The GW spectra correspond, from bottom to top, to increasing values of $\alpha$ as given in Table 1 of Huber and Konstandin 2008~\cite{arXiv:0709.2091}: $\alpha=0.128\,,~\alpha=0.201\,,~\alpha=0.311\,,~\alpha=0.586\,,~\alpha=1.197\,,~\alpha=2.268$. The values of $\beta/\HH_*$ and $T_*$ decrease accordingly (see discussion in the main text).  \label{HKdim6}}
\end{center}
\end{figure}

First of all, let us consider the analysis of Huber and Konstadin 2008~\cite{arXiv:0709.2091}. In this work, two models of the EWPT are studied: the nMSSM and the SM augmented with dimension 6 operators (c.f. section \ref{literature}). In the latter model, the PT can be considerably stronger with respect to the nMSSM, leading to a higher GW signal: we therefore choose this model in the present analysis. Ref.~\cite{arXiv:0709.2091} numerically evaluates the parameters of the PT $\alpha$, $R_*\HH_*$ and $T_*$, for different values of the parameters in the model: the suppression scale of the dimension 6 operator $M$, and the quartic coupling $\lambda$ - related to the physical Higgs mass. It then estimates the GW signal using the formulas presented in section \ref{literature}: here we update this analysis by employing Eqs.~\eqref{OmGbub}-\eqref{fturb} and the work of Espinosa et al 2010~\cite{Espinosa:2010hh} to relate $\kappa$ and $v_b$ to $\alpha$ and $\eta$. The results are given in Fig.~\ref{HKdim6}, for two different values of $\eta$. These figures are obtained as follows: each GW spectrum corresponds to a given set of parameters $\alpha$, $R_*\HH_*$ and $T_*$ (resulting from a choice of the energy scale $M$), according to table I of \cite{arXiv:0709.2091}. To translate the parameter $R_*\HH_*$ into $\beta/\HH_*$ one needs to know the bubble velocity $v_b$, since $R_*\simeq 3 v_b/\beta$ \cite{arXiv:0709.2091}: for each value of $\alpha$, and for two fixed values of the friction $\eta=0.2$ and $\eta=1$, we obtain the wall velocity $v_b(\alpha,\eta)$ by linear interpolation of the contour plot values given in Fig.~10 of Ref.~\cite{Espinosa:2010hh}. This allows a quite general estimation of $v_b$ which goes beyond the usual assumption of Jouguet detonations. Knowing $v_b$, we can relate $\beta/\HH_*=3v_b/(R_*\HH_*)$ directly to $\alpha$: again, we do a linear interpolation of the values of $R_*\HH_*$ given in table 1 of \cite{Espinosa:2010hh} for different values of $\alpha$. We can therefore determine, for a couple of values $(\alpha, \eta)$, the corresponding $\beta/\HH_*$ which one expects in the EWPT model under analysis, i.e. SM model extension by dimension 6 operators. To know the GW signal, it remains to evaluate the efficiency factor $\kappa=\kappa_\phi+\kappa_v$: for each value of $\alpha$, and for the two values of the friction $\eta=0.2$ and $\eta=1$, we determine it from figure 12 of \cite{Espinosa:2010hh}. As explained above, the GW signal from bubble collision, given in Eq.~\eqref{OmGbub}, is due to the total $\kappa=\kappa_\phi+\kappa_v$, since both the gradient energy of the Higgs field and the kinetic energy of the bulk fluid motions lead to anisotropic stresses when the bubbles collide. On the other hand, only the kinetic energy of the bulk fluid motions contributes to the GW signal from MHD turbulence: in Eq.~\eqref{OmGturb} we therefore substitute $\kappa=\kappa_v$. As mentioned above, we assume that after bubble collision all the bulk kinetic energy is transferred to MHD turbulent energy without any dissipative loss, due to the extremely low kinetic viscosity of the plasma. 

Fig.~\ref{HKdim6} shows the sum of the GW signal from bubble collision and MHD turbulence for $\eta=0.2$ and $\eta=1$. The peak of the GW spectrum from MHD turbulence occurs at higher frequencies, since it corresponds to the inverse characteristic length-scale of the PT, the bubble radius at collision $R_*$ (Eq.~\eqref{fturb}, as opposed to \Eqref{fbub}). It appears that as $\alpha$ increases, the signal from bubble collision gains importance with respect to the one from MHD turbulence. This can be explained as follows (see also Fig. 12 of \cite{Espinosa:2010hh}): in the deflagration or detonation regime, the same kinetic energy given by $\kappa_v$ determines both signals, from bubble collision and from MHD turbulence. But as soon as one enters the runaway regime, the gradient energy of the Higgs field $\kappa_\phi$ starts to play a role, and eventually, as the PT gets very strong, it completely dominates over $\kappa_v$. When this happens, there is no more GW production by MHD turbulence and the signal is only due to bubble collision. Clearly, the value of $\alpha$ for which this happens depends on the friction: if the friction is large, the transfer of the vacuum energy to bulk kinetic energy is more efficient and one needs a stronger PT (larger $\alpha$) to reach the point in which MHD turbulence is not any longer present. Therefore, in Fig.~\ref{HKdim6} the GW signal from MHD turbulence is still visible at large $\alpha\simeq 2.27$ for $\eta=1$; on the other hand, for $\eta=0.2$, when $\alpha\gtrsim 0.5$ the gradient energy of the Higgs field becomes dominant with respect to the bulk kinetic energy, and no MHD turbulence is present in the fluid any longer. Unfortunately, the GW signal from a first-order EWPT in the SM model plus dimension 6 operators never reaches the eLISA noise curve, not even in the most optimistic case in which the PT is the strongest $\alpha\simeq 2.27$, and the friction is the largest, $\eta=1$. Note that the PT in this model cannot get stronger than $\alpha\simeq 2.27$, since higher $\alpha$ corresponds to smaller values of the suppression scale of the dimension 6 operator $M$, for which the system becomes metastable and the Universe gets stuck in the symmetric phase \cite{arXiv:0709.2091}.

For small values of $\alpha$, both the GW signals from bubble collision and from the MHD turbulence are sourced only by the bulk kinetic energy fraction $\kappa_v$. However, the GW spectrum from MHD turbulence dominates the signal, determining the global peak. The reason is simply that the GW spectrum from MHD turbulence continues to rise after the bubble collision peak, because it peaks at a frequency which is higher by a factor $1/v_b$, see \Eqref{fturb}. We also remark that as $\alpha$ increases, $\beta/\HH_*$ and $T_*$ both decrease, so that the global peak of the GW spectrum is at smaller and smaller frequency (see Eq.~\eqref{fturb} and \Eqref{fbub}), and progressively gets shifted away from the region of maximal sensitivity of eLISA. This fact had already been remarked in Ref.~\cite{arXiv:0709.2091}, which for the first time correctly determined $\beta/\HH_*$ and $T_*$ from the Euclidean three action, without treating them as parameters independent of $\alpha$. This behaviour is not peculiar of the EWPT model under analysis here, but comes from the general shape of the three action $S_3(T)$ as a function of temperature: it is therefore to be generically expected for thermal first-order PTs. We finally remark that the double peak structure characteristic of the GW spectrum arising from the sum of the bubble collisions and MHD turbulence signals had already been pointed out in Ref.~\cite{gr-qc/0303084}: however, in the present analysis it is the peak due to the MHD turbulence which occurs at higher frequency, the one from bubble collisions being related to the duration of the PT $1/\beta$ while the one from MHD turbulence to the size of the bubbles $R_*$. This is different from what is found in \cite{gr-qc/0303084}, where \Eqref{Omkostur} was used, which states that the peak frequency for the GW spectrum from MHD turbulence is related to the turbulent eddy turnover time and not the inverse size of the bubbles $1/R_*$ as in Eqs.~\eqref{OmGturb} and \eqref{fturb}. 
  
\begin{figure}
\includegraphics[width=10cm]{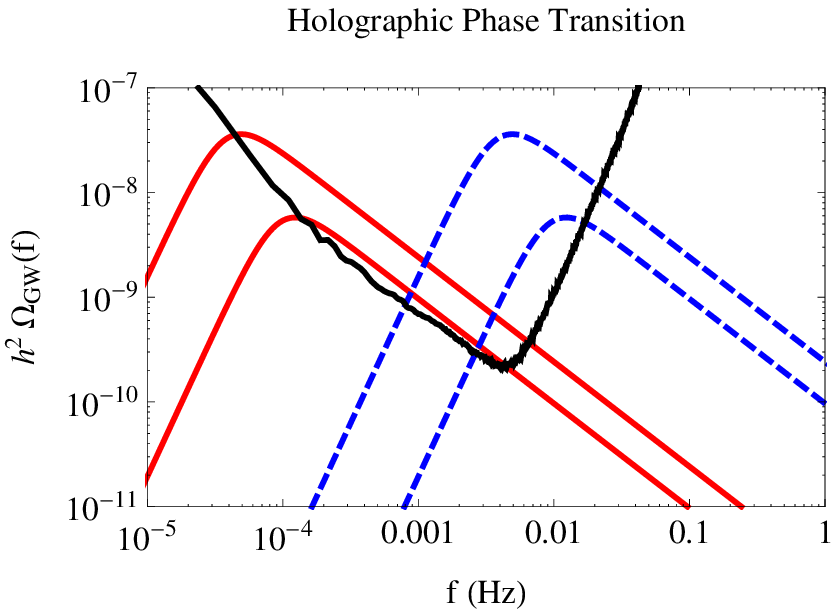}
\caption{GW spectra for the holographic PT \cite{hep-ph/0607158,arXiv:1007.1468}. The black line shows the eLISA noise curve. The solid lines are the GW spectra for $T_*= 100$ GeV, and from top to bottom $\beta/\HH_*=6$ and $\beta/\HH_*=15$; the dashed lines are for $T_*= 10^4$ GeV, and from top to bottom again $\beta/\HH_*=6$ and $\beta/\HH_*=15$.
\label{HolPT}}
\end{figure}

We have seen that the EWPT in the SM extended by dimension 6 operators is not strong enough to give rise to a GW signal which overcomes the eLISA noise curve. Moreover, to determine the GW spectrum by bubble collision, we have used the result of the numerical simulations \cite{arXiv:0806.1828}, which is in principle only valid in the envelope approximation. For a weakly first-order PT, the bubble wall velocity can be relatively small causing the bubbles to expand as deflagrations: for example, in the model under analysis, for $\alpha=0.128$ and $\alpha=0.201$ for $\eta=1$, the bubble wall velocity is respectively $v_b=0.09$ and $v_b=0.2$. It is not clear whether the numerical simulations are adapted to describe such a situation, since the validity of the envelope approximation most probably breaks down. Therefore, as a second example of a PT, which instead can give rise to a detectable GW signal for eLISA, we consider the holographic PT corresponding to the stabilisation of the radion in the Randall-Sundrum model, first analysed by Randall and Servant 2007~\cite{hep-ph/0607158} and then by Konstandin et al 2010~\cite{arXiv:1007.1468}. The appealing feature of this PT is its extreme supercooling: the PT is strongly first-order, so that $\alpha$ attains very large values leading to a high amplitude in the GW signal. The validity of the envelope approximation is not put into question in this case. The production of GW has been analysed already both in \cite{hep-ph/0607158} and \cite{arXiv:1007.1468}: in particular, Ref.~\cite{arXiv:1007.1468} uses Eqs.~\eqref{OmGbub} and \eqref{fbub} to estimate the GW signal. We show the corresponding GW spectrum in Fig.~\ref{HolPT}. In this case, the PT is so strong that the bubble propagation is always in the runaway regime \cite{arXiv:1007.1468}; for reasonable values of the friction, we can therefore assume that the gradient energy of the Higgs field strongly dominates the energy balance of the PT, and set $\kappa\simeq \kappa_\phi \simeq 1$ and $\kappa_v\simeq \rh{ther}/\rh{vac}\simeq 0$. Since there are no bulk motions of the fluid, no contribution from MHD turbulence is expected for such a strong PT, and the bubble wall velocity equals the speed of light $v_b=1$. Moreover, from Eq.~\eqref{OmGbub} we see that for $\alpha\gtrsim 10$, the amplitude of the GW spectrum becomes virtually independent from $\alpha$. The only parameters one is left with to estimate the GW signal are therefore $\beta/\HH_*$ and $T_*$. The parameter $\beta/\HH_*$ can take a wide range of values, depending on the rank of the gauge group in the dual CFT theory (c.f. table I in ~\cite{arXiv:1007.1468}). In Fig.~\ref{HolPT}, we show four examples of GW spectra for a choice of parameters for which the signal is much higher than the eLISA noise curve: respectively, $T_*=100 $ GeV, $\beta/\HH_*=6$ and $\beta/\HH_*=15$, and $T_*=10^4 $ GeV, $\beta/\HH_*=6$ and $\beta/\HH_*=15$.  

\begin{figure}
\begin{center}
\includegraphics[width=8cm]{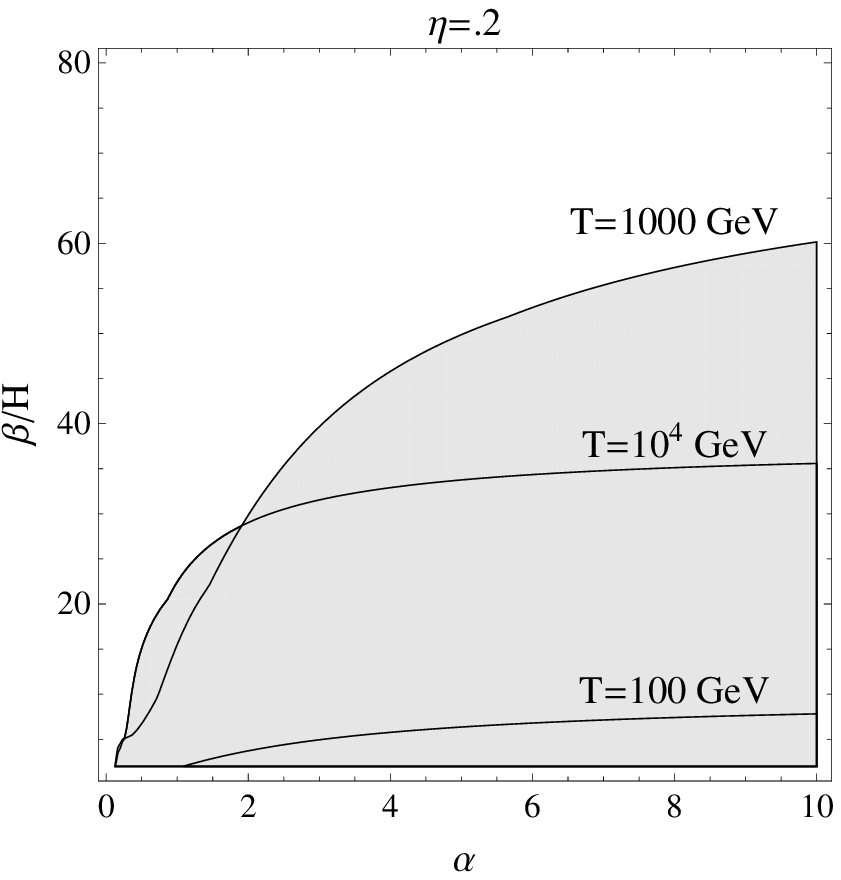}\hspace*{1cm}
\includegraphics[width=8cm]{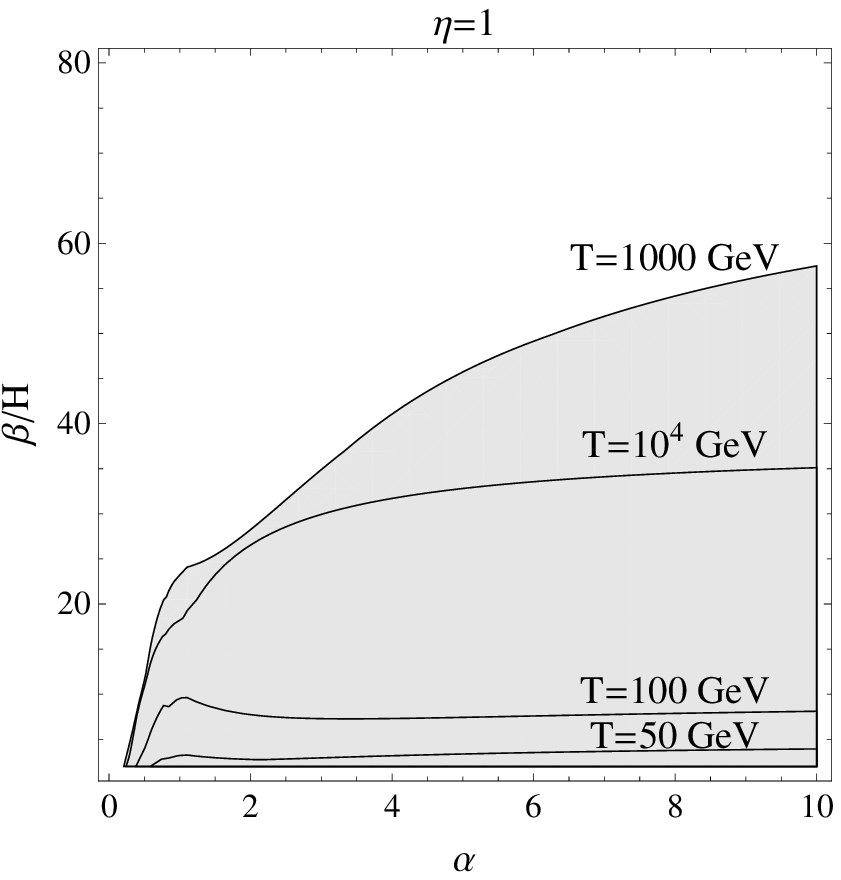}
\caption{Contour plots in the $(\alpha\,,\beta/\HH_*)$ parameter space for different values of $T_*$ and two values of the friction $\eta$. eLISA is most sensitive to PTs occuring at temperatures of the order of 10 TeV, for which it can detect a wider range of values of $\alpha$ and $\beta/\HH_*$. This means that if the PT takes place at high temperature, eLISA can detect its GW signal even if the PT is not very strong and occurs quite fast. The shadowed regions correspond to a signal to noise ratio of at least one.
\label{alphabetasurH}}
\end{center}
\end{figure}

From the two examples above, it appears that once a model for the PT is chosen, $\alpha$, $\beta/H_*$ and $T_*$ are in general not independent among each other; in particular, as $\alpha$ increases, $\beta/\HH_*$ and $T_*$ decrease. Indeed, this behaviour is due to the shape of the Euclidean three-action as a function of temperature: therefore, we can expect it to hold quite generally, and to occur in many different models of the PT. However, this relation among the parameters cannot be specified analytically or numerically in more precise terms without knowing the three-action in details, i.e. without restricting to a given model. Therefore, it is not possible to perform a model independent analysis which realistically accounts for the relation among $\alpha$, $\beta/H_*$ and $T_*$. As a consequence, in the following we have chosen to let them vary freely, and in Fig.~\ref{alphabetasurH} we present contour plots in the plane $(\alpha\,,~\beta/H_*)$ for different values of $T_*$. Note that $v_b(\alpha, \eta)$ and $\kappa(\alpha, \eta)$ are evaluated from the analysis of Espinosa et al. 2010~\cite{Espinosa:2010hh}, so we do not make use of the Jouguet detonation hypothesis. The criterium we adopted in Fig.~\ref{alphabetasurH} is that we consider a GW spectrum `detectable' (shadowed region) as soon as it overcomes the eLISA noise curve. However, the fact that for a given set of the parameters ($\alpha$, $\beta/H_*$, $T_*$) the corresponding GW signal is detectable does not mean that there can exist a realistic EWPT model in which this set of parameters is allowed. 

The first thing to notice about the contours in Fig.~\ref{alphabetasurH} is that small values of $\beta/\HH_*$ are easier to detect: this is because the amplitude of the GW spectrum is quadratic (linear for MHD turbulence) in $H_*/\beta$, c.f. Eqs.\eqref{OmGbub} and \eqref{OmGturb}. However, this happens only for high enough temperature: if $\beta/H_*$ is small and the temperature as well, then the signal peaks at too small frequency to hit the eLISA noise curve (we have seen that it is the case, for example, for the SM plus dimension 6 operators). Correspondingly, if $T_*$ becomes too high, the signal peaks at too large frequency and the same value of $\beta/\HH_*$ is no longer detectable: therefore, the shadowed region does not grow monotonically with $T_*$. Still, the detection region increases for increasing $T_*$ until $T_*\simeq 1000$ GeV, because of the shape of the eLISA noise curve: given a value for $\beta/H_*$, the amplitude of the spectrum can be too small for detection if the corresponding peak frequency is low; increasing the temperature does not affect the amplitude but shifts the peak to higher frequencies, causing it to enter the eLISA noise curve, which decreases as a function of frequency for small frequency.  For the smaller value of the friction $\eta=0.2$, the signal is less visible at small $T_*$: this is due to the fact that, for the same $\beta/H_*$, the GW spectrum from MHD turbulence peaks at higher frequencies with respect to the bubble collision one. If the friction is small, the transfer of the vacuum energy to bulk kinetic energy is less efficient and correspondingly for the same value of $\alpha$ MHD turbulence is less important, rendering the GW spectrum less visible. This is the same reason for which, for $\eta=1$, the curves show a peak at $\alpha\simeq 1$: for fixed $\beta/H_*$, when $\alpha$ reaches the value of about one, the MHD turbulent peak enters the eLISA noise curve. But if $\alpha$ grows further, the amplitude of the MHD turbulent part of the spectrum is less and less important ($\kappa_v\rightarrow 0$) and for the same value of $\beta/\HH_*$ the signal is no longer detectable. For $\eta=0.2$, the peak occurs at smaller $\alpha\simeq 0.3$. We also remark that the detectability of the signal stops increasing with $\alpha $ for sufficiently high values of $\alpha$, since the amplitude of the GW spectrum reaches saturation because of the dependence as $\al/(1+\al)$ in Eqs.~\eqref{OmGbub} and \eqref{OmGturb}. Note that this effect is absent for $T_*\simeq 1000$ GeV, because for the values of $\beta/H_*$ showed in the plot and for $T_*\simeq 1000$ GeV, the peak of the GW spectrum corresponds to the minimum of the eLISA noise curve: even if an increase in $\alpha$ causes a very small increase in the amplitude, this can be enough for the GW spectrum to enter the noise curve. 

\begin{figure}
\includegraphics[width=10cm]{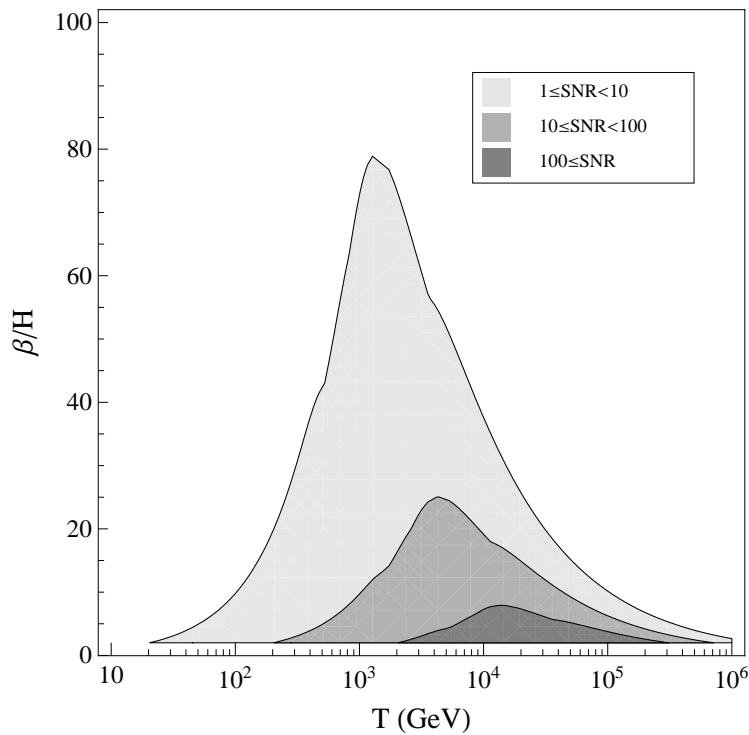}
\caption{Contour plots in the $(T_*\,,\beta/\HH_*)$ parameter space for very strong PTs, $\alpha\gtrsim 10$, and reasonable values of the friction $\eta\lesssim 1$. Again, we see that eLISA is most sensitive to PT at temperatures around 
$T_*\simeq 10$ TeV, where it can detect PTs occurring slowly (small $\beta/\HH_*$) with a signal to noise ratio higher than 10. 
\label{TbetasurH}}
\end{figure} 

From the analysis of \cite{Espinosa:2010hh} it appears that very strong PTs have bubbles propagating as runaway solutions: the bubble wall velocity is simply $v_b=1$, and the energy balance is such that practically all of the vacuum energy is transformed into gradient energy of the Higgs field instead of bulk motions of the plasma, $\kappa=\kappa_\phi=1$, at least for reasonable values of the friction, say $\eta\lesssim 1$. Moreover, only bubble collision can act as a source of GW, and the GW spectra become independent on $\alpha$ for $\alpha\gtrsim 10$. For very strong PTs, in summary, one is left only with the two parameters $\beta/\HH_*$ and $T_*$ to determine the GW signal (the holographic PT is an example of this). We analyse very strong first-order PTs $\alpha\gtrsim 10$ in Fig.~\ref{TbetasurH}, which shows contour plots in the parameter space ($\beta/\HH_*$, $T_*$) for different values of the signal to noise  (SN) ratio. The SN ratio has been evaluated by maximizing, for each frequency, the ratio among the value of the GW signal and the one of the eLISA noise curve at that frequency. The criterion adopted in Fig.~\ref{alphabetasurH} would correspond in this language to a SN ratio of at least one. We see that the SN grows for small values of $\beta/H_*$, since this corresponds to high amplitudes of the GW signal. The signal is visible for values of $\beta/H_*$ which first increase as a function of $T_*$, and then decrease for $T_*>1000$ GeV:  this is the same behaviour observed in Fig.~\ref{alphabetasurH} and it is due to the position of the peak of the GW spectrum, whether or not it falls in the eLISA frequency band. Moreover, it appears that the SN ratio is maximal for $T_*$ of the order of 10 TeV and small values of $\beta/\HH_*$, which maximize the amplitude: in this case, the GW signal falls in the minimum of the eLISA noise curve.


\section{Cosmic (Super-)Strings}
\label{Strings}

\subsection{Introduction and Overview of the Literature}
\label{IntroStrings}

Cosmic strings are linear concentrations of energy that have a cosmological size and are produced at early-universe phase transitions in a variety of high-energy physics scenarios~\cite{kibble,VS,HK}. They are relics of the spontaneous breaking of symmetries that can arise for instance in grand-unified and super-symmetric theories. A traditional example is provided by the Abrikosov-Nielsen-Olesen flux tubes associated to the breaking of a local $U(1)$ symmetry in the Abelian-Higgs 
model~\cite{NO}. More recently, following an original idea in \cite{superstring}, it has been appreciated~\cite{majumdar,tye,dvalenkin} that cosmic strings occur also as fundamental objects of string theory, known as cosmic super-strings, when the necessary stability requirements are satisfied~\cite{CMP}. The formation of cosmic strings depends on the kind of symmetries that are broken in the early universe (they form whenever the vacuum group manifold is not simply connected), but not on the dynamics of the corresponding phase transitions: the phase transitions can be first-order or second-order and they can occur equally well during the thermal evolution of the universe or at the end of inflation, e.g. at the end of hybrid inflation for field theory cosmic strings and at the end of brane inflation for cosmic super-strings. Contrary to stable domain walls and monopoles, which would come to dominate the expansion of the universe and lead to cosmological problems, cosmic strings evolve so as to contribute only to a relatively small fraction of the total energy density of the universe at any time. Yet they involve concentrations of very high energy densities with relativistic velocities, which can have a variety of astrophysical and cosmological consequences, notably GW production. The detection of cosmic string signatures would provide a spectacular probe of the physics at the highest energy scales, well above the reach of any accelerator experiment. 

After being produced in the early universe, and after a possible early friction-dominated evolution (where the motion of the strings is damped by their interactions with the thermal plasma), a network of cosmic strings evolves towards an attractor solution called "scaling regime". In this regime, the correlation length of long string segments divided by the Hubble radius, as well as their energy density divided by the total energy density of the universe, remain constant in time. In order to scale, the network of long strings must continuously loose a significant part of its energy, and it is widely believed that this is achieved through the production of closed loops - i.e. strings whose length is shorter than the Hubble radius. The typical size of the loops when they are produced, however, is still uncertain by many orders of magnitude (see below). Loops are produced when long string segments cross themselves and reconnect. The loops may then reconnect to the long string network or fragment into smaller loops, but eventually a population of isolated loops remains. These loops oscillate relativistically under the effect of their tension and shrink by emitting GW (together with other possible kind of radiation and particles, depending on the cosmic string model). The net result is that a significant fraction of the energy density of cosmic strings is continuously converted into GW, which then redshift until the present epoch. The resulting spectrum covers a wide frequency range and can be relevant for different kinds of GW probes, in particular pulsar timing observations, eLISA and ground-based interferometers. 

The observational signatures of cosmic strings vary in particular with two fundamental parameters that depend on the underlying high-energy physics scenario: their tension $\mu$ (which is equal to their energy per unit length) and their reconnection probability $p$. The gravitational effects of cosmic strings depend essentially on the dimensionless parameter $G \mu$, where $G$ is Newton constant. The reconnection probability is often $p = 1$ for field theory cosmic strings, but it can be smaller than unity for cosmic super-strings~\cite{psuper}. This is expected to increase the abundance of strings (although the precise dependence is still uncertain, see below) and the chances to detect their signatures. There are already a number of constraints on the string tension, which depend more or less strongly on the parameters that describe the network evolution. For instance, constraints from the Cosmic Microwave Background and the Large Scale Structures require typically $G \mu < \mathrm{few} \times 10^{-7}$ for $p = 1$, see e.g.~\cite{wyman}. This bound is conservative in the sense that it does not depend on the properties of the loops. As we will see, current pulsar timing data lead already to significantly stronger upper bounds on the string tension in the case of large initial loop sizes. Note that the production of particles by cosmic strings leads also to complementary constraints on small values of the string tension, see e.g.~\cite{KKmodes}.

The stochastic GW background from cosmic strings has been extensively studied in the 80's and in the beginning of the 
90's~\cite{early1,early2,early3,early4,early5,early6,caldwell,shellard}, see in particular Caldwell and Allen~\cite{caldwell} for a detailed review of early works. More recently, it has been pointed out by Damour and Vilenkin~\cite{DV} (see also \cite{hnatyk}) that the GW signal from cosmic string loops includes strong infrequent bursts that can be looked for individually and should not be included in the calculation of the stationary and nearly Gaussian background. These GW bursts are produced by string configurations such as cusps (highly boosted pieces of loops where the string doubles back on itself) and kinks (shape-discontinuities that propagate along the strings at the speed of light). The sensitivity of the GW signal to different cosmic string parameters has been further studied in \cite{DVsuper} and the detection of individual GW bursts in \cite{siemensbursts}. The stochastic background from loops was subsequently studied by Hogan et al~\cite{hogan} and Siemens et al~\cite{siemens}. The contribution of the long strings to the GW background has also been recently re-analyzed in \cite{longstrings}, where it was found to be sub-dominant compared to the contribution of the loops except for very small initial loop sizes and in certain frequency range (e.g. at CMB scales, where the GW spectrum from loops is negligible). 

In this Section, we will study the stochastic GW background produced by cosmic string loops. Our interest in this paper is on stationary and nearly Gaussian GW backgrounds, so we will not consider the detection of individual bursts. We will however study the effect of removing the rare bursts from the stochastic background. Our calculation method will be mostly similar to Siemens et al~\cite{siemens}, but with some differences that we will explain as we proceed.

Three general aspects must be modeled in order to compute the GW spectrum from cosmic strings. First, we need a model for the expansion history of the universe. The impact of this has been studied in a more and more sophisticated way in the early works. For instance, Caldwell and Allen~\cite{caldwell} took into account the variation of the number of relativistic species in the thermal plasma of the early universe as it cools, as well as the effect of the energy density in cosmic strings and in GW on the expansion of the universe. On the other hand, Refs.~\cite{siemens,hogan} used more recent values of the cosmological parameters, considering in particular the late-time acceleration of the expansion of the universe. Here we will consider both a $\Lambda$-CDM cosmological model and the variation of the number of relativistic species in the early universe (see section \ref{MethodStrings}). We will also discuss the dependence of the GW spectrum on the cosmological history. We will neglect the contribution of cosmic strings and GW to the total energy density of the universe. Given the improved upper bounds on the string tension, we expect this to be a good approximation. 

A second aspect that must be modeled is the number density of loops in the universe, which depends crucially on the typical size of the loops produced by the long string network. This issue is still under debate, with different simulations~\cite{simu1,simu2,simu3,simu4} and analytical works~\cite{ana1,ana2,ana3} obtaining different results. The initial length of a loop (i.e. before it shrinks by emitting GW) produced by the long string network at cosmic time $t$ can be written as
\be
\label{Li}
l_i = \alpha\,t
\ee
where the length is defined as the energy of the loop in its rest frame divided by its tension (the initial length 
in the cosmological frame is modified accordingly if the loops are produced with a large initial velocity). The first simulations indicated that loop production occurs at relatively small scales, close to their resolution limit 
(see \cite{VS} and references therein). This suggested that $l_i$ should be given by the size of the small scale structure on the long strings, which are damped by gravitational emission. This led to the estimate~\cite{back1} 
$\alpha \sim \Gamma G \mu$ where $\Gamma \sim 50$ measures the efficiency of GW emission. More recently, 
Ref.~\cite{back2} pointed out that gravitational emission may be much less efficient in damping the small scale structure, leading to much smaller initial loop sizes: $\alpha \propto (G \mu)^{\chi}$ with $\chi > 1$~\cite{ana1}. This was further investigated in \cite{ana2}, which obtained different values than \cite{ana1} for the exponent $\chi$. On the other hand, the latest simulations now indicate that $l_i$ is instead given by the large scale properties of the network, leading to much larger initial loop sizes: $\alpha \sim 0.1$~\cite{simu3,simu4}. Finally, it is possible that loop production occurs simultaneously at two very different scales, $\alpha < \Gamma G \mu$ and $\alpha \sim 0.1$, as advocated in 
e.g.~\cite{ana3}. 

Whatever the typical scale of loop production, there is in practice a distribution of different values of $\alpha$ around some peak, but this is usually approximated by a Dirac distribution in the calculations of the GW spectrum. Caldwell and Allen~\cite{caldwell} consider different possibilities with $\alpha \lesssim 10^{-3}$. Siemens et al~\cite{siemens} study both $\alpha \lesssim \Gamma G \mu$ and $\alpha \sim 0.1$. Hogan et al~\cite{hogan} consider $\alpha \sim 10^{-6} - 0.1$ and a case where loops are produced simultaneously at $\alpha \sim 0.1$ and $\alpha \sim 10^{-5}$. For the same values of the other parameters, the case $\alpha \sim 0.1$ leads to a much larger amplitude of the GW spectrum than the case $\alpha \lesssim \Gamma G \mu$, because the loops are long-lived (compared to the Hubble time) in the first case and short-lived in the second one. We will consider both cases separately below. For $\alpha \lesssim \Gamma G \mu$, we will use the same loop 
distribution as in \cite{siemens} (see section \ref{Sloops}). For $\alpha \sim 0.1$, our loop distribution will differ from the one used in \cite{siemens} in order to account for the smooth transition between the radiation and matter eras and for the effects of the variation of the number of relativistic species and of the late-time acceleration of the universe expansion on the loop number density (see section \ref{Lloops}). 

Finally, a third aspect that must be modeled in order to compute the GW background from cosmic strings is the GW spectrum emitted by each individual loop. As it oscillates quasi-periodically, a loop of length $l$ emits GW at frequencies $f_{em} = 2 n / l$, where $n = 1, 2, 3, ...$ denotes the harmonics of the loop oscillation. The total GW power emitted by the loop is given by
\be
\label{power}
P = \sum_n P_n \, G \mu^2 = \Gamma \, G \mu^2 
\ee
where $P_n$ and $\Gamma$ are dimensionless coefficients that depend on the loop oscillation. In practice these coefficients can take different values for different loops, but some average values (over the loop population) are assumed in order to calculate the GW spectrum. It is not clear, however, what the average values for each $P_n$ would be in a realistic population of loops. Different works have used different assumptions, typically all the power emitted in the fundamental $n = 1$ mode or a given asymptotic behavior $P_n \propto n^{-1 - q}$ for $n \rightarrow \infty$ (extrapolated down to $n=1$). The value of $q$ then depends on the presence of cusps or kinks, which are responsible for the emission at $n \gg 1$. Caldwell and Allen~\cite{caldwell} consider mostly $q = 1/3$, which was obtained analytically for loops with cusps in \cite{early3} and in the numerical simulations of \cite{simupower}. Ref.~\cite{shellard} considers also other values of $q$ 
as well as the effect of a cutoff for the power emitted at large $n$. As we already mentioned above, it was later emphasized by Damour and Vilenkin~\cite{DV} that cusps and kinks lead to strong infrequent GW bursts that should not be included in the computation of the stationary and nearly Gaussian background, in order not to over-estimate its amplitude. The GW spectrum from cusps and kinks after removing the rare bursts was then studied by Siemens et al~\cite{siemens}. On the other hand, Hogan et al~\cite{hogan} considered all the power to be emitted by the first few harmonics, mainly the fundamental mode $n = 1$. In the following, we will consider different models for the GW spectrum emitted by each individual loop, and find that they only lead to small differences in the GW background today. We will also find that, in the case of large initial loop sizes, removing the rare burst has practically no effect on the present-day GW spectrum, at least when the number of cusps and kinks per loop oscillation period is of order unity. 

In addition to the three general aspects discussed above, and depending on the kind of cosmic strings that one considers, other properties may affect the GW signal. For instance, global strings (associated to the breaking of a global symmetry) may radiate mainly Goldstone bosons, in addition to GW. The amplitude of the GW signal from global strings is then reduced, as estimated in e.g.~\cite{shellard}. If different kinds of strings are present and interact with each other (as for instance for cosmic super-string networks involving F-strings, D-strings and their bound states), the evolution of the network and its properties may also be more complicated. In particular, junctions between strings can form, leading to a drastic increase in the number of (sharp) kinks due to the fact that each kink gives rise to two transmitted and one reflected kink every time it reaches a junction. The impact of this phenomenon on the gravitational wave spectrum has been studied in \cite{kinkprolif}. It has also been found in \cite{gregory} that the GW signal from cusps and kinks is reduced for cosmic super-strings that can move in the extra-dimensions. On the other hand, it has been argued in \cite{psuper} that the position of the strings in the extra dimensions correspond to worldsheet moduli that should be stabilized, so that the classical motion of the strings should be effectively four-dimensional. Cosmic strings may also appear in hybrid defects, such as strings ending on domain walls or monopoles, in which case the network is typically unstable. The GW signal from such defects has been studied in \cite{hybdef1, hybdef2}. Finally, it has been argued in \cite{depies} that the gravitational clustering of loops in the galactic halo may lead to specific GW signatures. In the following, we consider only stable cosmic string networks, which are effectively four-dimensional and whose loops decay mostly into GW. We also assume that only one kind of string is present or dominate the loop number density, and we neglect any clustering of loops.

\subsection{Calculation of the GW Background}
\label{MethodStrings}

In this sub-section, we detail the procedure that we follow to compute the GW background from cosmic string loops. We derive an expression for the GW spectrum using two different methods. The first one is standard~\cite{VS} and is convenient to obtain the overall normalization of the GW spectrum. It can be applied to a variety of models for the GW spectrum emitted by each individual loop. The second method is similar to the one used in \cite{DV,siemens} and is convenient to 
remove the rare bursts from the GW background. It is based on the asymptotic waveform emitted by cusps and kinks. As we will see, except for the overall normalization and the removing of the rare bursts, the two methods are equivalent.

\subsubsection{Models for the GW Spectrum Emitted by Each Individual Loop and for the Cosmological Evolution}

As discussed above, a loop of length $l$ emits GW at frequencies $f_{em} = 2 n / l$ at the time of emission, with the power (\ref{power}) where $n = 1, 2, 3, ...$ denotes the harmonics of the loop oscillation. Approximating the discrete spectrum by a continuous one, we have
\be
\label{dPdf}
\frac{d P}{d f_{em}} = G \mu^2\,l\,S(f_{em} l) \hspace*{0.5cm} \mbox{ with } \hspace*{0.5cm} \int dx \, S(x) = \Gamma \, .
\ee
One common model of GW emission is provided by
\be
\label{Sq}
S_q(x) = \frac{A_q}{x^{q + 1}}\,\Theta(x - 2) \hspace*{0.5cm} \mbox{ with } \hspace*{0.5cm} A_q = 2^q\,q\,\Gamma
\ee
which corresponds to an asymptotic behavior $P_n \propto n^{-q - 1}$ at large $n$. The numerical pre-factor $A_q$ results from the normalization (\ref{dPdf}) and the Heaviside function ($\Theta(x) = 0$ for $x < 0$ and $1$ for $x > 0$) results from the condition $n \geq 1$ on the harmonics. As we will see, the contribution of cusps to the asymptotic power at large 
$n$ goes as $P_n \propto n^{-4/3}$, while the contribution of kinks goes as $P_n \propto n^{-5/3}$. Since the contribution from cusps decreases more slowly than the one from kinks when $n$ increases, one expects the total power at large $n$ to be dominated by the cusp contribution if the number of cusps is not strongly suppressed compared to the number of kinks. We will then use the model (\ref{Sq}) with $q = 1/3$ for loops with cusps, independently of the presence of kinks. We call these "cuspy loops" in the following. On the other hand, we use the model (\ref{Sq}) with $q = 2/3$ for loops that have kinks but no cusps. We call these "kinky loops". Note also that the model (\ref{Sq}) may be inaccurate for the low harmonics (it assumes that the asymptotic behavior of the power is valid down to $n = 1$, which need not be the case). It will therefore be interesting to also calculate the GW background in another model that assumes instead that all the power is emitted by the fundamental mode $n = 1$. This corresponds to
\be
\label{Sfun}
S_{fun}(x) = \Gamma\,\delta(x - 2)
\ee
where $\delta(x)$ is the Dirac distribution. 

Consider now the energy density $d \rho_{gw}$ emitted in GW with frequencies in the interval $[f_{em} , f_{em} + d f_{em}]$ during the time interval $[t , t + dt]$ by loops whose length belongs to the interval $[l , l + dl]$. The number density of such loops is denoted as $n(l, t) dl$. Each of them emits a GW energy $d E_{gw} = G \mu^2 l S(f_{em} l) dt \, df_{em}$. This gives 
\be
d \rho_{gw} = G \mu^2\,l\,S(f_{em} l)\,n(l, t)\,dt\,dl\,df_{em} \, .
\ee
After their production, the GW energy density redshifts as $1/a^4$ and their frequency as $1/a$, where $a$ is the scale factor. We denote by $a_0$ the value of the scale factor today and by $f = f_{em}\,a(t) / a_0$ the present-day GW frequency.
Integrating over production times and loop lengths, the present day GW spectrum is then given by
\be
\frac{d \rho_{gw}}{d f} = G \mu^2 \, \int_{t_*}^{t_0} dt\,\frac{a^3(t)}{a_0^3}\,
\int_0^{\alpha t} dl\,l\,n(l, t)\,S\left(\frac{a_0 f l}{a(t)}\right) \, .
\ee
Here $t_0$ denotes the present time and $t_*$ the time when GW start being produced, e.g. the end of the friction-dominated era. Assuming that all the loops produced at time $t$ have an initial length $l_i = \alpha\,t$ before shrinking by GW emission, there are only loops with length $0 \leq l \leq \alpha\,t$ at time $t$. The quantity of interest is 
(\ref{spectrum3}). Using $3 H_0^2 = 8 \pi G\,\rho_c$, where $H_0$ denotes the Hubble constant today, we have
\be
\label{spec1}
h^2\,\Omega_{gw} = \frac{8 \pi}{3}\,\frac{h^2}{H_0^2}\,(G \mu)^2\,f\,\int_{t_*}^{t_0} dt\,\frac{a^3(t)}{a_0^3}\,
\int_0^{\alpha t} dl\,l\,n(l, t)\,S\left(\frac{a_0 f l}{a(t)}\right) \, .
\ee

In the following, it will be more convenient to work in terms of redshift $z$, $1 + z = a_0 / a(t)$. We consider a 
$\Lambda$-CDM cosmological model where the Hubble rate at redshift $z$ is given by
\be
\label{Hz}
H(z) = H_0\,\mathcal{H}(z) \hspace*{0.5cm} \mbox{ with } \hspace*{0.5cm} 
\mathcal{H}(z) = \sqrt{\Omega_{\Lambda} + \Omega_M \, (1 + z)^3 + \Omega_R \, \mathcal{G}(z) \, (1 + z)^4} \, .
\ee
For the cosmological parameters, we use the values given in \cite{cosmopara}: $H_0 = 100\,h\,km / s/ Mpc$ with $h = 0.72$, 
$\Omega_{\Lambda} = 0.74$ for the cosmological constant abundance today and $\Omega_M = 0.26$ for the present-day abundance of cold dark matter and baryons. In the expression for $\mathcal{H}(z)$, the last term inside the square root dominates during the radiation era, when the energy density of the thermal plasma goes as $\rho_R(z) \propto g_*(z)\,T^4$ where 
$g_*(z)$ is the effective number of relativistic degrees of freedom at redshift $z$ and $T$ is the temperature of the photon background at that time~\cite{kolbturner}. Entropy conservation gives $T \propto g_S(z)^{-1/3}\,a^{-1}$ where 
$g_S(z)$ is the effective number of entropic degrees of freedom at redshift $z$. The energy density in radiation can then be written as
\be
\label{rhoz}
\rho_R(z) = \rho_R(0)\,\mathcal{G}(z)\,(1 + z)^4  \hspace*{0.5cm} \mbox{ with } \hspace*{0.5cm}  
\mathcal{G}(z) = \frac{g_*(z)\,g^{4/3}_S(0)}{g_*(0)\,g^{4/3}_S(z)} \, .
\ee
The function $\mathcal{G}(z)$ varies as the universe cools when species become non-relativistic and release their entropy to the relativistic species that are still in thermal equilibrium. The evolution of $g_*$ and $g_S$ with the photon temperature for the particle content of the Standard Model is given in e.g.~\cite{kolbturner}: $g_* = g_s = 106.75$ when all the Standard Model particles are relativistic, $g_* = g_s = 10.75$ in between the QCD phase transition and electron-positron annihilation, and $g_* = 3.36$, $g_S = 3.91$ after electron-positron annihilation and neutrinos decoupling. The neutrinos can be treated as if they were massless because they are relativistic during the radiation era and give a negligible contribution to the total energy density during the matter era. The Hubble rate during the radiation era is given by (\ref{Hz}) with an effective abundance of total radiation today (photons and neutrinos) given by 
$\Omega_R = \rho_R(0) / \rho_c \simeq 8 \times 10^{-5}$ for $T_0 = 2.725\,K$. The function $\mathcal{G}(z)$ varies mostly around $T = 200$ keV (electron-positron annihilation) and $T = 200$ MeV (QCD phase transition). We model it as a piecewise constant function which is discontinuous at these two moments of time. This approximation gives
\be
\label{Gz}
\mathcal{G}(z) = 
\left\{
\begin{array}{l}
1 \hspace*{1.5cm} \mbox{ for } z < 10^9\\
0.83 \hspace*{1.05cm} \mbox{ for } 10^9 < z < 2 \times 10^{12}\\
0.39 \hspace*{1.05cm} \mbox{ for } z > 2 \times 10^{12} \, .
\end{array}
\right.
\ee
The variation of $g_*$ and $g_S$ will introduce features in the present-day GW spectrum and reduce its amplitude at high frequencies. We consider this effect in a minimal way, based solely on the particle content of the Standard Model. Additional entropy release in the thermal plasma at high temperature, e.g. when other particles become non-relativistic, would further reduce the GW spectrum at high frequencies. This, however, is highly model-dependent. We will come back later to the dependence of the GW spectrum from cosmic strings on the thermal history of the universe. 

Integrating over redshift instead of time in Eq.~(\ref{spec1}), we have
\be
\label{spec2}
h^2\,\Omega_{gw} = \frac{8 \pi}{3}\,\frac{h^2}{H_0^3}\,(G \mu)^2\,f\,\int_{0}^{z_*} \frac{dz}{\mathcal{H}(z)\,(1 + z)^4}\,
\int_0^{\alpha t(z)} dl\,l\,n(l, z)\,S\left(f l (1 + z)\right)
\ee
where
\be
\label{tz}
t(z) = \frac{\phi_t(z)}{H_0} \hspace*{0.5cm} \mbox{ with } \hspace*{0.5cm} \phi_t(z) = 
\int_z^{+ \infty} \frac{dz'}{\mathcal{H}(z')\,(1 + z')} \, .
\ee
For the loop spectrum (\ref{Sq}), this gives
\be
\label{spec3}
h^2\,\Omega_{gw} = \frac{8 \pi}{3}\,\frac{h^2}{H_0^3}\,A_q\,(G \mu)^2\,
\int_{0}^{z_*} \frac{dz}{\mathcal{H}(z)\,(1 + z)^{5+q}}\,
\int_0^{\alpha t(z)} dl\,\frac{n(l, z)}{(f l)^q}\,\Theta\left(f l (1 + z) - 2\right) \, .
\ee

\subsubsection{Background and Rare Bursts from Cusps and Kinks}

We now consider another method to calculate the GW spectrum, which is based on the asymptotic waveform emitted by cusps and kinks. The logarithmic Fourier transform $h(f) = |f| \int h(t)\,e^{i 2 \pi f t}$ of the strain $h(t)$ produced by a cusp or a kink on a loop of length $l$ can be written as~\cite{DV}
\be
\label{hq}
h_q(f_{em}, l, r) = \frac{G \mu \, l}{r\,(f_{em} l)^q}
\ee
where $q = 1/3$ for a cusp and $q = 2/3$ for a kink, $f_{em}$ is the GW frequency at the time of emission and $r$ is the physical distance from the source. Eq.~(\ref{hq}) is derived in \cite{DV} for loops that are relatively smooth (i.e. loops that can be characterized by a single length scale $l$), up to factors of order unity that depend on the shape of the cusp or the kink, and in the limit $n = f_{em} l / 2 \gg 1$ (cusps and kinks contribute to the harmonics $n \gg 1$ of the loop oscillation). In this limit, the GW signal in the time domain corresponds to a burst with a characteristic time scale 
$\Delta t \sim 1 / f_{em} \ll l$. The GW emitted from a cusp is beamed into a cone of opening solid angle 
$d\Omega_q \approx \pi \theta_m^2$, where
\be
\label{thetam}
\theta_m(f_{em}, l) = \frac{1}{(f_{em} l)^{1/3}} \, .
\ee
On the other hand, a kink emits GW in a thick fan of directions, within a solid angle 
$d\Omega_q \approx 2 \pi \theta_m$. Although derived in the limit $n = f_{em} l / 2 \gg 1$, a common assumption is that these equations remain approximately valid down to the fundamental mode $n = 1$, up to numerical factors of order unity that depend on the particular motion of the loops. One can then consider an approximation where the signal is modeled by 
Eqs.~(\ref{hq}, \ref{thetam}) for all the frequencies with $n = f_{em} l / 2 \geq 1$. We continue to call the signal in this approximation a "burst", even if only its high-frequency part is burst-like.  

A GW emitted at redshift $z$ travels over a physical distance 
\be
r = \frac{\phi_r(z)}{H_0} \hspace*{0.5cm} \mbox{ with } \hspace*{0.5cm} 
\phi_r(z) = \int_0^{z} \frac{dz'}{\mathcal{H}(z')}
\ee
until today and has a present-day frequency $f = f_{em} / (1 + z)$. Its amplitude today can then be written as
\be
\label{hfzl}
h_q(f, z, l) = \frac{G \mu \, H_0 \, l}{\phi_r(z)\,\left(f l (1 + z)\right)^q} 
\hspace*{0.5cm} \mbox{ with } \hspace*{0.5cm} f l (1 + z) \geq 2
\ee
where the last condition results from the condition $n \geq 1$ on the harmonics. 
An important quantity is the rate $d^2 \dot{N} / dz dl$ of GW bursts observed today at the frequency $f$ and that were emitted between redshifts $z$ and $z + dz$ by loops of lengths between $l$ and $l + dl$. There are $n(l, z) dl dV$ of such loops in this redshift interval, where $dV = 4 \pi r^2 dr / (1 + z)^3 = 4 \pi \phi_r^2 dz / (H_0^3 (1 + z)^3 \mathcal{H}(z))$. Each of these loops emits $2 N_q / l$ bursts per unit emission time, i.e. $2 N_q / (l (1 + z))$ bursts per unit observation time, where $N_q$ is the average number of cusps per loop oscillation period 
$T = l / 2$ (for $q = 1/3$) or of kinks propagating on the loops (for $q = 2/3$) . Only a fraction $d\Omega_q / (4 \pi)$ of the bursts is emitted towards a given observer, where $d\Omega_q$ was given above. This gives
\be
\label{dNdzdl}
\frac{d^2 \dot{N}_q}{dz dl}(f, z, l) = 
\frac{2^{3 q}\,\pi\,N_q\,\phi^2_r(z)\,n(l, z)}{H_0^3\,l\,\mathcal{H}(z)\,(1 + z)^4\,\left(f l (1 + z)\right)^{1 - q}}
\,\Theta\left(f l (1 + z) - 2\right) \, .
\ee

The present-day spectrum of the GW background resulting from the superposition of many bursts can then be calculated 
as~\cite{DV,siemens,hybdef2}
\be
\label{spec4}
h^2\,\Omega_{gw} = \frac{4 \pi^2}{3}\,\frac{h^2}{H_0^2}\,f\,\int dz\,\int dl\,
h^2_q(f, z, l)\,\frac{d^2 \dot{N}_q}{dz dl}(f, z, l) \, .
\ee
As discussed above, Ref.~\cite{DV} pointed out that the strongest bursts may be rare enough to be observed individually and that these strong infrequent bursts should not be included in the computation of the stationary and nearly Gaussian background $h^2\,\Omega_{gw}$ (the background resulting from the superposition of many bursts is approximately Gaussian by the central limit theorem)~\footnote{The transition regime between the infrequent bursts and the Gaussian background has been recently studied in \cite{regimbau}.}. Thus only the bursts with an amplitude $h_q$ smaller than some threshold $h_b$ are effectively superimposed and the domain of integration in (\ref{spec4}) must be restricted to these bursts only. Following \cite{DV, siemens}, we define the threshold $h_b(f)$ at a given frequency $f$ (e.g. the characteristic GW frequency probed by a given observation) by the condition that the rate of occurrence of the bursts with amplitude 
$h_q > h_b$ is equal to $f$:
\be
\label{defhstar}
\int dz\,\int dl\,\frac{d^2 \dot{N}_q}{dz dl}(f, z, l)\,\Theta\left(h_q(f, l, z) - h_b(f)\right) = f \, .
\ee
The integration in (\ref{spec4}) is then restricted to the domain where $h_q(f, l, z) < h_b(f)$. 

Eq.~(\ref{spec4}) can be developed by using Eqs.~(\ref{hfzl}, \ref{dNdzdl}). Except for the condition 
$h_q(f, l, z) < h_b(f)$, the result then reduces to Eq.~(\ref{spec3}) with $A_q$ replaced by another pre-factor that depends on the average number $N_q$ of cusps or kinks per loop oscillation period. These two pre-factors are equal for an effective number of cusps or kinks given by
\be
\label{Nq}
N_q = \frac{2^q\,\Gamma}{3\,\pi^2} \, .
\ee
In that case, the GW spectrum is given by
\be
\label{spec5}
h^2\,\Omega_{gw} = \frac{8 \pi}{3}\,\frac{h^2}{H_0^3}\,A_q\,(G \mu)^2\,
\int_{0}^{z_*} \frac{dz}{\mathcal{H}(z)\,(1 + z)^{5+q}}\,
\int_0^{\alpha t(z)} dl\,\frac{n(l, z)}{(f l)^q}\,\Theta\left(f l (1 + z) - 2\right)\,
\Theta\left(h_b(f) - h_q(f, l, z)\right)
\ee
where the last factor removes the rare bursts with $h_q > h_b$. 

The models (\ref{Sq}) and (\ref{hfzl}) thus lead to the same GW spectrum when the condition (\ref{Nq}) is satisfied. In the following, we adopt this normalization for the effective number of cusps or kinks in terms of the factor $\Gamma$ that characterizes the total power emitted in GW by each individual loop, see (\ref{power}). Note that the parameter $\Gamma$ enters also in the expression for the loop number density $n(l, z)$ (see below), because it determines the lifetime of each loop. For consistency, the same value of $\Gamma$ should be used in the total power emitted by each loop and in the loop number density. In other words, if the number of cusps or kinks varies, one must take into account the resulting effect on the loop number density. Otherwise, one could naively conclude for instance that increasing the number of cusps or kinks would increase the amplitude of the GW background today, because it increases the coefficient $\Gamma$ in the total power emitted by each loop. However, this conclusion is incorrect because increasing $\Gamma$ also decreases the loop number density. As discussed below Eq.~(\ref{Sq}), we use $q = 1/3$ for loops with cusps (independently of the presence of kinks) and $q = 2/3$ for loops with kinks and no cusp. In the first (resp. second) case, variations of the average number of cusps (resp. kinks) can then be considered by varying the parameter $\Gamma$. For instance, for the value $\Gamma = 50$ that is often considered in the literature, Eq.~(\ref{Nq}) gives $N_q \approx 2 - 3$. Of course Eq.~(\ref{Nq}) is specific to the 
model (\ref{Sq}, \ref{hfzl}) for the GW spectrum emitted by each individual loop. The number of cusps or kinks on the loops could also be suppressed without decreasing the parameter $\Gamma$, by increasing the power emitted by the low harmonics. In this case, however, the model (\ref{Sq}, \ref{hfzl}) is not accurate. It will therefore be interesting to also calculate the GW spectrum with the model (\ref{Sfun}) where all the power is emitted in the fundamental mode.

There are few differences in Eq.~(\ref{spec5}) compared to the corresponding expression used by Siemens et al~\cite{siemens}. Ref.~\cite{siemens} considers one cusp and one kink per loop oscillation period, but uses an independent value of $\Gamma$ in the loop number density. For $\Gamma = 50$, this decreases the GW amplitude compared to the one we will obtain below by a factor of $2$ to $3$. Ref.~\cite{siemens} also considers all the frequencies with $f l (1 + z) \geq 1$, while we impose $f l (1 + z) \geq 2$ that follows from the condition $n \geq 1$ on the harmonics. This slightly increases the GW amplitude obtained in \cite{siemens} compared to the one we will obtain below, in particular at high frequencies where the GW spectrum is dominated by the contribution of the low harmonics. The second paper in Ref.~\cite{siemens} also uses an expression analog to Eq.~(\ref{spec5}) to calculate the contributions from cusps with $q = 1/3$ and from kinks with 
$q = 2/3$, and then sum the two contributions. However, both contributions involve also the low harmonics, whose effect on the GW spectrum is then effectively counted twice in the model (\ref{Sq},\ref{hfzl}). Here we consider instead that the parameter $q$ depends on the kind of loops that would be most abundant in a realistic population of loops. As discussed below Eq.~(\ref{Sq}), we then use $q = 1/3$ to model cuspy loops and $q = 2/3$ to model kinky ones, but we do not sum the two results. Other differences with respect to \cite{siemens} come from our different cosmological model (\ref{Hz}, \ref{Gz}) with the variation of the number of relativistic species, and from the different loop number density that we use in the case of large initial loop sizes (see Section \ref{Lloops}).

To proceed further, we need to specify the loop number density $n(l, z)$. As discussed below Eq.~(\ref{Li}), this depends on the typical size of the loops produced by the long string network. In the following, we consider separately the cases of small initial loop sizes ($\alpha \lesssim \Gamma G \mu$) and of large initial loop sizes ($\alpha \sim 0.1$).

\subsection{Small Initial Loop Size}
\label{Sloops}

\subsubsection{Loop Number Density}

As discussed in Section~\ref{IntroStrings}, the typical size of the loops when they are produced by the long string network is still uncertain. If their size is set by the small-scale structure on the long strings, one expects $\alpha \lesssim \Gamma G \mu$, where $\alpha$ is defined in Eq.~(\ref{Li}). When a loop of initial length $l = \alpha t_i$ is produced at time $t_i$, its instantaneous energy $\mu l$ decreases by GW emission at the rate given by Eq.~(\ref{power}). Its length at a later time $t > t_i$ is then given by
\be
\label{lti}
l = \alpha\,t_i - \Gamma G \mu \, (t - t_i) \, .
\ee
The lifetime of the loop is obtained by setting $l = 0$ in this expression, $\Delta t \sim \alpha t_i / (\Gamma G \mu)$. Thus for $\alpha \lesssim \Gamma G \mu$, the loops decay in less than one Hubble time after their production.

In that case, one expects the approximation where all the loops at time $t$ have the same length $l = \alpha t$ to be relatively accurate. We then use the same loop distribution as in \cite{DVsuper, siemens}
\be
\label{nlzSmall}
n(l, z) = \frac{C(z)\,\delta\left(l - \alpha t(z)\right)}{p \, \Gamma G \mu \, t^3(z)}
\ee
where $p$ is the reconnection probability, $t(z)$ is given in Eq.~(\ref{tz}) and
\be
\label{CSmall}
C(z) = C_M + \frac{(C_R - C_M)\,z}{z + z_{eq}}
\ee
with $C_R = 10$, $C_M = 1$ and $z_{eq} \simeq \Omega_M / \Omega_R$ is the redshift of radiation and matter equality. 

This expression for the loop number density is obtained by matching the energy density lost by the scaling network of long strings in the radiation and matter eras to the energy density produced in loops (assumed to have all the same size) and by taking into account the (short) lifetime of the loops, see e.g.~\cite{DVsuper}. One then finds that the loop number density is about ten times larger in the radiation era than in the matter era, which is modeled by the function $C(z)$. For small loops, we assume that the late-time acceleration of the universe expansion and the variation of the number of relativistic 
species in the early universe do not significantly affect the efficiency of loop production, so that Eq.~(\ref{nlzSmall}) remains a good approximation~\footnote{The evolution of the long string network and the efficiency of loop production are in principle affected by the late-time acceleration and by the variation of $g_*$, but to the best of our knowledge this has not been studied in numerical simulations. The variation of $g_*$ results in transient deviations from an exactly radiation-dominated expansion, which can only affect the loop number density for small loops in a small and transient way.}. 

Note that the dependence of the loop number density on the reconnection probability in Eq.~(\ref{nlzSmall}) is still uncertain. Simple arguments suggest $n(l, z) \propto 1 / p^{\beta}$ with $\beta = 1$~\cite{DVsuper} 
(see also \cite{sakel}), but the the values $\beta = 2$~\cite{tye} and $\beta = 0.6$~\cite{avgou} have also been obtained.
In Eq.~(\ref{nlzSmall}) and in the following, we focus on the case $n(l, z) \propto 1/p$ for definiteness. However, the results can be directly generalized to the case where $n(l, z) \propto 1/f(p)$ with an arbitrary function $f(p)$ by simply replacing $p$ by $f(p)$ in all the expressions below.

As discussed below Eq.~(\ref{Li}), in the case of small initial loop length, it is still unclear what this initial length should actually be. Following~\cite{DVsuper,siemens}, we use the parameter
\be
\label{defepsilon}
\epsilon = \frac{\alpha}{\Gamma G \mu} \leq 1
\ee
to parametrize this uncertainty. We will mainly treat $\epsilon$ as a free but constant-in-time parameter, as in 
\cite{DVsuper,siemens}. The latter assumption may however be inaccurate, as the results of \cite{ana1,ana2} indicate instead that $\epsilon$ takes different values during the radiation and matter eras. We will therefore also discuss below the GW spectrum computed in such a case.

\subsubsection{Calculation of the GW Spectrum}

The approximation that all the loops have the same length $l = \alpha t(z)$ at redshift $z$ significantly simplifies the expressions in Section~\ref{MethodStrings}. The RHS of Eq.~(\ref{hfzl}) decreases when $z$ increases, so the condition $h_q(f, l, z) < h_b(f)$ that removes the rare bursts in Eq.~(\ref{spec5}) corresponds to a lower bound on the redshift, 
$z > z_b(f)$. The value of $z_b(f)$ is found from the condition (\ref{defhstar}). Using (\ref{dNdzdl}) with 
(\ref{nlzSmall}), this gives
\be
\label{zb}
\int_0^{z_b(f)} dz \, \frac{C(z) \, \phi_r^2(z)}{\mathcal{H}(z) \, (1 + z)^{5-q} \, \phi_t^{5-q}(z)} = 
\frac{p \, \Gamma G \mu}{2^{3q}\,\pi\,N_q} \, \left(\frac{\alpha f}{H_0}\right)^{2-q} 
\ee
where $N_q$ is given in (\ref{Nq}). The bursts produced at redshifts $z < z_b(f)$ are too rare to contribute to 
$h^2 \Omega_{gw}(f)$. On the other hand, the condition $f l (1 + z) \geq 2$ on the harmonics of the loop oscillations in Eq.~(\ref{spec5}) leads to an upper bound on the redshift, $z < z_1(f)$ where
\be
\label{z1}
\left(1 + z_1(f)\right)\,\phi_t\left(z_1(f)\right) = \frac{2 H_0}{\alpha f} \, .
\ee
Compared to a given present-day frequency $f$, the GW produced at $z > z_1(f)$ contribute only to the GW spectrum at higher frequencies. For $z_1(f) < z_*$, Eq.~(\ref{spec5}) then gives
\be
\label{specSmall}
h^2\,\Omega_{gw} = \frac{8 \pi\,q}{3}\,h^2\,\frac{G \mu}{p}\,\left(\frac{2 H_0}{\alpha f}\right)^q \, 
\int_{z_b(f)}^{z_1(f)} dz \, \frac{C(z)}{\mathcal{H}(z)\,(1 + z)^{5+q}\,\phi_t^{3+q}(z)} \, .
\ee
This expression for the GW spectrum is valid for $f < f_*$, where $f_*$ is a high-frequency cutoff defined by the condition $z_1(f_*) = z_*$. Here $z_*$ is a high redshift above which the cosmological model (\ref{Hz},\ref{Gz}), or the loop number density (\ref{nlzSmall}), stops being applicable (see section \ref{thermalS}).

\begin{figure}[htb]
\begin{center}
\includegraphics[width=12cm]{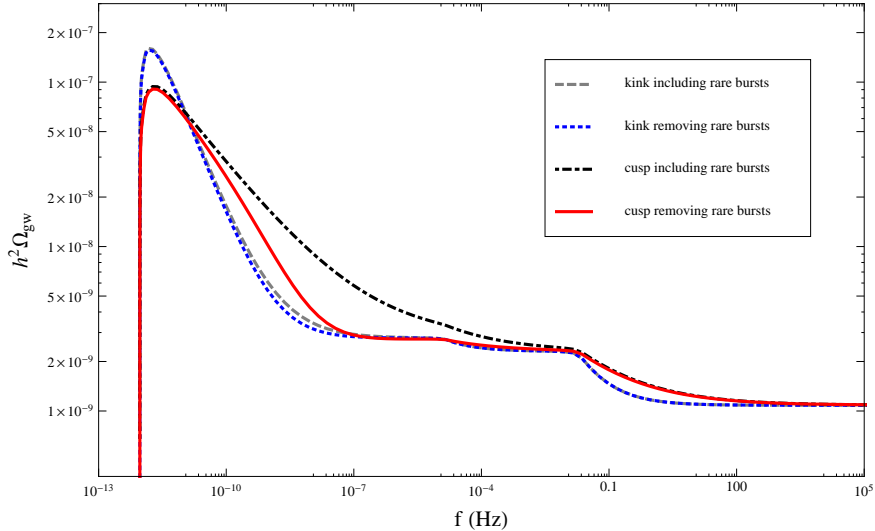}
\end{center}
\vspace*{-5mm}
\caption{Comparison of the GW spectra for cuspy loops ($q = 4/3$) and for kinky loops ($q = 5/3$), in the case of small initial loop size. In each case, we also show the spectrum that would be obtained without removing the rare bursts. The parameters are $\epsilon = 1$, $G \mu = 10^{-7}$, $p = 1$ and $\Gamma = 50$.}
\label{cuspkinkwithwithout}
\end{figure}

GW spectra computed with with Eqs.~(\ref{zb})-(\ref{specSmall}) for a given set of parameters in the case of small initial loop sizes are shown in Fig.~\ref{cuspkinkwithwithout}. The spectra are characterized by a nearly flat part at high frequency, corresponding to GW produced during the radiation-dominated era, and a low-frequency peak with a very steep infra-red tail. The high-frequency "steps" in the GW spectrum are due to the variation of the number of relativistic species in the thermal plasma of the early universe (see section \ref{thermalS}). In Fig.~\ref{cuspkinkwithwithout}, we compare the GW spectra obtained by using two different models for the GW spectrum emitted by each individual loops: 
Eq.~(\ref{Sq}) with $q = 4/3$ ("cuspy loops") and $q = 5/3$ ("kinky loops"). In both cases, we also show the GW spectra that would be obtained without removing the rare bursts. We see that, in the case of small initial loop sizes, the removing of the rare bursts is significant for cuspy loops, but it is barely noticeable for kinky loops~\footnote{This was also noticed analytically in the third paper of Ref.~\cite{kinkprolif}.}. Once the rare bursts are removed, the two models for the spectrum emitted by each loop lead to very similar GW spectra today. The differences occur mostly around the peak and around the high-frequency steps, but they always remain smaller than a factor of order $2$ in the amplitude.

\begin{figure}[htb]
\begin{center}
\includegraphics[width=12cm]{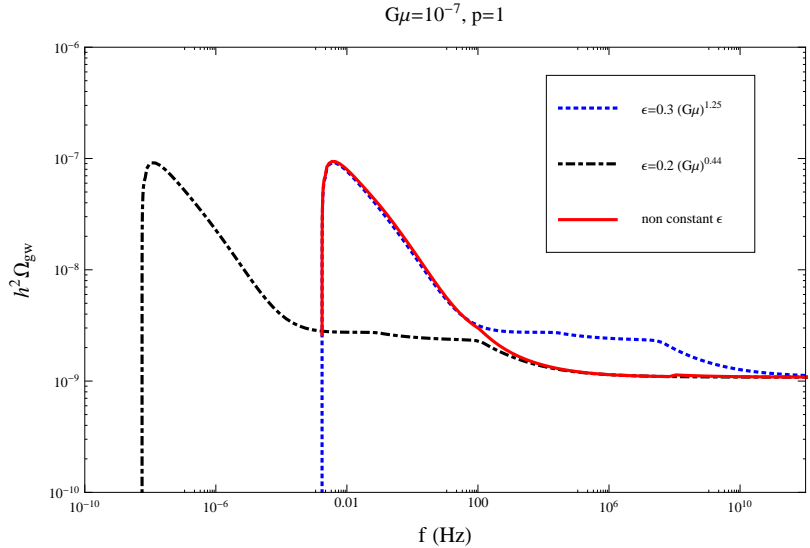}
\end{center}
\vspace*{-5mm}
\caption{GW spectrum for small cuspy loops with a different value of $\epsilon$ in the matter and radiation eras: respectively $\epsilon = \epsilon_M = 0.3 \, (G \mu)^{1.25}$ and $\epsilon = \epsilon_R = 0.2 \, (G \mu)^{0.44}$ (red, plain line). We also show for comparison the GW spectra for a constant value of $\epsilon = \epsilon_M$ (blue, dashed line at the right) and $\epsilon = \epsilon_R$ (black, dashed line at the left). The other parameters are 
$G \mu = 10^{-7}$, $p = 1$ and $\Gamma = 50$.}
\label{deuxepsGmu7}
\end{figure}

As mentioned above, the results of \cite{ana1,ana2} indicate that the parameter $\epsilon$ (determining the size of the loops) may take different values in the matter and radiation eras, although the values obtained in these two references differ from each other. According to \cite{ana2} (second paper in that reference), the effective values of $\epsilon$ that are relevant for the calculation of the GW spectrum (taking into account the high velocity of the loops in the cosmological frame) are $\epsilon_M \approx 0.3 \, (G \mu)^{1.25}$ in the matter era and $\epsilon_R \approx 0.2 \, (G \mu)^{0.44}$ in the radiation era. A GW spectrum calculated in this case is shown in Fig.~\ref{deuxepsGmu7}. As we will see, when 
$\epsilon$ decreases, the GW spectrum is shifted towards higher frequencies, while its overall amplitude remains unchanged. For a given value of $\epsilon$ during the matter era, a larger value of $\epsilon$ during the radiation era then shifts the almost flat part of the spectrum towards lower frequencies, effectively cutting out an intermediate part of the spectrum, while the GW spectrum around the peak is practically not modified, see Fig.~\ref{deuxepsGmu7}. The differences between the GW spectrum with $\epsilon$ varying from $\epsilon_R$ to $\epsilon_M$ and the GW spectrum with a constant value of $\epsilon = \epsilon_M$ occur essentially because of the high-frequency steps in the GW spectrum due to the variation of the number of relativistic species. However, the two spectra never differ by more than a factor of order two or so in the amplitude (this remains true for other values of $G \mu$ as well). If $\epsilon$ takes different values in the radiation and matter eras, one can then apply to a good approximation the results obtained for a constant value of $\epsilon = \epsilon_M$. We will therefore focus on the case where $\epsilon$ is constant in the following.

\subsubsection{Comparison With Observations}
\label{compaS}

\begin{figure}[htb]
\begin{center}
\includegraphics[width=15cm]{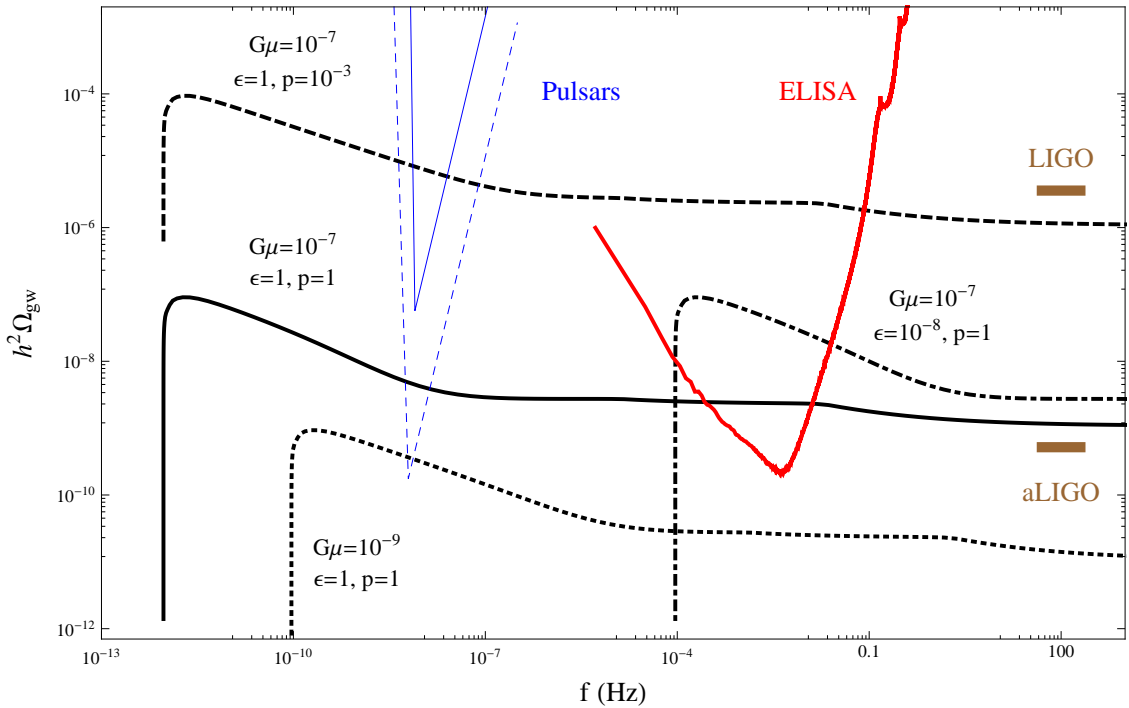}
\end{center}
\vspace*{-5mm}
\caption{GW spectra for cuspy loops compared to observational sensitivities for different values of the cosmic string parameters, in the case of small initial loop sizes with $\Gamma = 50$.}
\label{spectressmallsensitivites}
\end{figure}

The GW background from cosmic strings may be accessible to different observations. We now study the regions of the parameter space that can be probed eLISA and other experiments that will operate by the time eLISA flies, see 
Fig.~\ref{spectressmallsensitivites}. For ground-based interferometers, we consider the current limit set by the LIGO S5 run and the expected sensitivity of Advanced LIGO, both taken from \cite{ligo}. For pulsar timing experiments, we consider the current limit obtained by Jenet et al~\cite{jenet} and the expected sensitivity of the complete Parks Pulsar Timing Array (PPTA), both taken from \cite{jenet,sesana}. Regarding in particular the sensitivity of PPTA, we note that the background from cosmic strings may be hidden by the stochastic background from cosmic massive black hole 
binaries~\cite{sesana} - or vice versa, but in any case it will probably be very difficult to identify the source using only pulsars in the case of a detection. For eLISA, the confusion background from galactic binaries is already removed from the sensitivity curve.

Since we saw above that the GW spectrum does not depend strongly on the particular model for the power emitted by each individual loop once the rare bursts are removed, we focus for definiteness on the model (\ref{Sq}) with $q = 4/3$ 
("cuspy loops") in this sub-section. Fig.~\ref{spectressmallsensitivites} illustrates how the GW spectra vary with cosmic string parameters and how they compare to observational sensitivities. The dependence of the GW spectrum on the parameters will be further discussed in Section~\ref{thermalS}. When the string tension $\mu$ decreases, the GW spectrum moves towards higher frequencies as $f \propto 1 / (G \mu)$ (up to the small effect of removing the rare bursts), and its amplitude 
at high frequencies decreases as $\Omega_{gw} \propto G \mu$. Decreasing the reconnection probability $p$ simply increases the GW amplitude as $\Omega_{gw} \propto 1 / p$ (again neglecting the removal of the rare bursts), as for the loop number density, see the discussion below Eq.~(\ref{CSmall}). When the size $\epsilon$ of the loops decreases, the GW spectrum moves towards higher frequencies as $f \propto 1 / \epsilon$, and its overall amplitude remains unchanged.

\begin{figure}[htb]
\begin{center}
\includegraphics[width=8cm]{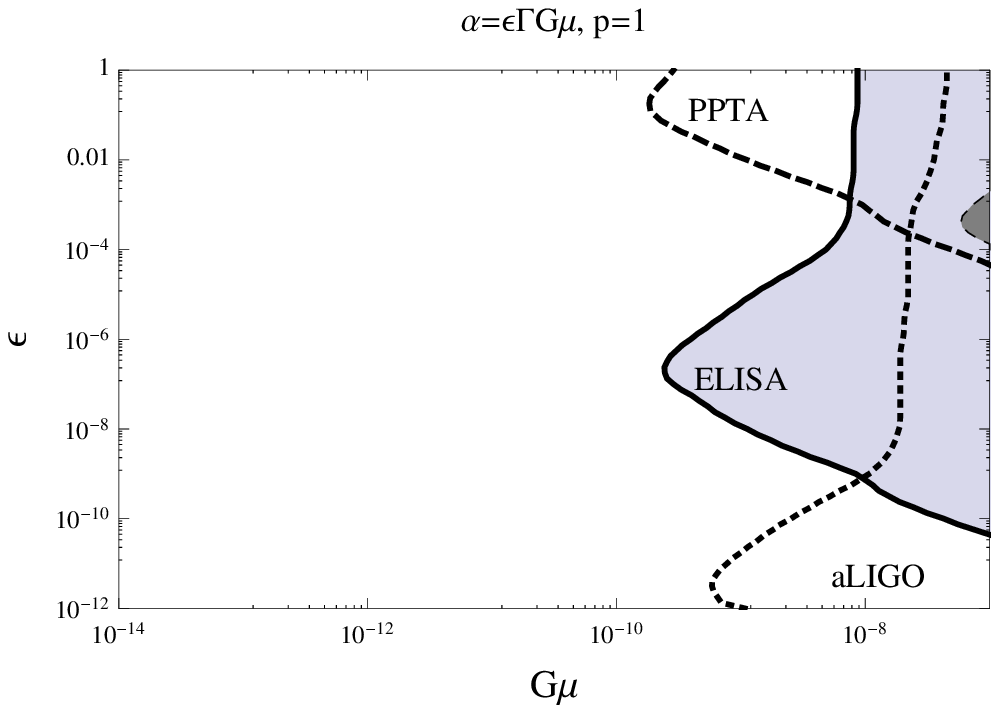} 
\hspace*{1cm}
\includegraphics[width=8cm]{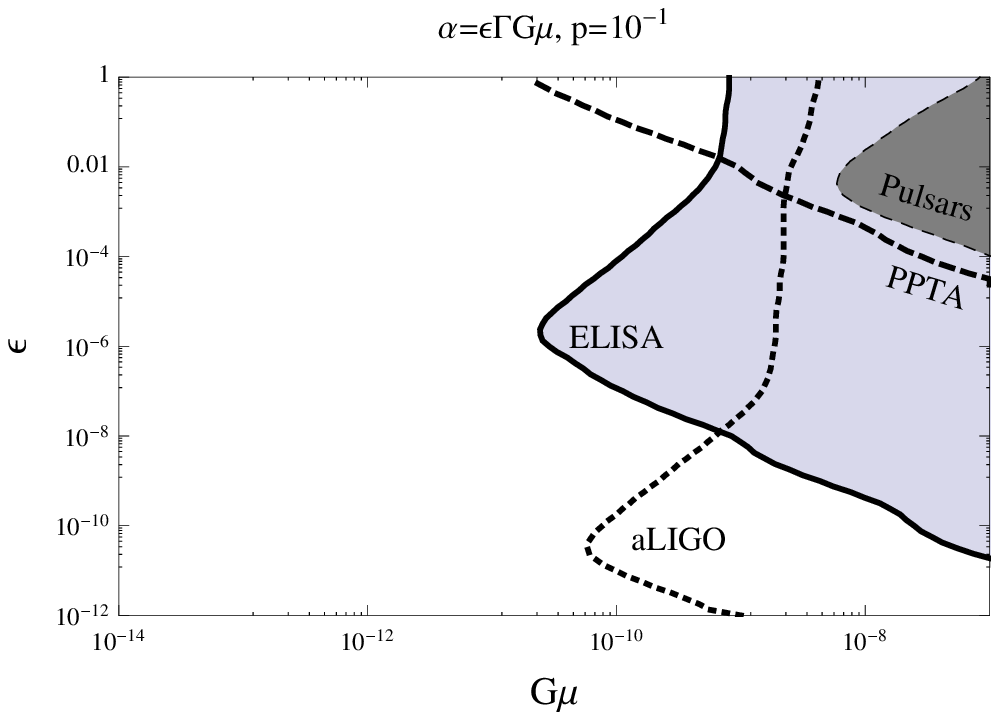}
\vspace*{0.5cm}\\
\includegraphics[width=8cm]{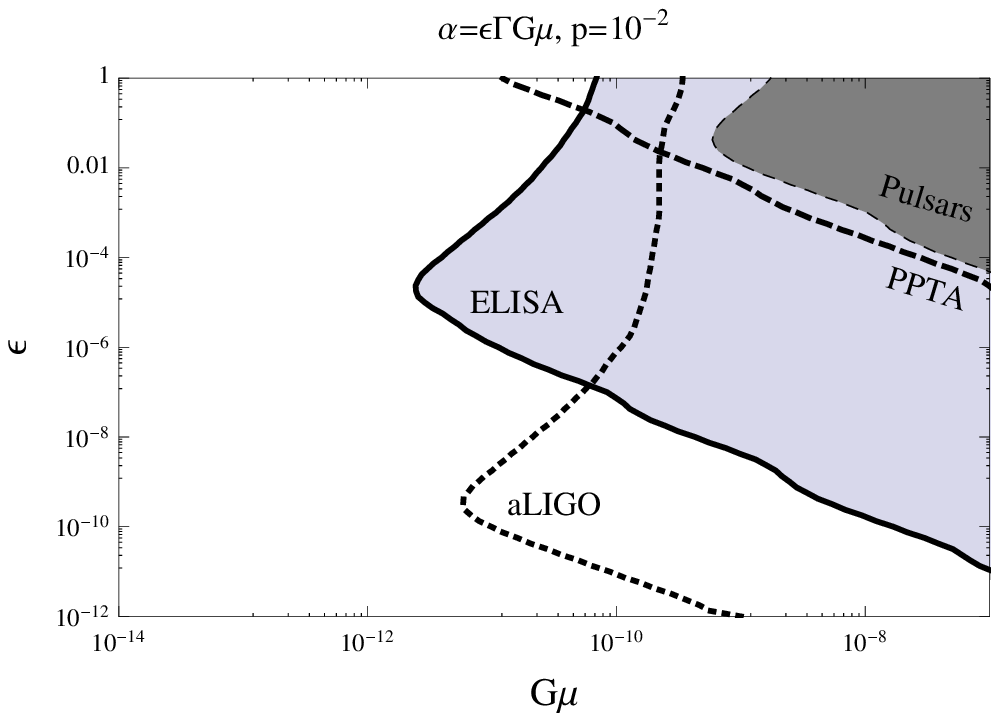} 
\hspace*{1cm}
\includegraphics[width=8cm]{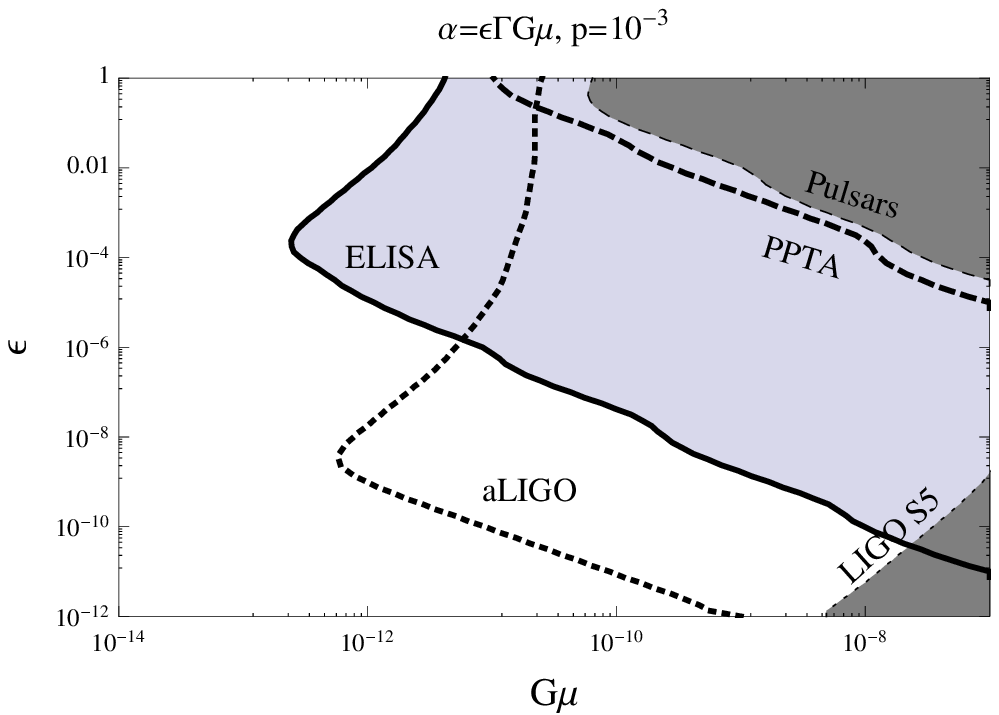}
\end{center}
\vspace*{-5mm}
\caption{Regions of the parameter space that can be probed by current and future observations, in the case of cuspy loops with small initial size and $\Gamma = 50$. We slice the parameter space in the $(G \mu, \epsilon)$-plane for $p = 1$ (top left), $p = 10^{-1}$ (top right), $p = 10^{-2}$ (bottom left) and $p = 10^{-3}$ (bottom right). The darkest regions are already ruled out by current observations (either pulsars or LIGO S5 run) for these values of the parameters. The results are shown in the case where the loop number density is inversely proportional to the reconnection probability, 
$n(l, z) \propto 1/p$, see the discussion above Eq.~(\ref{defepsilon}). In the case where $n(l, z) \propto 1/f(p)$ with an arbitrary function $f(p)$, the same results are obtained for $f(p) = 1, 10^{-1}, 10^{-2}$ and $10^{-3}$.}
\label{paramspacesmall}
\end{figure}

We show in Fig.~\ref{paramspacesmall} the regions of the parameter space that can be probed by each experiment in the case of small initial loop sizes. We set $\Gamma = 50$ and show the observational reaches in the $(G \mu, \epsilon)$-plane for $p = 1$, $p = 10^{-1}$, $p = 10^{-2}$ and $p = 10^{-3}$. We use the simple criterion that the signal $h^2 \Omega_{gw}(f)$ from cosmic strings is observable by a given experiment as soon as there is at least one frequency where the signal 
exceeds the sensitivity curve of that experiment. This criterion is accurate enough to determine the different possibilities, although the precise actual reach of different experiments will ultimately depend on both technological and data-analysis developments.

The shapes of the detectable regions for each experiment can be easily understood from Fig.~\ref{spectressmallsensitivites}. For $p=1$, which is expected for field theory cosmic strings, the high-frequency "flat" part of the spectrum is observable only for relatively large values of the string tension. For instance, the flat part of the spectrum falls always below the sensitivity of eLISA when $G\mu < 10^{-8}$. In that case, and provided that $G \mu$ is not too small, only the peak of the GW spectrum can be observable. This requires that the peak falls in an appropriate frequency range, so it is only possible for a limited range of values of $\epsilon$ . For larger tensions, the flat part is above the experimental sensitivity so the signal is detectable regardless of the value of $\epsilon$ (unless of course the whole spectrum gets shifted out of the frequency band, which happens for very small values of $\epsilon$). The regions of the parameter space that can be probed by Advanced LIGO and pulsar experiments are determined by similar considerations, except that pulsars probe such a low frequency band that the flat part of the GW spectrum always starts at higher frequencies, for any values of the parameters. As $p$ decreases, the amplitude of the whole signal increases and the main effect is to push the regions of detectability to the left: lower values of $G\mu$ now become observable. As a side effect, the observable regions are also shifted upwards: when the string tension decreases, the GW spectrum moves towards higher frequencies, which shifts the values of $\epsilon$ for which the peak of the spectrum falls in the frequency range of a given experiment.

We see that PPTA, eLISA and Advanced LIGO are complementary in the search for the stochastic background from cosmic strings. The signal can be simultaneously detected by the three experiments for sufficiently large values of the $G\mu$ and $\epsilon$. This would enable to characterize the signal with good precision and increase our ability to distinguish it from other stochastic backgrounds. For smaller values of $\epsilon$, but still large enough string tensions, the GW spectrum moves away from the frequency band of pulsar observations, but it remains observable by both eLISA and the advanced versions of the ground-based interferometers. For very small values of $\epsilon$, the spectrum also moves away from the eLISA frequency band, so that the only hope to detect the signal relies on ground-based interferometers. Finally, for smaller values of the string tension, each experiment probes a different range of loop sizes: pulsars are sensitive to values of $\epsilon$ that are few orders of magnitude smaller than $\epsilon \sim 1$, eLISA to an intermediate range of smaller values, and the advanced versions of the ground-based detectors to very small values $\epsilon$. Therefore, eLISA will also probe a region of the parameter space that will not have been accessible to any other GW experiment.

\subsubsection{Analytical Calculation and Dependence on the Cosmological History}
\label{thermalS}

In order to further study the dependence of the GW signal on the parameters and on the thermal history of the universe, it is useful to compute it analytically. This is easily done for present-day frequencies $f \gg f_m$, where
\be
\label{feq}
f_m \sim \frac{50}{\Gamma} \, \frac{10^{-8}}{G \mu} \, \frac{10^{-7} \, \mathrm{Hz}}{\epsilon} \, .
\ee
This corresponds to the high-frequency part of the GW spectrum, which is produced during the radiation era. In this limit, the integral in the RHS of Eq.~(\ref{specSmall}) is dominated by its upper limit $z = z_1(f) \gg z_{eq}$, and one can use the asymptotic behavior of the cosmological functions $\mathcal{H}(z)$, $\phi_t(z)$ and $\phi_r(z)$ for $z \gg z_{eq}$. For a given present-day frequency $f \gg f_m$, the redshift $z_1(f)$ given in Eq.~(\ref{z1}) corresponds to the temperature
\be
\label{TfSmall}
\frac{T(f)}{\mathrm{GeV}} \sim \frac{\Gamma}{50} \, \frac{G \mu}{10^{-8}} \, \frac{\epsilon\,f}{\mathrm{Hz}} 
\hspace*{0.5cm} \mbox{ for } \hspace*{0.5cm} f \gg f_m
\ee 
in the radiation era. Most of the GW observed today at a given frequency $f \gg f_m$ were produced when the early universe had a temperature of the order of $T(f)$. For instance, for $\Gamma = 50$ and $G \mu \sim 10^{-8}$, the GW that fall in the eLISA frequency range today were produced at temperatures of the order of $T(f) \sim \epsilon$ MeV or so. Note that this differs from e.g. GW from first-order phase transitions, which must be produced at significantly higher temperatures in order to fall in the eLISA frequency range. This is because the GW from cosmic strings have a much smaller wavelength at the time of their production, especially in the case of small initial loop sizes. 

Performing the integral in Eq.~(\ref{specSmall}) then gives
\be
\label{EstimSmall}
h^2\,\Omega_{gw} \simeq 1.1 \times 10^{-3} \, C_R \, \frac{G \mu}{p} \, \left(\frac{100}{g_*(T(f))}\right)^{1/3} 
\hspace*{0.5cm} \mbox{ for } \hspace*{0.5cm} f \gg f_m
\ee
where $g_*(T(f))$ is the number of relativistic degrees of freedom at the temperature (\ref{TfSmall}) for a given frequency $f$. This part of the spectrum is almost flat, except for the dependence in $g_*(T(f))$. Note also that it neither depends on the parameter $\Gamma$ nor on the parameter $q$ - describing respectively the total power (\ref{power}) and the spectrum (\ref{Sq}) emitted by each individual loop. Our numerical results agree very well with the analytical estimate 
(\ref{EstimSmall}). For instance, the "steps" in the high-frequency part of the GW spectrum in Fig.~\ref{cuspkinkwithwithout} are due to the variation of the number of relativistic species as the early universe cools down. The most significant step in Fig.~\ref{cuspkinkwithwithout} corresponds to the QCD phase transition at $T \sim 200$ MeV. It occurs around $f \sim 0.01$ Hz, in agreement with Eq.~(\ref{TfSmall}). For the particle content of the Standard Model, the variation of $g_*$ reduces the GW amplitude at high frequencies by a factor of order $2.6$, so the effect is relatively small. Depending on the physics beyond the Standard Model, additional entropy release in the thermal plasma at high temperature would further reduce the GW spectrum at high frequencies.  

As mentioned above, the GW spectrum is given by (\ref{specSmall}) only for frequencies $f < f_*$, where $z_1(f_*) = z_*$ and $z_*$ is a high redshift above which the cosmological model (\ref{Hz}, \ref{Gz}) or the loop number density 
(\ref{nlzSmall}) stops being applicable. For instance, $z_*$ can be the redshift at the end of the friction-dominated era or when the scaling regime for the evolution of the cosmic string network starts, or the redshift when the standard thermal 
evolution of the universe starts (e.g. the end of reheating after inflation), whichever is smaller. Thus in general $z_*$ and thus $f_*$ are highly model-dependent. However, a rather general constraint on the GW background, at least for field-theory cosmic strings, is that it should not involve GW produced at temperatures $T > T_d$, where $T_d \sim G^{1/2} \mu$ is the temperature at the end of the friction-dominated evolution of cosmic string networks~\cite{VS}~\footnote{This expression for $T_d$ applies only to field-theory cosmic strings. For cosmic super-strings, the friction-dominated epoch ends at higher temperatures, see \cite{KKmodes}. In that case, the high-frequency cutoff in the GW spectrum occurs at even higher frequencies.}. Indeed, even if cosmic strings are produced earlier in the radiation era, their motion is usually highly damped by their interactions with the thermal plasma at $T > T_d$, so much less GW (if any) are emitted during this epoch. From Eq.~(\ref{TfSmall}), this implies that the present-day GW spectrum would be cut off at the frequency 
$f_* \sim 10^{11}\,\mathrm{Hz} / \epsilon$ in the case $\epsilon \leq 1$. Even for $\epsilon = 1$, this is a very high frequency, well above the frequency range of e.g. ground-based interferometers. Thus, in the case of small initial loop sizes, the friction-dominated epoch is irrelevant for the direct search of GW from cosmic strings.

Let us now consider the following question: what is the maximal temperature in the early universe that could be probed by observations of GW backgrounds from cosmic strings? From Eq.~(\ref{TfSmall}), we see that the maximal temperature is obtained for $\epsilon = 1$, and $G \mu$ and $f$ as high as possible. Taking $G \mu \sim 10^{-7}$ as an upper bound on the string tension and $f \sim 10^2$ Hz for the frequency range of ground-based interferometers, this gives 
$T(f) \sim 1$ TeV. Thus in the case of small loops, the observation of GW from cosmic strings can provide informations about the early universe at temperatures as high as $T \sim $ TeV. Consider the case with $\epsilon = 1$ and $p = 1$ in more detail. In that case, the string tension cannot be much smaller than $G \mu \sim 10^{-7}$ if the resulting GW are to be observable by eLISA or Advanced LIGO. The GW that fall in the frequency range of Advanced LIGO were then produced around $T(f) \sim$ TeV, while those falling in the frequency range of eLISA were produced around $T(f) \sim 10 - 100$ MeV, i.e. a bit before the start of Big Bang Nucleosynthesis (BBN). The amplitude of the GW spectrum in the frequency range of ground-based interferometers may then be suppressed compared to its amplitude in the eLISA frequency range if the evolution of the universe is non-standard at temperatures between $10 - 100$ MeV and the TeV. This occurs for instance if a scalar field (e.g. a modulus) dominates the universe expansion, or more generally if any entropy is released in the thermal plasma, during that epoch. This would dilute any GW produced before that moment of time and suppress the amplitude of the GW spectrum in the frequency range of Advanced LIGO. This can in fact be a typical situation in super-gravity or string theory with moduli condensates decaying just before the onset of BBN, see e.g~\cite{acharya}. This example illustrates the kind of information we could get about the early universe if GW from cosmic strings were observed, and it further stresses the complementarity between eLISA and ground-based interferometers.

\subsection{Large Initial Loop Size}
\label{Lloops}

\subsubsection{Loop Number Density}

As discussed in Section~\ref{IntroStrings}, the most recent simulations~\cite{simu3,simu4} indicate that the initial size of the loops when they are produced is related to the large scale properties of the long string network, with 
$\alpha \sim 0.1$ in Eq.~(\ref{Li}). In that case, the loops are long-lived compared to the Hubble time and there is at any time $t$ a distribution of loops with a wide range of lengths, from $l = 0$ to $l \sim 0.1\,t$. From Eq.~(\ref{lti}), the loops that are present at time $t$ have been produced at times $t_i$ as small as $t_i \sim \Gamma G \mu t / \alpha$, which is much smaller than $t$ for $\alpha \sim 0.1$. The number density of loops of size $l$ at time $t$ in that case can be calculated as in \cite{VS}. It can be written as
\be
\label{nltL}
n(l, t) = \frac{C(t_i)}{p\,\alpha^2\,t_i^4} \, \frac{a^3(t_i)}{a^3(t)}
\ee
with
\be
\label{tilt}
t_i = \frac{l + \Gamma G \mu t}{\alpha} \,\,\, \in \,\,\, \left[\frac{\Gamma G \mu t}{\alpha} , t \right]
\ee
where the last expression follows from (\ref{lti}) with $\Gamma G \mu \ll \alpha$. The first factor in the RHS of Eq.~(\ref{nltL}) comes from matching the energy density lost by the scaling network of long strings to the energy density produced in loops of length $\alpha t_i$, with a different numerical factor $C(t_i)$ in the radiation and matter eras, as in the case of small initial loop sizes. The second factor in the RHS of Eq.~(\ref{nltL}), on the other hand, comes from the dilution of the loop number number density with the expansion of the universe between the time $t_i$ of their production and the time $t$ under consideration.  

In Eq.~(\ref{nltL}) and in the following, we again assume for definiteness that the loop number density varies with the reconnection probability $p$ as $n(l, z) \propto 1/p$, see the discussion above Eq.~(\ref{defepsilon}). Again, the results can be directly generalized to the case where $n(l, z) \propto 1/f(p)$ with an arbitrary function $f(p)$ by simply replacing $p$ by $f(p)$ in all the expressions below.

Using $a(t) \propto t^{1/2}$ in the radiation era and $a(t) \propto t^{2/3}$ in the matter era, and assuming a sharp transition between the two behaviors at time $t_{eq}$, Eq.~(\ref{nltL}) reduces to the standard result given in \cite{VS} for the loop number density in the radiation and matter eras. Here we use the more general expression (\ref{nltL}) for the loop number density, because it applies to a more realistic cosmological evolution. This allows us to account for the smooth transition between the radiation and matter eras and for the effects of the variation of the number of relativistic species and of the late-time acceleration of the universe expansion on the loop number density: these effects are encoded in the factor $a^3(t_i) / a^3(t)$ in Eq.~(\ref{nltL})~\footnote{The variation of $g_*$ and the late-time acceleration affect the cosmological evolution and therefore the dilution of the density of loops after their production: this is included in the second factor in the RHS of Eq.~(\ref{nltL}). They may also affect the efficiency of loop production, as we already mentioned in the case of small loops, and therefore the first factor in Eq.~(\ref{nltL}). We expect however these effects on the production of loops to be negligible for the GW background and we do not include them. Regarding the variation of $g_*$, this can only lead to small and transient modifications of loop production, as in the case of small initial loop sizes. Regarding the late-time acceleration of the universe expansion, its effect on loop production can only affect the number density of large loops with $l \sim 0.1 t$ today. These loops are negligible for the calculation of the GW spectrum.}. This leads to some differences in the GW spectra that we obtain compared to those of Siemens et al~\cite{siemens}, which used the loop number density given in \cite{VS}: mainly a more pronounced peak, similarly to the GW spectra obtained by Hogan et al~\cite{hogan} with another method. 

Numerically, it does not cost much to work with the variables $l$ and $z$ instead of $t_i$ and $t$. We define a function $Z_i(l,z) = Z\left(t_i(l, z)\right)$, where the function $t_i(l, z)$ is given by (\ref{tilt}) with $t(z)$ given in (\ref{tz}), and the function $Z(t)$ gives the redshift associated to any time $t$ (i.e. $Z(t) = t^{-1}(z)$). The number density of loops of length $l$ at redshift $z$ can then be written as
\be
\label{nlzL}
n(l, z) = \frac{C(l, z)\,\alpha^2}{p \, \left(l + \Gamma G \mu \,t(z)\right)^4}\,
\frac{(1 + z)^3}{\left(1 + Z_i(l, z)\right)^3}
\ee
where 
\be
\label{ClzL}
C(l, z) = C_M + \frac{(C_R - C_M)\,Z_i(l, z)}{Z_i(l, z) + z_{eq}}
\ee
with $C_R = 6$ and $C_M = 0.48$. We take these numerical values of $C_R$ and $C_M$ from \cite{VS}: 
$C_R \approx 0.4\,\zeta_R$ with $\zeta_R \approx 15$ and $C_M \approx 0.12\,\zeta_M$ with $\zeta_M \approx 4$ (more recent simulations~\cite{simu4} give similar results).

\subsubsection{Calculation of the GW Spectrum}
\label{CalL}

The GW spectrum can be calculated with Eqs.~(\ref{spec5}) where $n(l, z)$ is given by Eq.~(\ref{nlzL}). However, in order to impose the condition $h_q(f, l, z) < h_b(f)$ that removes the rare bursts, it is more convenient to work with the variables $h_q$ and $z$ instead of $l$ and $z$. We do that as Siemens et al~\cite{siemens}, using Eq.~(\ref{hfzl}) to express $l$ in terms of $h_q$ and $z$, and changing the variables $(l, z)$ into $(h_q, z)$ in the double integrals 
(\ref{spec4}, \ref{defhstar}). The value of $h_b(f)$ is then found from the condition
\be
\int_{h_b(f)}^{+ \infty} dh_q \, \int dz \, \frac{d^2 \dot{N}_q}{dz dh_q}(f, z, h_q) = f
\ee
and used as an upper limit in the calculation of the GW spectrum
\be
h^2\,\Omega_{gw} = \frac{4 \pi^2}{3}\,\frac{h^2}{H_0^2}\,f\,\int_{0}^{h_b(f)} dh_q \, \int dz \, h_q^2 \, 
\frac{d^2 \dot{N}_q}{dz dh_q}(f, z, h_q) \, .
\label{eqomegagwgrandes}
\ee

\begin{figure}[htb]
\begin{center}
\includegraphics[width=12cm]{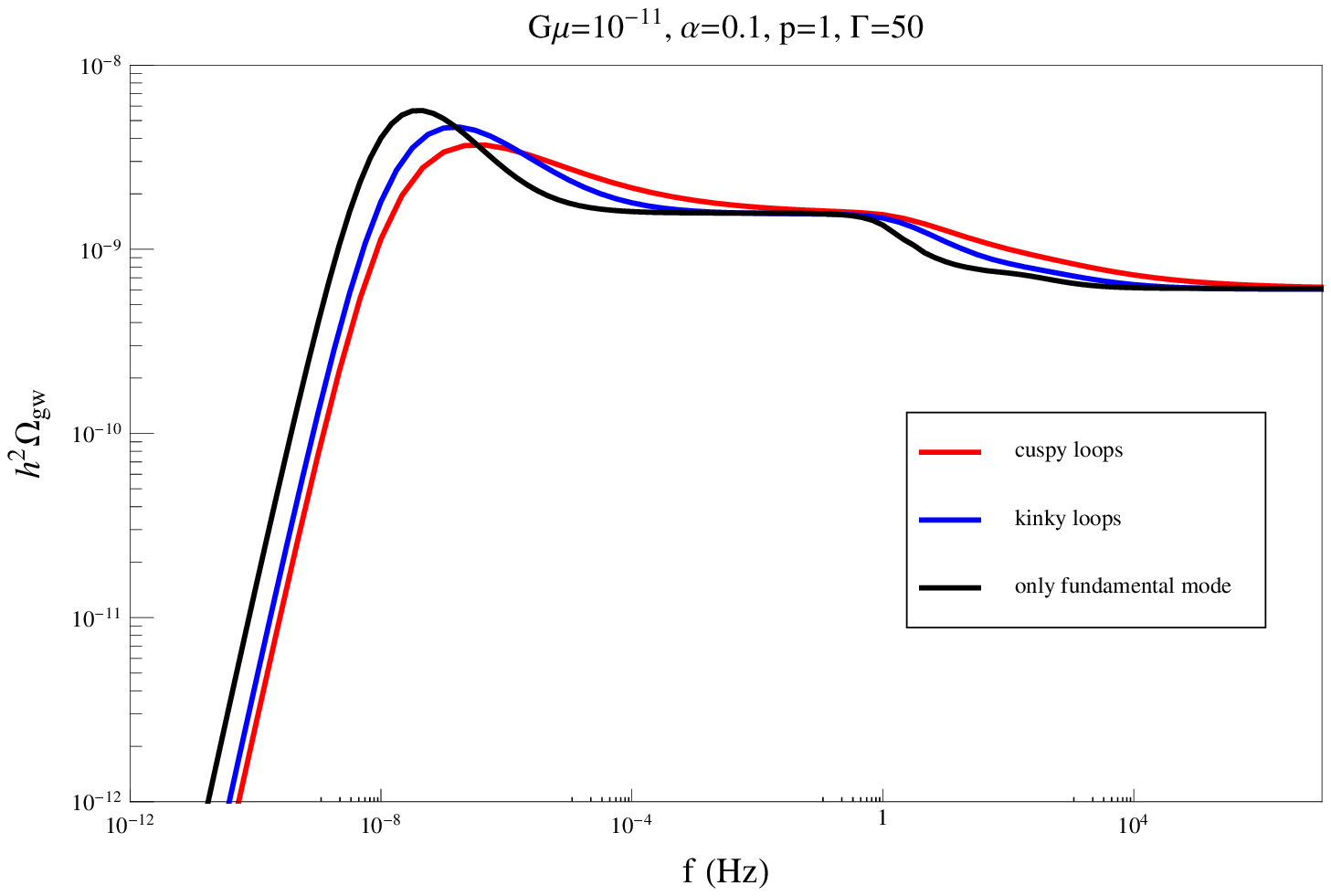}
\end{center}
\vspace*{-5mm}
\caption{Present-day GW spectra obtained with three different models for the GW spectrum emitted by each individual loop, in the case of large initial loop size with $\alpha = 0.1$, $G \mu = 10^{-11}$, $p = 1$ and $\Gamma = 50$. We compare the loop emission models (\ref{Sq}) with $q = 4/3$ (cuspy loops), (\ref{Sq}) with $q = 5/3$ (kinky loops) and (\ref{Sfun}) 
(loops emitting only in the fundamental mode). The effect of removing the rare bursts is so small that it cannot be seen on the plot.}
\label{comparaisoncuspkinkfund}
\end{figure}

GW spectra for a given set of parameters in the case of large initial loop sizes are shown in 
Fig.~\ref{comparaisoncuspkinkfund}. The spectra are again characterized by a nearly flat part at high frequencies and a 
peak at lower frequencies. However, compared to the case of small initial loop sizes, the peak of the GW spectrum is less pronounced and its infrared tail is less steep. The latter goes as $h^2 \Omega_{gw} \propto f^{3/2}$ and is emitted 
at redshifts $z \sim 1$ by loops that were produced during the radiation era~\footnote{At even smaller frequencies, not shown in Fig.~\ref{comparaisoncuspkinkfund}, the spectrum goes as $h^2 \Omega_{gw} \propto f$. That part of the spectrum is produced at $z \sim 1$ by loops that were produced during the matter era.}. The high-frequency "steps" are again due to the variation of $g_*$ in the early universe. In Fig.~\ref{comparaisoncuspkinkfund}, we compare the GW spectra for large loops obtained for three different models of the GW spectrum emitted by each individual loop: Eq.~(\ref{Sq}) with $q = 4/3$ ("cuspy loops"), Eq.~(\ref{Sq}) with $q = 5/3$ ("kinky loops") and Eq.~(\ref{Sfun}) (all the power emitted in the fundamental mode). In the first two cases, we also compared the GW spectra computed with and without removing the rare burst. We found no difference at all in the frequency range shown in Fig.~\ref{comparaisoncuspkinkfund}. The effect of removing the rare bursts starts to be noticeable (but remains small) only at lower frequencies, where the amplitude of the GW spectrum has already dropped significantly. Therefore, removing the rare bursts has practically no effect in the case of large loops. We also see in Fig.~\ref{comparaisoncuspkinkfund} that the three models of GW emission by each loop lead to very similar GW spectra today. As in the case of small initial loop sizes, the differences occur mostly around the peak and the steps, but they remain relatively small, and the amplitude at very-high frequencies is exactly the same in the three cases (see Section \ref{thermalL}).

\subsubsection{Comparison with Observations}
\label{compaL}

\begin{figure}[htb]
\begin{center}
\includegraphics[width=15cm]{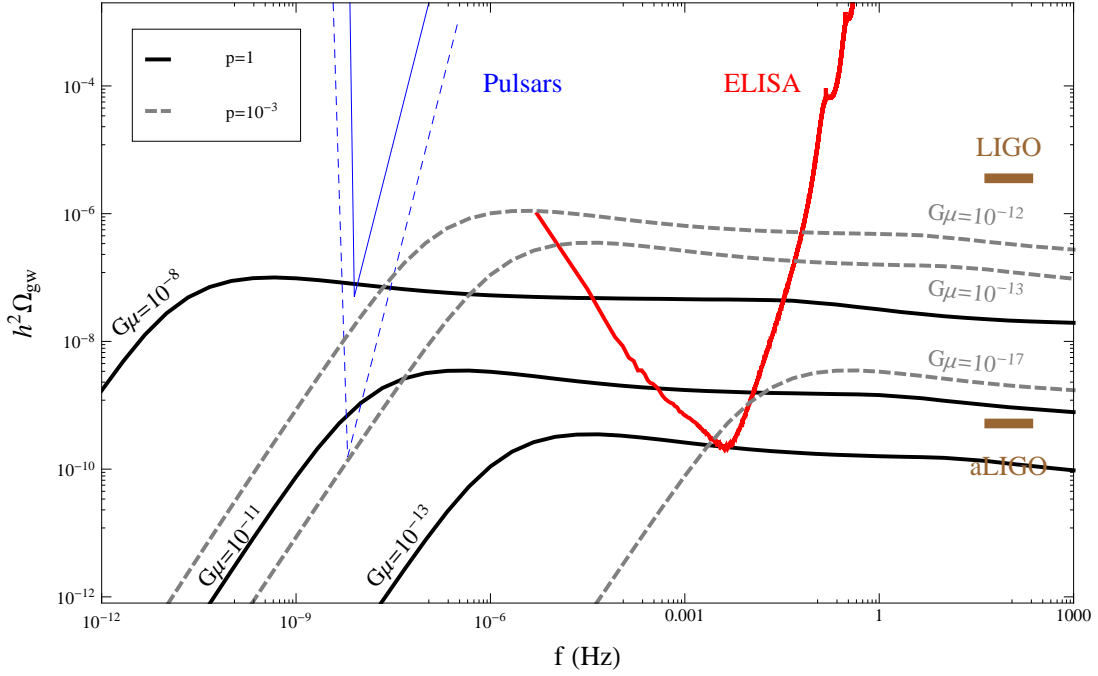}
\end{center}
\vspace*{-5mm}
\caption{GW spectra for cuspy loops compared to observational sensitivities for different values of the cosmic string parameters, in the case of large initial loop size with $\alpha = 0.1$ and $\Gamma = 50$.}
\label{spectreslargesensitivites}
\end{figure}

We now discuss the regions of the parameters that can be probed by observations in the case of large initial loop size. We consider the same experiments as in Section~\ref{compaS} and we focus again on cuspy loops since we saw above that other models for the GW spectrum emitted by each loop lead to similar results. Fig.~\ref{spectreslargesensitivites} illustrates how the GW spectrum varies with the parameters (see Section~\ref{thermalL} for more details) and how it compares 
with the observational sensitivities. When $G \mu$ decreases, the spectrum moves towards higher frequencies as 
$f \propto 1 / (G \mu)$, as in the case of small loops, but its amplitude now decreases only as 
$\Omega_{gw} \propto \sqrt{G \mu}$ at high frequencies. As a consequence, each experiment can probe smaller values of the string tension in the case of large initial loop sizes. For a reconnection probability $p=1$, eLISA will be able to reach string tensions as small as $G \mu \sim 10^{-13}$. Inside this range of accessible tensions, the GW spectrum is always nearly flat in the eLISA frequency band. On the other hand, for $p = 10^{-3}$, eLISA can reach much smaller values of the string tension, down to $G \mu \sim 10^{-17}$. In this case, eLISA probes the power-law infrared tail of the GW background, so the slope of the spectrum is very different, see Fig.~\ref{spectreslargesensitivites}. The infrared tail of the GW spectrum goes as $\Omega_{gw} \propto f^{3/2}$ (see Section \ref{CalL}) and it can only be observed by eLISA for reconnection probabilities smaller than unity, more precisely for $p \lesssim 10^{-2}$ or so. The observation of this part of the spectrum could therefore provide a signature of cosmic super-strings, as opposed to field-theory cosmic strings with 
$p = 1$.

\begin{figure}[htb]
\begin{center}
\includegraphics[width=12cm]{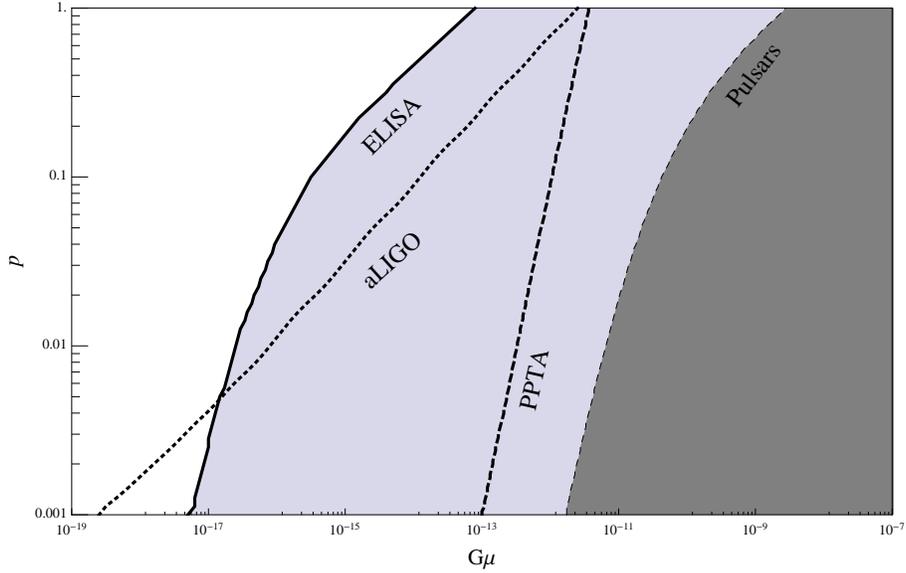}
\end{center}
\vspace*{-5mm}
\caption{Regions of the parameter space in the $(G \mu, p)$-plane that can be probed by current and future observations, in the case of cuspy loops and large initial size with $\alpha = 0.1$ and $\Gamma = 50$. The results are shown in the case where the loop number density is inversely proportional to the reconnection probability, $n(l, z) \propto 1/p$, see the discussion above Eq.~(\ref{defepsilon}). In the case where $n(l, z) \propto 1/f(p)$ with an arbitrary function $f(p)$, 
$p$ should be replaced by $f(p)$ on the y-axis of the plot.}
\label{paramspace}
\end{figure}

Fig.~\ref{paramspace} shows the regions of the parameter space in the $(G \mu, p)$-plane that can be probed by each experiment - since the GW amplitude increases when $G\mu$ increases and when $p$ decreases, the detectable regions are always located to the right of the curve corresponding to each experiment. In order to understand the shape of the observable regions, it is useful to decompose the GW spectrum into a nearly flat high-frequency part and a power-law infrared tail, ignoring the weak peak. The border of the region probed by eLISA exhibits roughly two different regimes. For 
$p \sim 0.1 - 1$ or so, the GW spectrum is nearly flat in the frequency band of interest for the the smallest value of the string tension that can be observed. On the other hand, for smaller values of $p$, eLISA probes the power-law infrared tail of the GW spectrum for the smallest value of $G\mu$ that it can observe, see e.g. the spectrum for $G \mu = 10^{-17}$ and 
$p = 10^{-3}$ in Fig.~\ref{spectreslargesensitivites}. Since that part of the spectrum varies more strongly with $G \mu$ - it varies as $\Omega_{gw} \propto (G \mu)^2$ - the border of the observable region in Fig.~\ref{paramspace} becomes steeper for small values of $p$. A similar argument explains the shape of the region accessible to current pulsar observations. 
On the other hand, in all the range of values of $p$ shown in Fig.~\ref{paramspace}, Advanced LIGO and PPTA always probe, respectively the nearly flat part of the GW spectrum and its power-law infrared tail for the smallest value of the string tension they can detect, so that the border of the observable region for each of these experiments is essentially a straight line.

Interestingly, the GW signal can be simultaneously observable by PPTA, eLISA and Advanced LIGO for a wide range of string tensions, down to $G \mu \sim 10^{-12}$. For smaller values of $G \mu$, the ability to observe the signal with PPTA quickly decreases, but there remains a wide region of the parameter space that can be probed by both eLISA and Advanced LIGO. Again, a joint detection would improve the characterization of the signal and strengthen the case for a background produced by cosmic strings. For very small values of $p$, Advanced LIGO can reach the smallest values of the string tension, because for these values the GW spectrum is significantly shifted towards higher frequencies. On the other hand, eLISA will be able to explore a new region of the parameter space for $p \sim 10^{-2} - 1$, gaining roughly one order of magnitude compared to the smallest value of the string tension that can be probed by Advanced LIGO for these values of the reconnection probability.

\subsubsection{Analytical Calculation and Dependence on the Cosmological History}
\label{thermalL}

As in the case of small initial loop sizes, the high-frequency part of the GW spectrum was produced during the radiation era and can be calculated analytically, see e.g.~\cite{VS}. For a given frequency today $f \gg f_m$, where
\be
\label{feqL}
f_m \sim \frac{50}{\Gamma} \, \frac{10^{-8}}{G \mu} \, 10^{-9} \, \mathrm{Hz} \, ,
\ee
the double integral in \eqref{eqomegagwgrandes} is dominated by the vicinity of the point $(z,l)$ defined by the conditions $(1 + z) f l = 2$ and $l = \Gamma G \mu t(z)$. The solution $z(f)$ of these conditions satisfies $z(f) \gg z_{eq}$ for 
$f \gg f_m$. Thus most of the GW observed today at a frequency $f \gg f_m$ were produced by the low harmonics 
$n = f l (1 + z) / 2 \sim 1$ of the loops with length around $l \sim \Gamma G \mu \, t(z)$, which are the most abundant ones and are going to die by GW emission in less than one Hubble time. These GW were produced at redshifts $z$ such that 
$t(z) \, z \sim 1 / (f \Gamma G \mu)$, which corresponds to temperatures of the order of
\be
\label{TfL}
\frac{T(f)}{\mathrm{GeV}} \sim \frac{\Gamma}{50} \, \frac{G \mu}{10^{-8}} \, \frac{f}{\mathrm{Hz}} 
\hspace*{0.5cm} \mbox{ for } \hspace*{0.5cm} f \gg f_m
\ee
in the radiation era. This is the same as Eq.~(\ref{TfSmall}) for small loops, but with $\epsilon = 1$. Note however that the estimate (\ref{TfL}) for large loops is less accurate than the estimate (\ref{TfSmall}) for small loops, in the sense that in the first case the spectrum observed today at frequency $f$ is dominated by GW that were emitted in a relatively wider range of times around the time $t(f)$ when the temperature was $T(f)$. Nevertheless, Eq.~(\ref{TfL}) still provides a useful order-of-magnitude estimate. The most abundant loops (those of length $l \sim \Gamma G \mu \, t$) at time $t(f)$ when the temperature is $T(f)$ were produced at times $t_i(f) \sim \Gamma G \mu\,t(f) / \alpha$, when the temperature was
\be
\label{Tif}
T_i(f) \sim \left(\frac{\alpha}{\Gamma G \mu}\right)^{1/2} \, T(f) \, .
\ee
To summarize: most of the GW observed today at a given frequency $f \gg f_m$ were emitted in the early universe when the temperature was of the order of (\ref{TfL}), by loops that were produced from the long string network when the temperature was of the order of (\ref{Tif}). For large loops, the GW spectrum observed at frequency $f$ is sensitive to the thermal history of the universe up to temperatures $T_i(f)$, instead of only $T(f)$ in the case of small loops.

The GW spectrum at $f \gg f_m$ can be calculated analytically as in e.g.~\cite{VS,siemens}. Taking into account the variation of the number of relativistic degrees of freedom, we obtain
\be
\label{EstimL}
h^2\,\Omega_{gw} \simeq 7.3 \times 10^{-4} \, C_R \, \frac{\sqrt{\alpha}}{\sqrt{\Gamma}} \, \frac{\sqrt{G \mu}}{p} \, 
\left(\frac{100}{g_*(T(f))}\right)^{1/3} \, \left(\frac{g_*(T(f))}{g_*(T_i(f))}\right)^{1/4}
\hspace*{0.5cm} \mbox{ for } \hspace*{0.5cm} f \gg f_m
\ee
where $g_*(T(f))$ and $g_*(T_i(f))$ are the numbers of relativistic degrees of freedom at the temperatures (\ref{TfL}) and (\ref{Tif}), respectively, and we used the values of the cosmological parameters given below Eq.~(\ref{Hz}). This result applies for any value of $\alpha \gg \Gamma G \mu$ (large initial loop size). Again, the spectrum is flat except for the frequency-dependence in $g_*(T(f))$ and $g_*(T_i(f))$. The last factor in the RHS of Eq.~(\ref{EstimL}) comes from the effect of the variation of $g_*$ on the loop number density. It reduces to one for $g_*(T(f)) = g_*(T_i(f))$, i.e. when 
$g_*$ does not vary in between the moment where GW are emitted by a given population of loops and the moment where these loops were born. It is interesting to note that Eq.~(\ref{EstimL}) is independent of the particular model that is chosen to describe the GW spectrum emitted by each individual loop: it applies to the model (\ref{Sq}) for any value of $q > 1$, as well as to the model (\ref{Sfun}) where all the power is emitted by the fundamental harmonic. However, contrary to the 
result (\ref{EstimSmall}) for small loops, Eq.~(\ref{EstimL}) for large loops now depends on the parameter $\Gamma$ that describes the total power emitted by each loop. Specifically, increasing $\Gamma$ decreases the amplitude of the GW spectrum because, although it increases the power emitted by each individual loop, it also decreases the lifetime of the loops and therefore their number density. Note also that $\Omega_{gw} \propto \sqrt{G \mu}$ for large loops, while 
$\Omega_{gw} \propto G \mu$ for small loops.   

Again, the analytical estimate (\ref{EstimL}) agrees very accurately with our numerical results for the high-frequency part of the GW spectrum, except around the location of the steps due to the variation of the number of relativistic species. Compared to the case of small initial loop sizes, these steps are smoother and extend over a wider frequency range, see e.g. Fig.~\ref{comparaisoncuspkinkfund}. In particular, a feature in the cosmological evolution at a given temperature $T$ now affects the GW spectrum in an interval around two different frequencies, those given by Eqs.~(\ref{TfL}) and 
(\ref{Tif}). As in the case of small loops, the variation of $g_*$ leads to a total reduction of the GW amplitude at high frequencies by a factor of order $2.6$ for the particle content of the Standard Model. However, when this reduction occurs at frequencies smaller than the ones where a given experiment operates, it now increases by almost an order of magnitude the minimal value of the string tension that this experiment can probe, because $\Omega_{gw} \propto \sqrt{G \mu}$ at 
high-frequency for large loops. This applies mostly to Advanced LIGO, which operates at higher frequencies. 

As in the case of small loops, we can determine the maximal temperature in the early universe that can be probed by observations of GW today at a given frequency $f$. For large loops, this is not given by the temperature $T(f)$ in 
Eq.~(\ref{TfL}) when most of these GW were emitted, but by the temperature $T_i(f)$ in Eq.~(\ref{Tif}) when most of the loops that emit these GW were produced from the long string network. For instance, if there is a significant release of entropy in the thermal plasma between the temperatures $T(f)$ and $T_i(f)$, the number density of the most abundant loops when the temperature is $T(f)$ would be suppressed, which would in turn reduce the amplitude of the GW emitted at that moment of time. Taking $f \sim 10^2$ Hz for ground-based experiments and $G \mu < 10^{-8}$ as required by current pulsars observations, Eqs.(\ref{TfL}, \ref{Tif}) with $\alpha \sim 0.1$ and $\Gamma \sim 50$ then give 
$T_i(f) \sim \mathrm{few} \times 10^4$ GeV for the maximal temperature that can be probed by observations of the GW background from cosmic strings. Again, if the standard thermal history of the universe starts only at a lower temperature, the GW spectrum can be suppressed in the frequency range of Advanced LIGO. 

Let us also check that the friction-dominated epoch for field-theory cosmic strings does not affect the regions of the parameter space that can be probed by each experiment in Fig.~\ref{paramspace}. This requires that 
$T_i(f) < T_d \sim G^{1/2} \mu$. The strongest constraint is obtained for ground-based interferometers: 
$G \mu \gtrsim 10^{-21}$ for $f \sim 10^2$ Hz, $\alpha \sim 0.1$ and $\Gamma \sim 50$. This is actually quite close to 
the minimal value of the string tension $G \mu \sim 10^{-19}$ shown in Fig.~\ref{paramspace}, but this value can only be reached by Advanced LIGO for $p \sim 10^{-3}$. Such a small reconnection probability is more typical of cosmic super-strings, for which the friction-dominated epoch can end at much higher temperatures~\cite{KKmodes}. For field-theory cosmic strings with $p = 1$, Advanced LIGO can only reach $G \mu \sim 10^{-12}$ or so. We therefore conclude that the friction-dominated epoch does not affect the regions of the parameter space that can be probed by each experiment in 
Fig.~\ref{paramspace}.

\begin{figure}[htb]
\begin{center}
\includegraphics[width=15cm]{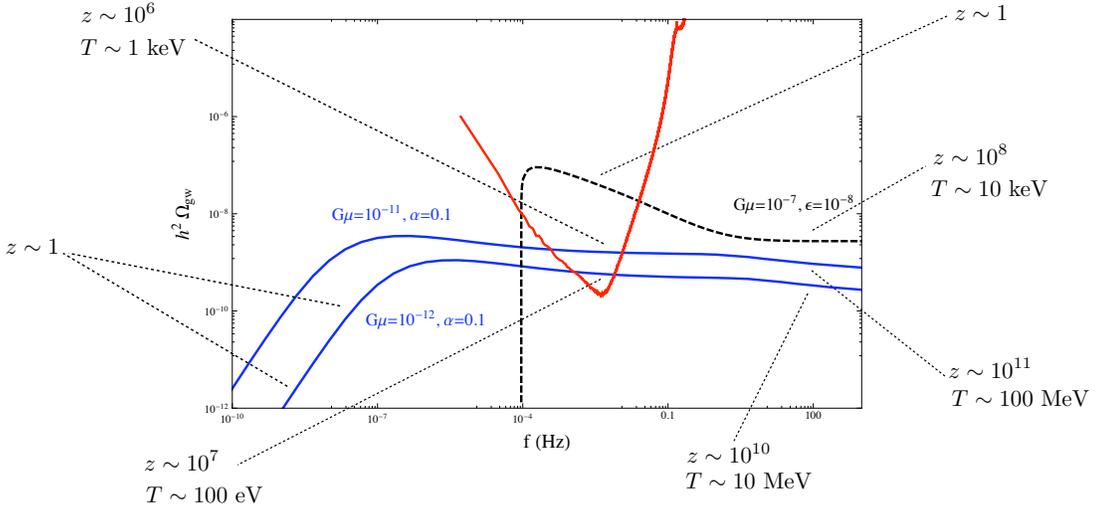}
\end{center}
\vspace*{-5mm}
\caption{Examples of GW spectra from large and small loops, together with the redshift and the temperature when most of the GW contributing to different parts of the spectrum were produced. Most of the loops emitting these GW were themselves produced much earlier from the string netowk in the case of large initial loop sizes, see the main text for details.}
\label{figureenergiesannotee}
\end{figure}

As a final illustration, we show in Fig.~\ref{figureenergiesannotee} examples of GW spectra in the cases of large and small loops, together with the redshift and the temperature when most of the GW contributing to different parts of the spectrum were produced. At high frequencies, this corresponds to the temperatures (\ref{TfSmall}) for small loops and (\ref{TfL}) for large loops. In the case of small loops, the GW produced at redshifts $z \sim 1$ contribute mostly to the region of the spectrum around the peak. In the case of large loops, the low-frequency tail of the GW spectrum is produced at $z \sim 1$, as discussed in section~\ref{CalL}.


\section{Other Cosmological Sources}
\label{Other}

In addition to first-order phase transitions and cosmic strings, many other cosmological sources of GW have been discussed in the literature. We review several of them in this Section, with a particular attention to the predictions in the eLISA frequency band. Other reviews with more details about some of these sources are given in 
\cite{allen, hoganRev, maggiore, buonanno}. We consider GW sources that operate during inflation (section~\ref{during}), just after inflation (section~\ref{justafter}) and during the subsequent thermal evolution (section \ref{thermevol}). Of course, since we have currently no direct probe of the universe at energies relevant for the eLISA frequency band, one must also keep in mind the possibility of GW produced by as yet unforeseen cosmological sources.

\subsection{Gravitational Waves from Inflation and Extensions Thereof}
\label{during}

One GW background that is expected with high confidence is the one generated during inflation. What is much more uncertain is its amplitude, which varies as $h^2 \Omega_{\mathrm{gw}} \propto G^2\,E^4_{\mathrm{infl}}$ with the (unknown) energy scale $E_{\mathrm{infl}}$ of inflation. During inflation, quantum fluctuations of the graviton field are parametrically amplified~\cite{grishchuk} into tensor perturbations at super-Hubble scales by the quasi-exponential expansion of the universe~\cite{starobinsky}. The amplitude of each Fourier mode is directly proportional to the Hubble rate at horizon exit and remains constant during all the time spent by the mode outside the Hubble radius. When the tensor modes re-enter the Hubble radius during the post-inflationary evolution, they become standard gravitational waves whose energy density is diluted as radiation with the expansion of the universe. The resulting spectrum today covers a very wide range of scales, from the Hubble scale just after inflation (corresponding to modes produced at the end of inflation) to the present-day one (corresponding to modes produced earlier during inflation and that only re-entered the Hubble radius today). The amplitude of the GW spectrum today at a given frequency is directly proportional to the almost constant (but slowly decreasing) energy density when the corresponding mode left the Hubble radius during inflation. All the modes that re-entered the Hubble radius during the radiation-dominated era are subsequently diluted by the same amount compared to the background energy density at that time and lead to an almost flat (but slightly red) spectrum. After radiation and matter equality, the later a mode re-enters the Hubble radius, the less it is diluted compared to the background energy density. As a consequence, the present-day spectrum increases when the frequency decreases as $h^2 \Omega_{\mathrm{gw}} \propto f^{-2}$ (again up to a small correction due to the slow decrease of the energy density during inflation) in the frequency range 
$10^{-18}\,\mathrm{Hz} < f < 10^{-16}\,\mathrm{Hz}$. 

If inflation occurs at sufficiently high energy scales (i.e. not much below the GUT scale, $E_{\mathrm{infl}} \sim 10^{16}$ GeV), the resulting GW in the frequency range $f \sim 10^{-18} - 10^{-16}\,\mathrm{Hz}$ may be indirectly detected in the future through their effect on the CMB polarization~\cite{polarization}. Except in some circumstances to be discussed below, the current CMB bound on inflationary GW leads already to the upper bound $h^2 \Omega_{\mathrm{gw}} \lsim 10^{-15}$ on their amplitude at frequencies accessible by direct detection experiments (the precise bound depends on the inflationary model, see e.g.~\cite{smith}). This is much below the reach of eLISA and ground-based experiments. Post-eLISA missions in space, like the Big Bang Observer (BBO)~\cite{bbo} and the Deci-Hertz Interferometer Gravitational-wave Observer 
(DECIGO)~\cite{decigo}, might reach the required sensitivity in a frequency band around $f \sim 0.1 - 1\,\mathrm{Hz}$, where the astrophysical foreground is expected to be easier to remove.  

A couple of ways to increase the inflationary GW spectrum in the direct detection frequency range, while satisfying the CMB constraint at low frequencies, have been discussed in the literature. One possibility is that the early universe was dominated during some epoch after inflation by a fluid stiffer than radiation, i.e. with an equation of state $w > 1/3$, e.g. a scalar field with a potential energy that is negligible compared to its kinetic energy 
($w \approx 1$)~\cite{stiff1, stiff2}. During such an epoch, the background energy density is diluted by the expansion of the universe as $a^{-\nu}$ with $\nu > 4$, while the GW energy density still redshifts as $a^{-4}$. Thus the fraction of energy density in GW increases. The earlier a mode re-enters the Hubble radius the more important this relative growth is, so the resulting inflationary GW spectrum today increases at high frequencies. This can significantly increase the amplitude of the inflationary GW in the frequency band of direct detection experiments, although one must check that the amplitude at higher frequencies satisfies the BBN bound (see section \ref{notations}). Depending on the inflationary model, on the equation of state during the non-standard epoch, on the duration of this epoch and on when it occurs, this effect can be observable by interferometric experiments, preferentially ground-based experiments since they operate at higher frequencies~\cite{stiff2}. It might also be observable by eLISA, e.g. if the GW at higher frequencies are sufficiently diluted by a matter-dominated stage taking place before the non-standard era with $w > 1/3$. Of course, if GW are produced by any other source between inflation and the end of the non-standard era, their amplitude would be amplified too and they would also be constrained by BBN.

The predictions for the inflationary GW may also be modified if, in addition to the parametric amplification of quantum fluctuations, GW are also emitted classically by a non-zero anisotropic stress during inflation. GW produced inside the Hubble radius during inflation are first diluted by the exponential expansion of the universe, but their amplitude remains frozen afterwards once they leave the Hubble radius. Therefore, in order to minimize the amount of dilution, the typical scale of GW production by a non-zero anisotropic stress during inflation must be sufficiently close to the Hubble radius.

One example that has been recently discussed is the emission of GW by particles produced during inflation~\cite{axion1, axion2, lorenzo, silverstein, axion3}. As the inflaton rolls down its potential, it provides a time-dependent effective mass to fields coupled to it. If such a field becomes effectively massless during inflation, particles of this field can be produced efficiently in a non-perturbative way. If particles are produced in sudden way at a given moment of time during inflation, they emit GW during a short amount of time before being rapidly diluted by the quasi-exponential expansion of the universe. This adds a bump to the standard inflationary GW spectrum, which can be located at any frequency today 
depending on the moment of time when the particles are produced during inflation~\cite{lorenzo}. In the models studied 
in \cite{lorenzo}, this effect is however too weak to be observable by eLISA or ground-based interferometers. On the other hand, if the inflaton has derivative couplings to another field, this field may remain light throughout inflation and its particles can be efficiently produced in a continuous way. This occurs in particular when the inflaton $\phi$ couples to a gauge field through an interaction term of the form $\phi F_{\mu \nu} \tilde{F}_{\mu \nu}$, where $F_{\mu \nu}$ is the field strength and $\tilde{F}_{\mu \nu}$ its dual. Such an interaction is natural in models where the inflaton field is an axion. In that case, the continuous production of particles of the gauge field during inflation can significantly enhance the inflationary GW spectrum~\cite{lorenzo, axion3}. However, these particles also lead to a highly non-gaussian contribution to the scalar perturbations~\cite{axion1}, so their production is constrained by the bounds on non-gaussianities from observations of the CMB and the Large Scale Structures (LSS). In the models studied in 
\cite{lorenzo, axion3}, the GW emitted by the gauge field may nevertheless be observable by ground-based interferometers, but not by eLISA, which probes scales that are closer to the ones probed by the CMB and LSS.

Another example of inflationary GW produced by a non-zero anisotropic stress that has been discussed in the literature occurs if a first-order phase transition takes place during inflation~\cite{amendola, chialva}. Again, one must check that the phase transition does not spoil the successful predictions of inflation. Different sources of GW from an inflationary first-order phase transition have been considered in \cite{chialva}, where it was concluded that the signal is dominated by the collision of vacuum bubbles. Ref.~\cite{chialva} also finds that the phase transition must complete relatively quickly, so that the parameter $\beta$ introduced in section~\ref{general} is significantly larger than the Hubble rate $H$ during inflation. The GW wavelength at the time of production, which is set by $\beta^{-1}$, is therefore well inside the Hubble radius. As discussed above, the GW are then significantly diluted by the inflationary expansion before they leave the Hubble radius. A further suppression of the GW amplitude comes from the fact that the energy density in the bubble walls is much smaller than the total energy density driving inflation. Ref.~\cite{chialva} then concludes that a negligible amount of GW is produced if the number of phase transitions is not very large. On the other hand, it is found in that paper that 
in inflationary models involving many first-order phase transitions, the GW could be marginally observable by LISA, which can perhaps be extended to the case of eLISA as well. Note that, because of the peculiar redshift of the modes during inflation, the shape of the GW spectrum differs from the one generated by first-order phase transitions occurring after inflation.

GW produced by the parametric amplification of quantum fluctuations, but with an enhanced amplitude at high frequencies, have also been discussed in the context of the so-called pre-big-bang~\cite{prebigbang} and ekpyrotic/cyclic 
universe~\cite{cyclic} scenarios, which have been proposed as alternatives to the standard inflationary paradigm. Although qualitatively different, these two kinds of scenarios usually involve (in the Einstein frame) an epoch of accelerated contraction of the universe before the expanding era. Throughout the contracting phase, during which the tensor modes leave the Hubble radius, the background energy density increases with time, as opposed to the slow decrease taking place during slow-roll inflation. As a consequence, the GW spectrum today increases with the frequency. The amplitude is then negligible at CMB scales and is constrained by BBN at high frequencies. Depending on the evolution that is assumed in between the contracting and expanding phases, the GW may be observable by direct detection experiments, see~\cite{maggiore, buonanno} for details. In between these two phases, the effective description breaks down and the dynamics is unknown. A better understanding of this regime is required in order to obtain firm predictions for the GW spectrum, as well as to assess the viability of these scenarios as alternatives to inflation.

\subsection{Gravitational Waves from the End of Inflation}
\label{justafter}

We now move on to GW produced after inflation. In fact, another privileged epoch for GW production is the end of inflation itself, or just after it, when the potential energy density driving inflation is converted into the thermal bath of the Hot Big Bang in the course of reheating. 

In many inflationary models, reheating starts with an explosive and non-perturbative decay of the inflaton condensate into large, non-thermal fluctuations of itself and other bosonic fields coupled to it - a process called preheating~\cite{preheating}. The subsequent dynamics is dominated by a highly non-linear and turbulent-like evolution towards thermal equilibrium. The large field fluctuations amplified by preheating source a GW background that has been studied in several papers, see in particular \cite{PreGW1, PreGW2, PreGW3, GWMeth, GWHyb, GWVec, GWLatt}. Similarly to Eq.~(\ref{kstarg}), preheating emits GW with a characteristic physical wave-number $k_* = H_* / \epsilon_*$ at the time of production, where 
$\epsilon_* \leq 1$ depends on the preheating mechanism and $H_*$ is the Hubble rate during preheating, which is directly related to the energy scale $E_{\mathrm{infl}}$ during inflation~\footnote{Eq.~(\ref{kstarg}) is not directly applicable to preheating since by definition the universe is not in a thermal state and there is no definite temperature $T_*$. In models where the full reheating process completes in less than one Hubble time, the reheat temperature is of the order of the energy scale at the end of inflation. Even if the full reheating process takes a much longer time to complete, the equation of state of the universe jumps usually quickly towards $w \approx 1/3$ at the beginning of preheating~\cite{eos}. In both cases, Eq.~(\ref{kstarg}) with $T_*$ replaced by the energy scale at the end of inflation gives the correct order of magnitude for the peak frequency of the GW spectrum today.}. One then finds that the inflationary energy scale 
must be smaller than about $E_{\mathrm{infl}} \sim 10^{11}$ GeV for the peak of the spectrum to fall into the frequency range of ground-based interferometers ($f \sim 10^2$ Hz), and smaller than $E_{\mathrm{infl}} \sim 10^{7}$ GeV for the peak to fall into the eLISA frequency band ($f \sim 10^{-2}$ Hz). Thus GW from preheating could be observable if inflation occurs at sufficiently small energy scales, which is complementary to the case of the GW generated during inflation itself. Note that an inflationary phase at low energies (possibly following a first stage of inflation at higher energies) may have several cosmological advantages, e.g. preheating-induced baryo-genesis~\cite{PreBaryo} and magneto-genesis~\cite{PreMagneto}, and the avoidance of dangerous relics such as moduli and gravitinos~\cite{randall, thermal}. 

GW from preheating have been extensively studied with numerical lattice simulations as well as analytical methods, see 
\cite{GWMeth} for details. Similarly to Eq.~(\ref{OmgwPT}), the peak amplitude depends essentially on 
$\epsilon_*$ and on the fraction of the total energy density that contributes to GW production. These two quantities can both be close to unity, which then maximizes the amount of GW produced. The peak frequency and the shape of the spectrum depend on the detail of the mechanism that is responsible for preheating, which in turn depends on the particular inflationary model that is considered. They also depend on the nature of the fields that are produced during preheating, in particular gauge fields lead to specific features in the GW spectrum~\cite{GWVec}. Two main classes of models have been studied so far: preheating after chaotic inflation and preheating after hybrid inflation. Another important class of models that may be interesting from the perspective of GW observation is preheating after small-field inflation~\cite{PreSmall}, but the resulting GW spectrum has not been studied with lattice simulations yet. In the following, we discuss in more detail the example of the GW from preheating after hybrid inflation. Chaotic inflation occurs typically at high energy scales, in which case the GW from preheating have a present-day frequency that is too high to be observable. 

In hybrid inflation models, the scalar field potential is of the form
\be
\label{PotHyb}
V = \frac{\lambda}{4}\,\left(|\chi|^2 - v^2\right)^2 + \frac{g^2}{2}\,|\chi|^2\,\phi^2 + V_{sr}(\phi)
\ee
where $\phi$ is the inflaton and $\chi$ is another scalar field coupled to it. These fields can be complex or have an arbitrary number of components, but this has little effect on the GW produced by preheating itself. For 
$\phi > \phi_c \equiv \sqrt{\lambda}\,v/g$, where $\phi_c = \sqrt{\lambda}\,v/g$ is called the critical point, the potential has a valley at $\chi = 0$. Inflation occurs when the inflaton slowly decreases in this valley due to the uplifting slow-roll term $V_{sr}(\phi)$. When $\phi = \phi_c$, the curvature of the potential becomes negative and the fields quickly roll towards the true minimum at $\phi = 0$ and $|\chi| = v$. During this rolling, quantum fluctuations of the fields have a negative effective mass squared and are exponentially amplified by a tachyonic instability~\cite{TachPre, TachPre2}. This quickly converts the homogeneous field energy into large fluctuations of the fields, which in turn emit GW. Three different regimes of preheating and GW production in the model (\ref{PotHyb}) were identified in \cite{GWHyb}, depending on the couplings $\lambda$ and $g^2$ and on the inflaton velocity at the critical point (the latter depends on the particular inflationary potential $V_{sr}(\phi)$). For instance, in the regime $g^2 \ll \lambda$, Ref.~\cite{GWHyb} obtains the following estimates for the peak frequency and amplitude of the GW spectrum today
\be
\label{g2lam}
\mbox{For } \, g^2 \ll \lambda: \hspace*{0.5cm} 
\left\{
\begin{array}{l}
f_{\mathrm{peak}} \, \sim \, \frac{g}{\lambda^{1/4}} \, 10^{10} \, \mathrm{Hz} \vspace*{0.4cm}\\
\left. h^2\,\Omega_{gw} \right|_{peak} \, \sim \, 10^{-5} \, \frac{\lambda}{g^2} \, G v^2
\end{array}
\right.
\ee
where in that case the condition $G v^2 < g^2 / \lambda$ must be satisfied. 

GW from preheating in the model (\ref{PotHyb}) are observable in different regions of the parameter space, but this requires very small values of the coupling constants: typically $g \sim \sqrt{\lambda} < 10^{-5}$ or $\lambda \sim 1$ and 
$g < 10^{-7}$, see \cite{GWHyb} for details. References to explicit particle-physics models of hybrid inflation were this may occur are given in \cite{GWHyb}. An example is provided by a variant~\cite{fastroll} of brane inflation in a warped throat~\cite{braninf}, which is a prototype of hybrid inflation in string theory. According to \cite{neil}, the end of inflation in this class of models can be described by the effective potential (\ref{PotHyb}) with $\lambda \sim 1$ and 
$g \sim e^{- A_b}$, where $e^{- A_b} \ll 1$ is the warp factor at the bottom of the throat. In that case, a very small value of $g$ may emerge naturally from the exponential warping of the throat, as small as $g \sim e^{- A_b} \sim 10^{-16}$ if the hierarchy problem of the Standard Model is addressed \emph{\`{a} la} Randall-Sundrum~\cite{RS}. The original model of \cite{braninf} uses only a moderate warping, but it also suffers from a conformal-coupling problem. A model that avoids this problem has been proposed in \cite{fastroll}, where $e^{- A_b} \sim 10^{-16} - 10^{-13}$ and $v \sim 1 - 1000$ TeV. In such a case, Eqs.~(\ref{g2lam}) indicates that the GW spectrum from preheating can be observable by eLISA.
 
In other inflationary models, inflation may also end with a vacuum (as opposed to thermal) first-order phase transition.  
In that case, bubble collision is an efficient source of GW, as already pointed out in \cite{Turner:1990rc}. Again, the peak frequency of these GW depends on the energy scale of inflation and can fall into an observable range if inflation occurs at sufficiently small energies. Recently, Ref.~\cite{wanil} studied the GW produced by bubble collision at the end of thermal inflation - an explicit inflationary model at low energy that may solve the cosmological moduli problem~\cite{thermal}. In this specific model, inflation is followed by a rather long matter-dominated stage before reheating completes and the radiation-dominated era starts. GW from bubble collision are then diluted by this matter-dominated stage and their peak amplitude today is found to be $h^2 \Omega_{\mathrm{gw}} \sim 10^{-18}$ in \cite{wanil}, which is too small to be observable by eLISA or ground-based interferometers.

\subsection{Gravitational Waves from the Hot Big Bang}
\label{thermevol}

As discussed above, the present-day frequency of the GW produced by the non-perturbative decay of the inflaton condensate (i.e. preheating) depends on the energy scale of inflation. However, the inflaton is not the only scalar field condensate that can decay in such a violent and highly inhomogeneous way. Another example is provided by scalar fields that acquire a large amplitude along flat directions of the potential in super-symmetric theories. The GW produced by the non-perturbative decay of super-symmetric flat direction condensates were studied in \cite{GWFlat}, where it was found that they can be observable by ground-based interferometers. Indeed these condensates start to oscillate and decay when the Hubble rate becomes of the order of their soft super-symmetry breaking mass $m$. For $m \sim$ TeV, this corresponds to a temperature $T_* \sim 10^{10}$ GeV and Eq.~(\ref{kstarg}) shows that the resulting GW fall naturally in the frequency range of 
ground-based interferometers ($\epsilon_*$ is close to unity in these models~\cite{GWFlat}). One difference with respect to preheating is that, while the inflaton condensate dominates the total energy density before it decays, flat direction condensates are typically sub-dominant. This may significantly reduce the GW amplitude, see the last factor in the RHS of Eq.~(\ref{OmgwPT}). The GW are then observable for flat direction condensates that have a sufficiently large initial amplitude, which occurs for directions that are indeed sufficiently flat. 

Condensates along super-symmetric flat directions may also fragment into non-topological solitons, called Q-balls. The GW produced by this process were studied in \cite{QBalls1, QBalls2}. A significant amount of GW can be produced for Q-balls with a large conserved charge. However, the decay rate of such Q-balls is small and they may come to dominate the energy density of the universe~\cite{QBalls2}. This leads to a matter-dominated stage that further dilutes the GW produced from 
Q-ball formation. Taking this effect into account, Ref.~\cite{QBalls2} obtains GW amplitudes that are too weak to be observable by eLISA or ground-based interferometers. 

GW can also be produced by unstable domain walls~\cite{DW1, DW2, DW3, DW4}. Domain walls originate from the spontaneous breaking of discrete symmetries. Once produced, they evolve towards a scaling regime, similarly to cosmic strings. However, stable domain walls would come to dominate the expansion of the universe and lead to cosmological disasters. This may be avoided if the discrete symmetry is only approximate. An example is provided by a real scalar field 
$\phi$ with low-energy potential
\be
\label{VDW}
V = \frac{\lambda}{4}\,\left(\phi^2 - v^2\right)^2 + \epsilon \, v \, \phi \, \left(\frac{\phi^2}{3} - v^2\right) \, ,
\ee
where $\epsilon$ is a numerical coefficient called ''bias'' and the second term in the potential lifts the degeneracy between the two minima at $\phi = \pm v$. This generates a pressure force on the walls that makes them collapse. The GW background from collapsing domain walls in the model (\ref{VDW}) has been studied with lattice simulations in 
\cite{DW1, DW3}. The GW spectrum can cover a very wide range of scales, from the width of the domain walls to the Hubble radius when they decay. This wide range of scales cannot be probed in lattice simulations, which must then be supplemented with analytical estimates. 

Ref.~\cite{DW2} estimates the amount of GW produced by the collapse of domain walls in the following way. In a scaling regime, one expects the curvature radius of the walls and the average distance between them to be set by the Hubble radius, with the domain wall energy density evolving as $\rho_{DW} \sim \sigma\,H$ where $\sigma \sim \sqrt{\lambda} v^3$ is the domain wall tension. Stable domain walls would then dominate the total energy density at the time $t_{dom}$ when the Hubble rate is $H_{dom} \sim G\,\sigma$. On the other hand, the bias term in Eq.~(\ref{VDW}) makes the domain wall collapse when the bias energy density $V_{\epsilon} \sim \epsilon\,v^4$ becomes of the order of $\rho_{DW}$, which occurs at the time 
$t_{dec}$ when the Hubble rate is $H_{dec} \sim V_{\epsilon} / \sigma$. The condition $H_{dec} > H_{dom}$ is then required in order for the domain walls to collapse before they come to dominate the total energy density. On dimensional grounds, one expects that, in one Hubble time around $t_{dec}$, the collapse of the domain walls emits GW with physical wave-numbers 
$k_* \sim H_{dec}$ and energy density $\Delta \rho_{gw} \sim G\,\sigma^2$, which corresponds to the fraction 
$\Delta \rho_{gw} / \rho_{tot} \, \sim \, G \, \sigma^2 / (H_{dec}^2 / G) \, \sim \, (H_{dom} / H_{dec})^2$ of the total energy density at the time of production. If the walls collapse during the radiation era, the amplitude of the GW spectrum today is then roughly estimated as
\be
h^2 \Omega_{gw}(f_{dec}) \, \sim \, 10^{-5} \, \left(\frac{H_{dom}}{H_{dec}}\right)^2
\ee
where the present-day frequency $f_{dec}$ is given by 
\be
f_{dec} \sim \left(\frac{T_{dec}}{\mathrm{TeV}}\right) \, 10^{-4} \, \mathrm{Hz} \, ,
\ee
see Eq.~(\ref{kstarg}). These estimates agree qualitatively with the numerical results of \cite{DW3}, where it is also found that the GW spectrum is approximately flat at frequencies $f > f_{dec}$. This suggests that GW from unstable domain walls could be observable by eLISA for $T_{dec} \lsim 10^3$ TeV if $H_{dom} / H_{dec}$ is not too small. The results depend strongly on the bias term in the potential, which in turn depends on the underlying particle-physics model. This has been studied in the context of gaugino condensation in \cite{DW2} and in the context of thermal inflation in \cite{DW4}.  

Another source of GW that has been considered in several papers~\cite{hoganRev, self1, self2, fenu, selfGiblin} is the 
self-ordering of massless (or light) scalar  degrees of freedom that may follow a global symmetry-breaking phase transition in the early universe. When a global symmetry is spontaneously broken, the massless Nambu-Goldstone bosons may lead to the production of GW at large, Hubble scales. An example is provided by a $N$-component scalar field $\bar{\phi}$ with $O(N)$ symmetry and low-energy potential
\be
\label{selfpot}
V = \frac{\lambda}{4}\,\left(|\bar{\phi}|^2 - v^2\right)^2 \, .
\ee  
During the phase transition, the scalar field settles in the true vacuum $|\bar{\phi}| = v$, but with different values of 
$\bar{\phi}$ in different Hubble patches because the process cannot be correlated on distances larger than the horizon distance at that time. As the Hubble radius grows during the subsequent cosmological evolution, the scalar field relaxes into a uniform configuration over the expanding Hubble patches. This self-ordering can lead to a continuous production of GW at the Hubble scale from the anisotropic stress generated by the gradients of the Nambu-Goldstone bosons. This process shares similarities with the continuous "production" of GW when the tensor modes generated by inflation re-enter the Hubble radius. The GW produced by the self-ordering in the radiation-dominated era lead to a flat spectrum today with 
$h^2 \Omega_{\mathrm{gw}} \propto G^2\,v^4$. Note that, compared to the case of inflation, the inflationary energy scale 
$E_{\mathrm{infl}}$ is replaced by the vacuum expectation value (VEV) $v$ of the scalar field.  

Because the spectrum is proportional to $G^2 \, v^4$, a very large VEV is required for these GW to be observable. The GW produced by this mechanism have been calculated analytically in \cite{self2, fenu} for the model (\ref{selfpot}) in the large $N$ limit, and with lattice simulations in \cite{selfGiblin}. Their results indicate that these GW could be observable by eLISA only if $v > 10^{17}$ GeV, which is larger than the maximal energy scale of inflation allowed by the current CMB bound. Nevertheless, the phase transition initiating the process could still occur after inflation if the potential is sufficiently flat ($\lambda \ll 1$ in Eq.~(\ref{selfpot})). On the other hand, as in the case of inflation, the spectrum can cover a very wide range of scales. If the source is active until GW at CMB scales are produced, then the spectrum can be subject to the COBE bound (see section~\ref{notations}). This can be avoided for instance if the relevant scalar modes are light instead of massless (e.g. pseudo-Nambu-Goldstone bosons if the broken global symmetry is only approximate). In such a case, we can expect that the source stops being active when the Hubble scale becomes of the order of the mass $m$ of the light modes. This would introduce a low-frequency cutoff in the GW spectrum today, at a frequency corresponding to the wave-number $k \sim m$ at the time of production. 

Finally, GW are also produced at second order in cosmological perturbation theory by scalar modes re-entering the Hubble radius after inflation, see \cite{scal1, scal2, scal3} and references therein. Since the scalar perturbations at CMB scales are observed, the GW they produce are "guaranteed". These GW have however a small amplitude since they are produced at second order by small scalar perturbations. On the other hand, the scalar perturbations can have a larger amplitude at smaller scales, where they are much less constrained. This would increase the amplitude of the GW spectrum they produce, which in turn can provide complementary constraints on the scalar perturbations and on inflationary models. The GW produced by large density perturbations leading to primordial black holes (PBH) was then studied in \cite{scalBH1}, see also 
\cite{scalBH2}. It is found in particular that pulsar timing observations can probe PBH with $\sim 10^2$ solar masses (which are candidates for intermediate-mass black holes), while eLISA could probe PBH in the mass range 
$10^{22} - 10^{25}$ g (which can play the role of dark matter).

\section{Conclusion}
\label{conclu}

In this paper, we have investigated the scientific potential of eLISA in the 
area of cosmological backgrounds. We have considered several potential sources 
of gravitational waves that could fall in the frequency band of the instrument, 
focusing on the ones that are best understood: first-order phase transitions 
and cosmic strings. 
We have also reviewed several other sources that operate at different epochs 
in the cosmological history. In most cases, the frequency dependence of the gravitational 
wave spectrum can be estimated reliably, which allows to disentangle different cosmological 
signals from each other and from astrophysical or instrumental backgrounds.


In Section~\ref{PT}, we analyzed the GW spectrum generated by first-order phase transitions, through the collision of bubbles and MHD turbulence. The analysis combines in a consistent way the most recent results of the literature, and provides therefore an up-to-date and realistic estimate of the GW signal. The analytic formulas of the GW spectra from bubble collisions and MHD turbulence, as a function of frequency and of the PT parameters $\alpha$, $\beta/H_*$, $\eta$ and $T_*$, are taken from the latest analyses of Huber and Konstandin 2008 \cite{arXiv:0806.1828} and Caprini et al. 2009 \cite{arXiv:0909.0622}. For the first time the bubble wall velocity $v_b$ and the efficiency factor $\kappa$ have been evaluated beyond the Jouguet detonation hypothesis, adopting the complete model for bubble propagation developed in Espinosa et al. 2010 \cite{Espinosa:2010hh}. This causes the GW signal to depend also on the friction $\eta$, acting on the bubble wall because of its interactions with the surrounding plasma. Moreover, the efficiency factor $\kappa$ has been constructed to account for both the gradient energy of the expanding wall and the bulk kinetic energy of the plasma in a continuous way, and it is dominated by one or the other depending on the strength of the PT: therefore, the contribution to the GW signal from bubble collisions and/or MHD turbulence is modeled consistently for each value of $\alpha$. We also point out that, in a realistic model of the PT, the parameters $\alpha$ and $\beta/H_*$ cannot be considered as independent (see for example Huber and Kostandin 2008 \cite{arXiv:0709.2091}). We account for this fact when we evaluate the GW signal due to specific models of the PT, and show that it can decrease considerably the chances for detection. 

From our analysis of the GW spectra we provide detection forecasts in the case of a moderately strong 
(Fig.~\ref{alphabetasurH}) and of a very strong (Fig.~\ref{TbetasurH}) first order PT. A key result is that eLISA will probe the presence of strongly first order phase transitions in the energy range from $100$ GeV to $10$ TeV or more, therefore complementing and even superseding the searches at colliders such as the LHC. We have given explicit examples, such as the holographic phase transition, where the signal lies well above the sensitivity of eLISA. There is of course no guaranteed detection but any such signal would point towards a strongly first-order phase transition: the frequency dependence is well understood, with a $f^3$ behavior at low frequency (due to the causal nature of process generating the GW), and a high-frequency tail given by the combination of the $f^{-1}$ slope from bubble collisions and of the $f^{-5/3}$ slope from MHD turbulence. If the phase transition occurs in the $10$ to $100$ TeV temperature range, eLISA might be the only way to identify it, and with a large signal-to-noise ratio if the transition is slow enough (lasting for a tenth of the Hubble time or more).

In Section \ref{Strings}, we focused on the gravitational wave spectrum produced by a network of cosmic string loops. We used an improved model for the cosmological evolution and we studied the dependence of the results on the thermal history of the early universe. We checked that the gravitational wave background does not depend much on the spectrum of emission by individual loops (see Figs.~\ref{cuspkinkwithwithout} and \ref{comparaisoncuspkinkfund}). We also found that removing the rare bursts is numerically relevant only for cuspy loops with a small initial size and in the region around the peak. As in the case of first-order phase transitions, the spectral dependance of the signal from cosmic strings is well determined. The high-frequency part of the spectrum (which is produced during the radiation era) is always almost flat over many decades of frequency (with small deviations from a perfectly flat spectrum arising from the variation of the number of relativistic species during the expansion of the universe). In the case of small initial loop sizes 
($\alpha \lesssim \Gamma G \mu$), there is a well-distinct peak at low frequencies, with a very steep infrared tail and a tail at higher frequencies that depends (relatively weakly) on the spectrum of emission by individual loops. In the case of large initial loop sizes ($\alpha \sim 0.1$), the peak is less pronounced and its infrared tail varies more slowly, as 
$\Omega_{gw} \propto f^{3/2}$. 

An important feature of the background from cosmic strings is that it extends over many decades of frequency, because the source remains active during a long period of time. This may enable to detect the spectrum simultaneously with different experiments (pulsars timing measurements, eLISA, ground-based interferometers), which would improve the characterization of the signal and the ability to distinguish it from other stochastic backgrounds. We addressed the question of the detectability of the cosmic string signal by different experiments and in particular by eLISA. As shown in Figs.~\ref{paramspacesmall} and \ref{paramspace}, eLISA will be able to probe a wide portion of the parameter space. In a large part of the parameter space, the signal should be simultaneously accessible to at least one other experiment. In other regions, eLISA will be the only instrument capable of detecting the signal, thus opening a new window on the parameter space of cosmic strings. In the case of small loops, eLISA should be able to reach string tensions down to 
$G\mu \sim \text{few} \times 10^{-10}$ for $p = 1$ and $G\mu \sim \text{few} \times 10^{-13}$ for $p=10^{-3}$, in an appropriate range of loop sizes. For large loops, eLISA will be sensitive to $G\mu \sim 10^{-13}$ for $p = 1$, and even 
$G\mu \sim 10^{-17}$ for $p = 10^{-3}$.

Our main conclusion is that the scientific potential of eLISA regarding the 
detection of cosmological backgrounds is not significantly decreased from the 
original LISA mission. Indeed, the decrease in sensitivity is much lower than 
the uncertainties that remain on the magnitude of the overall signal, due mainly to model dependence. This is why we have been careful in specifying the models chosen and the hypotheses made. As discussed in section \ref{Other}, eLISA has also the potential to probe a rich variety of other physical phenomena and new ideas.

\section*{Acknowledgments}

It is a pleasure to thank Thomas Konstandin for helping in the determination of the EWPT parameters, Andrea Lommen and Marco Peloso for useful correspondance, Antoine Petiteau for useful discussions and for providing the eLISA sensitivity curve, and G\'{e}raldine Servant for reading part of the manuscript. C.C. also thanks C\'{e}dric Delaunay, Ruth Durrer and Jos\'{e}-Miguel No for interesting discussions. Part of this work was supported by funds for the ANR LISAScience and 
from CNES through the LISA-France consortium.




\newpage



\begin{thebibliography}{99}

\bibitem{yellowbook}
 P.~Amaro-Seoane {\it et al.},
  arXiv:1201.3621 [astro-ph.CO].

\bibitem{allen}
  B.~Allen,
  arXiv:gr-qc/9604033.

\bibitem{hoganRev}
C.~J.~Hogan,
  arXiv:astro-ph/9809364. 
C.~J.~Hogan,
  AIP Conf.\ Proc.\  {\bf 873}, 30 (2006)
  [arXiv:astro-ph/0608567].

\bibitem{maggiore}
  M.~Maggiore,
  Phys.\ Rept.\  {\bf 331} (2000) 283
  [arXiv:gr-qc/9909001].

\bibitem{buonanno}
  A.~Buonanno,
  arXiv:gr-qc/0303085.

\bibitem{CMP}
  E.~J.~Copeland, R.~C.~Myers and J.~Polchinski,
  JHEP {\bf 0406} (2004) 013
  [arXiv:hep-th/0312067].

\bibitem{jenet}
  F.~A.~Jenet {\it et al.},
  Astrophys.\ J.\  {\bf 653}, 1571 (2006)
  [arXiv:astro-ph/0609013].

\bibitem{ligo}
  B.~P.~Abbott {\it et al.}  [LIGO Scientific Collaboration and VIRGO
                  Collaboration],
  Nature {\bf 460}, 990 (2009)
  [arXiv:0910.5772 [astro-ph.CO]].

\bibitem{Hogan:2001jn}
  C.~J.~Hogan and P.~L.~Bender,
  Phys.\ Rev.\  D {\bf 64} (2001) 062002
  [arXiv:astro-ph/0104266].





    
\bibitem{arXiv:0906.3434}
  D.~J.~Schwarz and M.~Stuke,
  JCAP\ {\bf 0911} (2009) 025
   [Erratum-ibid.\ \ {\bf 1010} (2010) E01]
  [arXiv:0906.3434 [hep-ph]].
  
\bibitem{arXiv:0909.0622}
  C.~Caprini, R.~Durrer and G.~Servant,
  JCAP\ {\bf 0912} (2009) 024
  [arXiv:0909.0622 [astro-ph.CO]].

\bibitem{arXiv:1001.3694}
  T.~Stevens and M.~B.~Johnson,
  arXiv:1001.3694 [Unknown].
  
\bibitem{hep-ph/9507429}
  G.~Baym, D.~Bodeker and L.~D.~McLerran,
  Phys.\ Rev.\ D\ {\bf 53} (1996) 662
  [hep-ph/9507429].

\bibitem{arXiv:1007.1218}
  C.~Caprini, R.~Durrer and X.~Siemens,
  Phys.\ Rev.\ D\ {\bf 82} (2010) 063511
  [arXiv:1007.1218 [astro-ph.CO]].

\bibitem{arXiv:1005.5291}
  C.~Caprini,
  arXiv:1005.5291 [astro-ph.CO].

\bibitem{Witten:1984rs}
  E.~Witten,
  Phys.\ Rev.\  D {\bf 30}, (1984) 272.

\bibitem{hogan86}
C.~J.~Hogan,
M.~N.~R.~A.~S., {\bf 218}, (1986) 629

\bibitem{Turner:1990rc}
  M.~S.~Turner and F.~Wilczek,
  Phys.\ Rev.\ Lett.\  {\bf 65} (1990) 3080.

\bibitem{FERMILAB-PUB-91-323-A}
  A.~Kosowsky, M.~S.~Turner and R.~Watkins,
  Phys.\ Rev.\ D\ {\bf 45} (1992) 4514.

\bibitem{astro-ph/9211004}
  A.~Kosowsky and M.~S.~Turner,
  Phys.\ Rev.\ D\ {\bf 47} (1993) 4372
  [astro-ph/9211004].

\bibitem{astro-ph/9310044}
  M.~Kamionkowski, A.~Kosowsky and M.~S.~Turner,
  Phys.\ Rev.\ D\ {\bf 49} (1994) 2837
  [astro-ph/9310044].
  
\bibitem{Steinhardt:1981ct}
  P.~J.~Steinhardt,
  Phys.\ Rev.\  D {\bf 25} (1982) 2074.
  
\bibitem{Espinosa:2010hh}
  J.~R.~Espinosa, T.~Konstandin, J.~M.~No and G.~Servant,
  JCAP {\bf 1006} (2010) 028
  [arXiv:1004.4187 [hep-ph]].

\bibitem{Caprini:2007xq}
  C.~Caprini, R.~Durrer and G.~Servant,
  Phys.\ Rev.\  D {\bf 77}, 124015 (2008)
  [arXiv:0711.2593 [astro-ph]].
  
\bibitem{arXiv:0901.1661}
  C.~Caprini, R.~Durrer, T.~Konstandin and G.~Servant,
  Phys.\ Rev.\ D\ {\bf 79} (2009) 083519
  [arXiv:0901.1661 [astro-ph]].
  
\bibitem{arXiv:0806.1828}
  S.~J.~Huber and T.~Konstandin,
  JCAP\ {\bf 0809} (2008) 022
  [arXiv:0806.1828 [hep-ph]].
  
\bibitem{astro-ph/0111483}
  A.~Kosowsky, A.~Mack and T.~Kahniashvili,
  Phys.\ Rev.\ D\ {\bf 66} (2002) 024030
  [astro-ph/0111483].

\bibitem{astro-ph/0206461}
  A.~D.~Dolgov, D.~Grasso and A.~Nicolis,
  Phys.\ Rev.\ D\ {\bf 66} (2002) 103505
  [astro-ph/0206461].
  
\bibitem{gr-qc/0303084}
  A.~Nicolis,
  Class.\ Quant.\ Grav.\ \ {\bf 21} (2004) L27
  [gr-qc/0303084].

\bibitem{astro-ph/0603476}
  C.~Caprini and R.~Durrer,
  Phys.\ Rev.\ D\ {\bf 74} (2006) 063521
  [astro-ph/0603476].
  
\bibitem{astro-ph/0607651}
  C.~Caprini, R.~Durrer and R.~Sturani,
  Phys.\ Rev.\ D\ {\bf 74} (2006) 127501
  [astro-ph/0607651].
  
\bibitem{arXiv:0705.1733}
  G.~Gogoberidze, T.~Kahniashvili and A.~Kosowsky,
  Phys.\ Rev.\ D\ {\bf 76} (2007) 083002
  [arXiv:0705.1733 [astro-ph]].
  
\bibitem{arXiv:0804.0391}
  A.~Megevand,
  Phys.\ Rev.\ D\ {\bf 78} (2008) 084003
  [arXiv:0804.0391 [astro-ph]].

\bibitem{arXiv:0802.3524}
  T.~Kahniashvili, G.~Gogoberidze and B.~Ratra,
  Phys.\ Rev.\ Lett.\ \ {\bf 100} (2008) 231301
  [arXiv:0802.3524 [astro-ph]].

\bibitem{arXiv:0809.1899}
  T.~Kahniashvili, L.~Campanelli, G.~Gogoberidze, Y.~Maravin and B.~Ratra,
  Phys.\ Rev.\ D\ {\bf 78} (2008) 123006
   [Erratum-ibid.\ D\ {\bf 79} (2009) 109901]
  [arXiv:0809.1899 [astro-ph]].

\bibitem{astro-ph/0505628}
  T.~Kahniashvili, G.~Gogoberidze and B.~Ratra,
  Phys.\ Rev.\ Lett.\ \ {\bf 95} (2005) 151301
  [astro-ph/0505628].
  
\bibitem{hel}J.~M.~Cornwall,
  Phys.\ Rev.\  D {\bf 56}, 6146 (1997)
  [arXiv:hep-th/9704022]; 
  M. Joyce, M. Shaposhnikov
Phys. Rev. Lett. {\bf 79}, 1193 (1997) [arXiv:astro-ph/9703005]; 
T. Vachaspati,
Phys. Rev. Lett. {\bf 87}, 251302 (2001) [arXiv:astro-ph/0101261]; 
C.J. Copi, F. Ferrer, T. Vachaspati and A. Achucarro
 Phys. Rev. Lett.{\bf 101}, 171302 (2008)  [arXiv:0801.3653].

\bibitem{arXiv:0906.4976}
  C.~Caprini, R.~Durrer and E.~Fenu,
  JCAP\ {\bf 0911} (2009) 001
  [arXiv:0906.4976 [astro-ph.CO]].

\bibitem{hep-ph/0102140}
  R.~Apreda, M.~Maggiore, A.~Nicolis and A.~Riotto,
  Class.\ Quant.\ Grav.\ \ {\bf 18} (2001) L155
  [hep-ph/0102140].
  
\bibitem{gr-qc/0107033}
  R.~Apreda, M.~Maggiore, A.~Nicolis and A.~Riotto,
  Nucl.\ Phys.\ B\ {\bf 631} (2002) 342
  [gr-qc/0107033].
  
\bibitem{arXiv:0709.2091}
  S.~J.~Huber and T.~Konstandin,
  JCAP\ {\bf 0805} (2008) 017
  [arXiv:0709.2091 [hep-ph]].
  
\bibitem{Grojean:2004xa}
  C.~Grojean, G.~Servant and J.~D.~Wells,
  Phys.\ Rev.\ D {\bf 71} (2005) 036001
  [hep-ph/0407019].
  
\bibitem{arXiv:0711.2511}
  C.~Delaunay, C.~Grojean and J.~D.~Wells,
  JHEP\ {\bf 0804} (2008) 029
  [arXiv:0711.2511 [hep-ph]].
  
\bibitem{hep-ph/0701145}
  J.~R.~Espinosa and M.~Quiros,
  Phys.\ Rev.\ D\ {\bf 76} (2007) 076004
  [hep-ph/0701145].

\bibitem{arXiv:0809.3215}
  J.~R.~Espinosa, T.~Konstandin, J.~M.~No and M.~Quiros,
  Phys.\ Rev.\ D\ {\bf 78} (2008) 123528
  [arXiv:0809.3215 [hep-ph]].
  
\bibitem{arXiv:0911.0687}
  J.~Kehayias and S.~Profumo,
  JCAP\ {\bf 1003} (2010) 003
  [arXiv:0911.0687 [hep-ph]].

\bibitem{hep-ph/0607158}
  L.~Randall and G.~Servant,
  JHEP\ {\bf 0705} (2007) 054
  [hep-ph/0607158].
  
\bibitem{arXiv:1007.1468}
  T.~Konstandin, G.~Nardini and M.~Quiros,
  Phys.\ Rev.\ D\ {\bf 82} (2010) 083513
  [arXiv:1007.1468 [hep-ph]].
  
\bibitem{arXiv:1104.4791}
  T.~Konstandin and G.~Servant,
  arXiv:1104.4791 [hep-ph].
  
\bibitem{hep-ph/0607107}
  C.~Grojean and G.~Servant,
  Phys.\ Rev.\ D\ {\bf 75} (2007) 043507
  [hep-ph/0607107].
  
\bibitem{Leitao:2010yw}
  L.~Leitao and A.~Megevand,
  Nucl.\ Phys.\ B {\bf 844} (2011) 450
  [arXiv:1010.2134 [astro-ph.CO]].

\bibitem{Turner:1992tz}
  M.~S.~Turner, E.~J.~Weinberg and L.~M.~Widrow,
  Phys.\ Rev.\ D {\bf 46} (1992) 2384.
  
\bibitem{Megevand:2009ut}
  A.~Megevand and A.~D.~Sanchez,
  Nucl.\ Phys.\ B {\bf 820} (2009) 47
  [arXiv:0904.1753 [hep-ph]].

\bibitem{Moore:1995ua}
  G.~D.~Moore and T.~Prokopec,
  Phys.\ Rev.\ Lett.\  {\bf 75} (1995) 777
  [hep-ph/9503296].
  
\bibitem{Moore:1995si}
  G.~D.~Moore and T.~Prokopec,
  Phys.\ Rev.\ D {\bf 52} (1995) 7182
  [hep-ph/9506475].

\bibitem{Ignatius:1993qn}
  J.~Ignatius, K.~Kajantie, H.~Kurki-Suonio and M.~Laine,
  Phys.\ Rev.\ D {\bf 49} (1994) 3854
  [astro-ph/9309059].

\bibitem{Megevand:2009gh}
  A.~Megevand and A.~D.~Sanchez,
  Nucl.\ Phys.\ B {\bf 825} (2010) 151
  [arXiv:0908.3663 [hep-ph]].
  
\bibitem{John:2000zq}
  P.~John and M.~G.~Schmidt,
  Nucl.\ Phys.\ B {\bf 598} (2001) 291
   [Erratum-ibid.\ B {\bf 648} (2003) 449]
  [hep-ph/0002050].




\bibitem{kibble}
T.~W.~B.~Kibble,
  J.\ Phys.\ A  {\bf 9}, 1387 (1976).

\bibitem{VS}
A.~Vilenkin and E.~P.~S.~Shellard, \emph{Cosmic Strings and Other Topological Defects} 
  (Cambridge University Press, Cambridge, 1994).

\bibitem{HK}
M.~B.~Hindmarsh and T.~W.~B.~Kibble,
  Rept.\ Prog.\ Phys.\  {\bf 58}, 477 (1995)
  [arXiv:hep-ph/9411342].

\bibitem{NO}
H.~B.~Nielsen and P.~Olesen,
  Nucl.\ Phys.\  B {\bf 61} (1973) 45.

\bibitem{superstring}
E.~Witten,
  Phys.\ Lett.\  B {\bf 153}, 243 (1985).

\bibitem{majumdar}
M.~Majumdar and A.~Christine-Davis,
  JHEP {\bf 0203}, 056 (2002)
  [arXiv:hep-th/0202148].

\bibitem{tye}
N.~T.~Jones, H.~Stoica and S.~H.~H.~Tye,
  JHEP {\bf 0207}, 051 (2002)
  [arXiv:hep-th/0203163]. 
S.~Sarangi and S.~H.~H.~Tye,
  Phys.\ Lett.\  B {\bf 536}, 185 (2002)
  [arXiv:hep-th/0204074]. 
N.~T.~Jones, H.~Stoica and S.~H.~H.~Tye,
  Phys.\ Lett.\  B {\bf 563}, 6 (2003)
  [arXiv:hep-th/0303269].

\bibitem{dvalenkin}
G.~Dvali and A.~Vilenkin,
  JCAP {\bf 0403}, 010 (2004)
  [arXiv:hep-th/0312007].

\bibitem{psuper}
M.~G.~Jackson, N.~T.~Jones and J.~Polchinski,
  JHEP {\bf 0510}, 013 (2005)
  [arXiv:hep-th/0405229].

\bibitem{wyman}
  M.~Wyman, L.~Pogosian and I.~Wasserman,
  Phys.\ Rev.\  D {\bf 72}, 023513 (2005)
  [Erratum-ibid.\  D {\bf 73}, 089905 (2006)]
  [arXiv:astro-ph/0503364].

\bibitem{KKmodes}
  J.~F.~Dufaux,
  arXiv:1109.5121 [hep-th].
  J.~F.~Dufaux,
  arXiv:1201.4850 [hep-th].

\bibitem{early1}
A.~Vilenkin,
  Phys.\ Lett.\  B {\bf 107}, 47 (1981).

\bibitem{early2}
C.~J.~Hogan and M.~J.~Rees,
  Nature {\bf 311}, 109 (1984).

\bibitem{early3}
T.~Vachaspati and A.~Vilenkin,
  Phys.\ Rev.\  D {\bf 31}, 3052 (1985).

\bibitem{early4}
R.~H.~Brandenberger, A.~Albrecht and N.~Turok,
  Nucl.\ Phys.\  B {\bf 277}, 605 (1986).

\bibitem{early5}
F.~S.~Accetta and L.~M.~Krauss,
  Nucl.\ Phys.\  B {\bf 319}, 747 (1989).

\bibitem{early6}
D.~P.~Bennett and F.~R.~Bouchet,
  Phys.\ Rev.\  D {\bf 43}, 2733 (1991).

\bibitem{caldwell}
R.~R.~Caldwell and B.~Allen,
  Phys.\ Rev.\  D {\bf 45}, 3447 (1992).

\bibitem{shellard}
R.~R.~Caldwell, R.~A.~Battye and E.~P.~S.~Shellard,
  Phys.\ Rev.\  D {\bf 54}, 7146 (1996)
  [arXiv:astro-ph/9607130].

\bibitem{DV}
T.~Damour and A.~Vilenkin,
  Phys.\ Rev.\ Lett.\  {\bf 85}, 3761 (2000)
  [arXiv:gr-qc/0004075]. 
T.~Damour and A.~Vilenkin,
  Phys.\ Rev.\  D {\bf 64}, 064008 (2001)
  [arXiv:gr-qc/0104026].

\bibitem{hnatyk}
V.~Berezinsky, B.~Hnatyk and A.~Vilenkin,
  arXiv:astro-ph/0001213.
V.~Berezinsky, B.~Hnatyk and A.~Vilenkin,
  Phys.\ Rev.\  D {\bf 64}, 043004 (2001)
  [arXiv:astro-ph/0102366].

\bibitem{DVsuper}
T.~Damour and A.~Vilenkin,
  Phys.\ Rev.\  D {\bf 71}, 063510 (2005)
  [arXiv:hep-th/0410222].

\bibitem{siemensbursts}
X.~Siemens, J.~Creighton, I.~Maor, S.~Ray Majumder, K.~Cannon and J.~Read,
  Phys.\ Rev.\  D {\bf 73}, 105001 (2006)
  [arXiv:gr-qc/0603115].

\bibitem{hogan}
C.~J.~Hogan,
  Phys.\ Rev.\  D {\bf 74}, 043526 (2006)
  [arXiv:astro-ph/0605567]. 
M.~R.~DePies and C.~J.~Hogan,
  Phys.\ Rev.\  D {\bf 75}, 125006 (2007)
  [arXiv:astro-ph/0702335].

\bibitem{siemens}
X.~Siemens, V.~Mandic and J.~Creighton,
  Phys.\ Rev.\ Lett.\  {\bf 98}, 111101 (2007)
  [arXiv:astro-ph/0610920]. 
S.~Olmez, V.~Mandic and X.~Siemens,
  Phys.\ Rev.\  D {\bf 81}, 104028 (2010)
  [arXiv:1004.0890 [astro-ph.CO]].

\bibitem{longstrings}
M.~Kawasaki, K.~Miyamoto and K.~Nakayama,
  Phys.\ Rev.\  D {\bf 81}, 103523 (2010)
  [arXiv:1002.0652 [astro-ph.CO]].

\bibitem{simu1}
C.~J.~A.~Martins and E.~P.~S.~Shellard,
  Phys.\ Rev.\  D {\bf 73}, 043515 (2006)
  [arXiv:astro-ph/0511792].

\bibitem{simu2}
C.~Ringeval, M.~Sakellariadou and F.~Bouchet,
  JCAP {\bf 0702}, 023 (2007)
  [arXiv:astro-ph/0511646].

\bibitem{simu3}
V.~Vanchurin, K.~D.~Olum and A.~Vilenkin,
  Phys.\ Rev.\  D {\bf 74}, 063527 (2006)
  [arXiv:gr-qc/0511159].
K.~D.~Olum and V.~Vanchurin,
  Phys.\ Rev.\  D {\bf 75}, 063521 (2007)
  [arXiv:astro-ph/0610419]. 

\bibitem{simu4}
J.~J.~Blanco-Pillado, K.~D.~Olum and B.~Shlaer,
  Phys.\ Rev.\  D {\bf 83}, 083514 (2011)
  [arXiv:1101.5173 [astro-ph.CO]].

\bibitem{ana1}
X.~Siemens, K.~D.~Olum and A.~Vilenkin,
  Phys.\ Rev.\  D {\bf 66}, 043501 (2002)
  [arXiv:gr-qc/0203006].

\bibitem{ana2}
J.~Polchinski and J.~V.~Rocha,
  Phys.\ Rev.\  D {\bf 74}, 083504 (2006)
  [arXiv:hep-ph/0606205].
J.~Polchinski and J.~V.~Rocha,
  Phys.\ Rev.\  D {\bf 75}, 123503 (2007)
  [arXiv:gr-qc/0702055].

\bibitem{ana3}
F.~Dubath, J.~Polchinski and J.~V.~Rocha,
  Phys.\ Rev.\  D {\bf 77}, 123528 (2008)
  [arXiv:0711.0994 [astro-ph]].

\bibitem{back1}
D.~P.~Bennett and F.~R.~Bouchet,
  ``Evidence for a Scaling Solution in Cosmic String Evolution,''
  Phys.\ Rev.\ Lett.\  {\bf 60}, 257 (1988).

\bibitem{back2}
X.~Siemens and K.~D.~Olum,
  Nucl.\ Phys.\  B {\bf 611}, 125 (2001)
  [Erratum-ibid.\  B {\bf 645}, 367 (2002)]
  [arXiv:gr-qc/0104085].

\bibitem{simupower} 
B.~Allen and E.~P.~S.~Shellard,
  Phys.\ Rev.\ D\ {\bf 45}, 1898  (1992).

\bibitem{kinkprolif}
P.~Binetruy, A.~Bohe, T.~Hertog and D.~A.~Steer,
  Phys.\ Rev.\  D {\bf 80}, 123510 (2009)
  [arXiv:0907.4522 [hep-th]].
P.~Binetruy, A.~Bohe, T.~Hertog and D.~A.~Steer,
  Phys.\ Rev.\  D {\bf 82}, 083524 (2010)
  [arXiv:1005.2426 [hep-th]].
P.~Binetruy, A.~Bohe, T.~Hertog and D.~A.~Steer,
  Phys.\ Rev.\  D {\bf 82}, 126007 (2010)
  [arXiv:1009.2484 [hep-th]].
A.~Bohe,
  Phys.\ Rev.\  D {\bf 84}, 065016 (2011)
  [arXiv:1103.0768 [hep-th]].

\bibitem{gregory}
E.~O'Callaghan, S.~Chadburn, G.~Geshnizjani, R.~Gregory and I.~Zavala,
  Phys.\ Rev.\ Lett.\  {\bf 105}, 081602 (2010)
  [arXiv:1003.4395 [hep-th]].
E.~O'Callaghan, S.~Chadburn, G.~Geshnizjani, R.~Gregory and I.~Zavala,
  JCAP {\bf 1009}, 013 (2010)
  [arXiv:1005.3220 [hep-th]].
E.~O'Callaghan and R.~Gregory,
  JCAP {\bf 1103}, 004 (2011)
  [arXiv:1010.3942 [hep-th]].

\bibitem{hybdef1}
X.~Martin and A.~Vilenkin,
  Phys.\ Rev.\ Lett.\  {\bf 77}, 2879 (1996)
  [astro-ph/9606022].

\bibitem{hybdef2}
L.~Leblond, B.~Shlaer and X.~Siemens,
  Phys.\ Rev.\  D {\bf 79}, 123519 (2009)
  [arXiv:0903.4686 [astro-ph.CO]].

\bibitem{depies} 
M.~RDePies and C.~JHogan,
  arXiv:0904.1052 [astro-ph.CO].

\bibitem{cosmopara}
O.~Lahav and A.~R.~Liddle,
  arXiv:1002.3488 [astro-ph.CO].

\bibitem{kolbturner}
E.~Kolb and M.~Turner, \emph{The Early Universe} (Addison-Wesley, 1990).

\bibitem{regimbau}
T.~Regimbau, S.~Giampanis, X.~Siemens and V.~Mandic,
  arXiv:1111.6638 [astro-ph.CO].

\bibitem{sakel}
  M.~Sakellariadou,
  JCAP {\bf 0504}, 003 (2005)
  [arXiv:hep-th/0410234].

\bibitem{avgou}
  A.~Avgoustidis and E.~P.~S.~Shellard,
  Phys.\ Rev.\  D {\bf 73}, 041301 (2006)
  [arXiv:astro-ph/0512582].

\bibitem{sesana}
  A.~Sesana, A.~Vecchio and C.~N.~Colacino,
  arXiv:0804.4476 [astro-ph].

\bibitem{acharya} 
  B.~S.~Acharya, G.~Kane and E.~Kuflik,
  arXiv:1006.3272 [hep-ph].


\bibitem{grishchuk}
L.~P.~Grishchuk,
  Sov.\ Phys.\ JETP {\bf 40}, 409 (1975)
  [Zh.\ Eksp.\ Teor.\ Fiz.\  {\bf 67}, 825 (1974)].

\bibitem{starobinsky}
A.~A.~Starobinsky,
  JETP Lett.\  {\bf 30} (1979) 682
  [Pisma Zh.\ Eksp.\ Teor.\ Fiz.\  {\bf 30} (1979) 719].

\bibitem{polarization}
M.~Kamionkowski, A.~Kosowsky and A.~Stebbins,
  Phys.\ Rev.\ Lett.\  {\bf 78}, 2058 (1997)
  [arXiv:astro-ph/9609132]. 
U.~Seljak and M.~Zaldarriaga,
  Phys.\ Rev.\ Lett.\  {\bf 78}, 2054 (1997)
  [arXiv:astro-ph/9609169].

\bibitem{smith}
T.~L.~Smith, M.~Kamionkowski and A.~Cooray,
  Phys.\ Rev.\  D {\bf 73}, 023504 (2006)
  [arXiv:astro-ph/0506422].

\bibitem{bbo}
S.Phinney \emph{et al.}, \emph{The big bang observer: direct detection of gravitational waves from the birth of the universe to the present}, NASA Mission Concept Study.

\bibitem{decigo}
N.~Seto, S.~Kawamura and T.~Nakamura,
  Phys.\ Rev.\ Lett.\  {\bf 87}, 221103 (2001)
  [arXiv:astro-ph/0108011].

\bibitem{stiff1}
P.~J.~E.~Peebles and A.~Vilenkin,
  Phys.\ Rev.\  D {\bf 59}, 063505 (1999)
  [arXiv:astro-ph/9810509]. 
M.~Giovannini,
  Phys.\ Rev.\  D {\bf 60}, 123511 (1999)
  [arXiv:astro-ph/9903004]. 
A.~Riazuelo and J.~P.~Uzan,
  Phys.\ Rev.\  D {\bf 62}, 083506 (2000)
  [arXiv:astro-ph/0004156].

\bibitem{stiff2}
L.~A.~Boyle and A.~Buonanno,
  Phys.\ Rev.\  D {\bf 78}, 043531 (2008)
  [arXiv:0708.2279 [astro-ph]].

\bibitem{axion1}
N.~Barnaby and M.~Peloso,
  Phys.\ Rev.\ Lett.\  {\bf 106}, 181301 (2011)
  [arXiv:1011.1500 [hep-ph]]. 
N.~Barnaby, R.~Namba and M.~Peloso,
  JCAP {\bf 1104}, 009 (2011)
  [arXiv:1102.4333 [astro-ph.CO]].

\bibitem{axion2}
L.~Sorbo,
  JCAP {\bf 1106}, 003 (2011)
  [arXiv:1101.1525 [astro-ph.CO]].

\bibitem{lorenzo}
J.~L.~Cook and L.~Sorbo,
  arXiv:1109.0022 [astro-ph.CO].

\bibitem{silverstein}
L.~Senatore, E.~Silverstein and M.~Zaldarriaga,
  arXiv:1109.0542 [hep-th].

\bibitem{axion3}
N.~Barnaby, E.~Pajer and M.~Peloso,
  arXiv:1110.3327 [astro-ph.CO].

\bibitem{amendola}
C.~Baccigalupi, L.~Amendola, P.~Fortini and F.~Occhionero,
  Phys.\ Rev.\  D {\bf 56}, 4610 (1997)
  [arXiv:gr-qc/9709044].

\bibitem{chialva}
D.~Chialva,
  Phys.\ Rev.\  D {\bf 83}, 023512 (2011)
  [arXiv:1004.2051 [astro-ph.CO]].

\bibitem{prebigbang}
M.~Gasperini and G.~Veneziano,
  Phys.\ Rept.\  {\bf 373}, 1 (2003)
  [arXiv:hep-th/0207130].

\bibitem{cyclic}
J.~Khoury, B.~A.~Ovrut, P.~J.~Steinhardt and N.~Turok,
  Phys.\ Rev.\  D {\bf 64}, 123522 (2001)
  [arXiv:hep-th/0103239]. 
J.~Khoury, B.~A.~Ovrut, N.~Seiberg, P.~J.~Steinhardt and N.~Turok,
  Phys.\ Rev.\  D {\bf 65}, 086007 (2002)
  [arXiv:hep-th/0108187].
P.~J.~Steinhardt and N.~Turok,
  Phys.\ Rev.\  D {\bf 65}, 126003 (2002)
  [arXiv:hep-th/0111098]. 

\bibitem{preheating}
L.~Kofman, A.~D.~Linde and A.~A.~Starobinsky,
  Phys.\ Rev.\ Lett.\  {\bf 73}, 3195 (1994).
L.~Kofman, A.~D.~Linde and A.~A.~Starobinsky,
  Phys.\ Rev.\ D {\bf 56}, 3258 (1997).

\bibitem{PreGW1}
S.~Y.~Khlebnikov and I.~I.~Tkachev,
  Phys.\ Rev.\ D {\bf 56}, 653 (1997).

\bibitem{PreGW2}
R.~Easther and E.~A.~Lim,
  JCAP {\bf 0604}, 010 (2006)
  [arXiv:astro-ph/0601617].
R.~Easther, J.~T.~.~Giblin and E.~A.~Lim,
  Phys.\ Rev.\ Lett.\  {\bf 99}, 221301 (2007)
  [arXiv:astro-ph/0612294].

\bibitem{PreGW3}
J.~Garc\'ia-Bellido and D.~G.~Figueroa,
  Phys.\ Rev.\ Lett.\  {\bf 98}, 061302 (2007)
  [arXiv:astro-ph/0701014].
J.~Garc\'ia-Bellido, D.~G.~Figueroa and A.~Sastre,
  Phys.\ Rev.\  D {\bf 77}, 043517 (2008)
  [arXiv:0707.0839 [hep-ph]].

\bibitem{GWMeth}
J.~F.~Dufaux, A.~Bergman, G.~N.~Felder, L.~Kofman and J.~P.~Uzan,
  [arXiv:0707.0875 [astro-ph]].

\bibitem{GWHyb}
J.~F.~Dufaux, G.~N.~Felder, L.~Kofman and O.~Navros,
  JCAP {\bf 0903}, 001 (2009)
  [arXiv:0812.2917 [astro-ph]].

\bibitem{GWVec}
J.~F.~Dufaux, D.~G.~Figueroa and J.~Garcia-Bellido,
  Phys.\ Rev.\  D {\bf 82}, 083518 (2010)
  [arXiv:1006.0217 [astro-ph.CO]].

\bibitem{GWLatt}
Z.~Huang,
  Phys.\ Rev.\  D {\bf 83}, 123509 (2011)
  [arXiv:1102.0227 [astro-ph.CO]].
D.~G.~Figueroa, J.~Garcia-Bellido and A.~Rajantie,
  JCAP {\bf 1111}, 015 (2011)
  [arXiv:1110.0337 [astro-ph.CO]].

\bibitem{eos}
D.~I.~Podolsky, G.~N.~Felder, L.~Kofman and M.~Peloso,
  Phys.\ Rev.\  D {\bf 73}, 023501 (2006)
  [arXiv:hep-ph/0507096].
J.~F.~Dufaux, G.~N.~Felder, L.~Kofman, M.~Peloso and D.~Podolsky,
  JCAP {\bf 0607}, 006 (2006)
  [arXiv:hep-ph/0602144].

\bibitem{PreBaryo}
J.~Garcia-Bellido, D.~Y.~Grigoriev, A.~Kusenko and M.~E.~Shaposhnikov,
  Phys.\ Rev.\  D {\bf 60}, 123504 (1999)
  [arXiv:hep-ph/9902449].
J.~Garcia-Bellido, M.~Garcia-Perez and A.~Gonzalez-Arroyo,
  Phys.\ Rev.\  D {\bf 69}, 023504 (2004)
  [arXiv:hep-ph/0304285].

\bibitem{PreMagneto}
A.~Diaz-Gil, J.~Garcia-Bellido, M.~Garcia Perez and A.~Gonzalez-Arroyo,
  Phys.\ Rev.\ Lett.\  {\bf 100}, 241301 (2008)
  [arXiv:0712.4263 [hep-ph]].
A.~Diaz-Gil, J.~Garcia-Bellido, M.~G.~Perez and A.~Gonzalez-Arroyo,
  JHEP {\bf 0807}, 043 (2008)
  [arXiv:0805.4159 [hep-ph]].

\bibitem{randall}
L.~Randall and S.~D.~Thomas,
  Nucl.\ Phys.\  B {\bf 449}, 229 (1995)
  [arXiv:hep-ph/9407248].

\bibitem{thermal}
D.~H.~Lyth and E.~D.~Stewart,
  Phys.\ Rev.\ Lett.\  {\bf 75}, 201 (1995)
  [arXiv:hep-ph/9502417].
D.~H.~Lyth and E.~D.~Stewart,
  Phys.\ Rev.\  D {\bf 53}, 1784 (1996)
  [arXiv:hep-ph/9510204].

\bibitem{PreSmall}
P.~Brax, J.~F.~Dufaux and S.~Mariadassou,
  Phys.\ Rev.\  D {\bf 83}, 103510 (2011)
  [arXiv:1012.4656 [hep-th]].

\bibitem{TachPre}
G.~N.~Felder, J.~Garcia-Bellido, P.~B.~Greene, L.~Kofman, A.~D.~Linde and I.~Tkachev,
  Phys.\ Rev.\ Lett.\  {\bf 87}, 011601 (2001)
  [arXiv:hep-ph/0012142].
G.~N.~Felder, L.~Kofman and A.~D.~Linde,
  Phys.\ Rev.\  D {\bf 64}, 123517 (2001)
  [arXiv:hep-th/0106179].

\bibitem{TachPre2}
  J.~Garcia-Bellido, M.~Garcia Perez and A.~Gonzalez-Arroyo,
  Phys.\ Rev.\  D {\bf 67}, 103501 (2003)
  [arXiv:hep-ph/0208228].

\bibitem{fastroll}
  L.~Kofman and S.~Mukohyama,
  Phys.\ Rev.\  D {\bf 77}, 043519 (2008)
  [arXiv:0709.1952 [hep-th]].

\bibitem{braninf}
  S.~Kachru, R.~Kallosh, A.~D.~Linde, J.~M.~Maldacena, L.~P.~McAllister and S.~P.~Trivedi,
  JCAP {\bf 0310}, 013 (2003)
  [arXiv:hep-th/0308055].

\bibitem{neil}
  N.~Barnaby and J.~M.~Cline,
  Phys.\ Rev.\  D {\bf 75}, 086004 (2007)
  [arXiv:astro-ph/0611750].

\bibitem{RS}
L.~Randall and R.~Sundrum,
  Phys.\ Rev.\ Lett.\  {\bf 83}, 3370 (1999)
  [arXiv:hep-ph/9905221].

\bibitem{wanil}
R.~Easther, J.~T.~.~Giblin, E.~A.~Lim, W.~I.~Park and E.~D.~Stewart,
  JCAP {\bf 0805}, 013 (2008)
  [arXiv:0801.4197 [astro-ph]].

\bibitem{GWFlat}
J.~F.~Dufaux,
  Phys.\ Rev.\ Lett.\  {\bf 103}, 041301 (2009)
  [arXiv:0902.2574 [astro-ph.CO]].

\bibitem{QBalls1}
A.~Kusenko and A.~Mazumdar,
  Phys.\ Rev.\ Lett.\  {\bf 101}, 211301 (2008)
  [arXiv:0807.4554 [astro-ph]].
A.~Kusenko, A.~Mazumdar and T.~Multamaki,
  Phys.\ Rev.\  D {\bf 79}, 124034 (2009)
  [arXiv:0902.2197 [astro-ph.CO]].

\bibitem{QBalls2}
  T.~Chiba, K.~Kamada and M.~Yamaguchi,
  Phys.\ Rev.\  D {\bf 81}, 083503 (2010)
  [arXiv:0912.3585 [astro-ph.CO]].

\bibitem{DW1}
  M.~Gleiser and R.~Roberts,
  Phys.\ Rev.\ Lett.\  {\bf 81}, 5497 (1998)
  [arXiv:astro-ph/9807260].

\bibitem{DW2}
  F.~Takahashi, T.~T.~Yanagida and K.~Yonekura,
  Phys.\ Lett.\  B {\bf 664}, 194 (2008)
  [arXiv:0802.4335 [hep-ph]].

\bibitem{DW3}
T.~Hiramatsu, M.~Kawasaki and K.~Saikawa,
  JCAP {\bf 1005}, 032 (2010)
  [arXiv:1002.1555 [astro-ph.CO]].
M.~Kawasaki and K.~Saikawa,
  JCAP {\bf 1109}, 008 (2011)
  [arXiv:1102.5628 [astro-ph.CO]].

\bibitem{DW4}
  T.~Moroi and K.~Nakayama,
  Phys.\ Lett.\  B {\bf 703}, 160 (2011)
  [arXiv:1105.6216 [hep-ph]].

\bibitem{self1}
  L.~M.~Krauss,
  Phys.\ Lett.\  B {\bf 284}, 229 (1992).

\bibitem{self2}
K.~Jones-Smith, L.~M.~Krauss and H.~Mathur,
  Phys.\ Rev.\ Lett.\  {\bf 100}, 131302 (2008)
  [arXiv:0712.0778 [astro-ph]].
L.~M.~Krauss, K.~Jones-Smith, H.~Mathur and J.~Dent,
  Phys.\ Rev.\  D {\bf 82}, 044001 (2010)
  [arXiv:1003.1735 [astro-ph.CO]].

\bibitem{fenu}
E.~Fenu, D.~G.~Figueroa, R.~Durrer and J.~Garcia-Bellido,
  JCAP {\bf 0910}, 005 (2009)
  [arXiv:0908.0425 [astro-ph.CO]].

\bibitem{selfGiblin}
J.~T.~.~Giblin, L.~R.~Price, X.~Siemens and B.~Vlcek,
  arXiv:1111.4014 [astro-ph.CO].

\bibitem{scal1}
  K.~N.~Ananda, C.~Clarkson and D.~Wands,
  Phys.\ Rev.\  D {\bf 75}, 123518 (2007)
  [arXiv:gr-qc/0612013].

\bibitem{scal2}
  D.~Baumann, P.~J.~Steinhardt, K.~Takahashi and K.~Ichiki,
  Phys.\ Rev.\  D {\bf 76}, 084019 (2007)
  [arXiv:hep-th/0703290].

\bibitem{scal3}
  H.~Assadullahi and D.~Wands,
  Phys.\ Rev.\  D {\bf 81}, 023527 (2010)
  [arXiv:0907.4073 [astro-ph.CO]].

\bibitem{scalBH1}
R.~Saito and J.~Yokoyama,
  Phys.\ Rev.\ Lett.\  {\bf 102}, 161101 (2009)
  [Erratum-ibid.\  {\bf 107}, 069901 (2011)]
  [arXiv:0812.4339 [astro-ph]].
R.~Saito and J.~Yokoyama,
  Prog.\ Theor.\ Phys.\  {\bf 123}, 867 (2010)
  [arXiv:0912.5317 [astro-ph.CO]].

\bibitem{scalBH2}
E.~Bugaev and P.~Klimai,
  Phys.\ Rev.\  D {\bf 81}, 023517 (2010)
  [arXiv:0908.0664 [astro-ph.CO]].
E.~Bugaev and P.~Klimai,
  Phys.\ Rev.\  D {\bf 83}, 083521 (2011)
  [arXiv:1012.4697 [astro-ph.CO]].

\end{thebibliography}
\end{document}